\documentclass[preprint,aps,amsmath,nofootinbib]{revtex4-1}
\usepackage{graphicx,array,dcolumn}
\usepackage{calc,tabularx, epsfig,mathrsfs}
\usepackage{hyperref}
\usepackage{subfigure}

\usepackage{amsmath,verbatim,enumerate}
\usepackage{amssymb}
\usepackage{amsmath}
\usepackage{wasysym}
\usepackage{xcolor}
\usepackage{multirow}
\usepackage{xspace}
\usepackage{slashed}
\usepackage{mathtools}
\usepackage{cancel}
\usepackage{tikz}
\usepackage{mathtools}
\allowdisplaybreaks[1]
\newlength{\figurewidth}
\newcommand{\beq}{\begin{equation}}
\newcommand{\eeq}{\end{equation}}
\newcommand{\bea}{\begin{eqnarray}}
\newcommand{\eea}{\end{eqnarray}}
\newcommand{\ba}{\begin{array}}
\newcommand{\ea}{\end{array}}

\newcommand{\mn}{{\mu\nu}}

\newcommand{\pt}{\partial}

\newcommand{\Bl}{\biggl}
\newcommand{\Br}{\biggr}
\newcommand{\bl}{\bigl}
\newcommand{\br}{\bigr}
\newcommand{\g}{\gamma}
\newcommand{\ep}{\epsilon}
\newcommand{\ta}{\theta}
\newcommand{\lam}{\lambda}
\newcommand{\Lam}{\Lambda}

\newcommand{\de}{\delta}
\newcommand{\D}{\Delta}

\newcommand{\om}{\omega}
\newcommand{\sg}{\sigma}

\DeclareMathOperator\arctanh{arctanh}
\makeatother
\begin{document}
%
\title{
Initial conditions leave IR imprints in quantum cosmology 
}
\setlength{\figurewidth}{\columnwidth}
%
\author{Manishankar Ailiga}
\email{manishankara@iisc.ac.in}

\author{Akash Mondal}
\email{akashmondal1@iisc.ac.in}

\author{Gaurav Narain}
\email{gnarain@iisc.ac.in}

\affiliation{
Center for High Energy Physics, Indian Institute of Science,
C V Raman Road, Bangalore 560012, India.
}
\vspace{3em}
\begin{abstract}
We investigate the gravitational path integral of $\mathbb{R} \times S^3$ geometries in Lorentzian signature along with the metric fluctuations which live on the $S^3$ hypersurfaces. Each metric configuration is characterised by lapse, scale factor and fluctuations, and is weighted by the exponential of the Einstein-Hilbert action. A particular boundary requirement for the scale-factor is achievable via an infinite family of covariant boundary actions, each restricting the boundary behaviour of the fluctuation field which cannot be arbitrarily chosen. The paper explores the effect of this infinite family of covariant boundary actions on the wavefunction of Universe to one-loop. These boundary actions are characterized by parameters $c$ and $\omega$. Dominant contribution of initial momentum in the path-integral restricts $\omega$ to be half-integer, leading to an infinite family of $c$-dependent conjugate momenta. In special case for fixed $\omega$, four such $c$-dependent dominant branches of conjugate momenta emerge: two lead to deSitter, one leads to Hartle-Hawking and one to the tunnelling Universe. For all these $c$-dependent initial conjugate momenta, the lapse-saddles are relevant, however only, one of deSitter saddle and Hartle Hawking saddle have well-behaved perturbations. The one-loop wavefunction at these well-behaved saddles is computed systematically by isolating the UV-divergences, where the covariant boundary parameter $c$ acts as a UV-regulator and also helps in shifting the poles of one-loop determinant, away from the real $q_f$ axis for dS saddles, where $q_f$ is the final size of the Universe. UV-finite wavefunction for no-boundary and dS is obtained by adding suitable counterterms. The infrared behaviour of these wavefunctions show not only IR-divergences but also sensitivity toward the initial boundary choices.

\end{abstract}
\vspace{5mm}


\maketitle

\tableofcontents


\section{Introduction}
\label{intro}
Boundary conditions play a crucial role in determining the physical dynamics of a system. In gravitational theories, for instance, the choice of boundary conditions imposed in the asymptotic region determines the corresponding thermodynamic ensemble. A well-known example in general relativity is the Gibbons–Hawking–York (GHY) boundary term, which is associated with fixing the induced metric at the asymptotic boundary, known as Dirichlet boundary condition. This leads to the canonical partition function describing the thermodynamics of the gravitational system \cite{Gibbons:1976ue}. Other choice of boundary conditions, such as Neumann, Robin, and Brown-York boundary conditions, correspond to the Neumann, mixed, and microcanonical ensembles, respectively \cite{Brown:1992bq, Krishnan:2016mcj, Krishnan:2017bte, Krishnan:2016tqj}. Thus, boundary terms encode essential physical information about the system.

In context of quantum cosmology, understanding the quantum origin of the universe is of
central significance. In particular, boundary conditions on space-like boundaries are of great importance, as they define the initial and final configurations of the spatial geometry between which the universe evolves. The well known boundary conditions in quantum cosmology are the “no-boundary proposal” , the “tunneling proposal” and `` Wine-glass" proposal \cite{Hartle:1983ai, Vilenkin:1987kf, Betzios:2024oli}. The associated functional form of the boundary action of the boundary conditions is of central importance in this context, since it contributes to the wavefunction and can lead to significant changes in its physical behavior. In particular, it was shown that, by imposing Neumann or Robin boundary condition, one obtains a stable saddle point geometry \cite{DiTucci:2019bui}. This is in contrast to the Dirichlet case, for which the resulting saddle-point geometries exhibit unstable perturbations on the final spatial hypersurface \cite{Lehners:2018eeo}. This shows the non-trivial role of boundary conditions in dictating the physical behavior of universe.

Generically, the wavefunction of the universe can be computed using the covariant path-integral formulation with appropriate boundary conditions. Alternatively, one may adopt the canonical approach by solving the Wheeler–DeWitt equation. In this work, we take the covariant approach, as it provides elegant manifestation of the underlying symmetries. Moreover, the gravitational path integral has proven to be an important diagnostic tool for understanding various aspects of quantum gravity. In particular, the path-integral framework provides a natural description of the semiclassical properties of the black hole thermodynamics, offers a good understanding of quantum chaos of gravitational system through spectral form factor, and in supersymmetric index computations that strongly support the bulk–boundary duality \cite{Gibbons:1976ue, Hawking:1982dh, Saad:2018bqo, Saad:2019lba, Benini:2015eyy}. Furthermore, in context of cosmology, loop corrections and contributions from non-trivial topologies can be easily captured utilizing gravitational path integral. These naturally supports our choice of covariant path-integral approach over canonical formalism, in which replicating the aforementioned features is considerably more involved. In this work, we compute the 1-loop wavefunction of the late-time deSitter universe, starting from various initial conditions utilizing Lorentzian gravitational path integral. From perspective of the path-integral the wavefunction is obtained by summing over all geometries satisfying the prescribed boundary conditions.

The gravitational path integral in the Lorentzian signature, on a manifold $\mathcal{M}$ with the space-like boundary $\partial\cal{M}$ is given by:
\beq
\label{eq:PI_LI_g}
\Psi[{{\rm Bd}_f},{\rm{Bd}_i}] = \int_{{\cal M} + \pt {\cal M}} {\cal D} g_\mn \, e^{iS[g_\mn]/\hbar} \, ,
\eeq
where $S[g_\mn]$ is the gravitational action and  ${{\rm Bd}_{i,f}}$ denotes the initial and final boundary conditions on $\partial \cal{M}$, respectively. In this work, we restrict the path integral over complex geometries in four dimensions with topology $\mathbb{R}\times S^3$, thereby defining the corresponding minisuperspace model. In this work, we consider the $\Lambda>0$ Einstein gravity with the gravitational action is given by:
\bea
\label{eq:grav_act}
S_{\rm grav}[g_\mn] = \frac{1}{16 \pi G} \int d^4x \sqrt{-g}\bigl[R
-2\Lambda  \bigr]
+ S_{\rm bd},
\eea
where $G$ is the Newton's gravitational constant, $\Lam$ is the cosmological constant, $R$ is the Ricci scalar of manifold $\cal{M}$ and $S_{\rm bd}$ is the boundary action needed for a consistent variational problem with the prescribed boundary conditions on $\partial \cal{M}$. 

Despite the appealing features of geometrical formulation, the gravitational path integral requires appropriate addressing of gauge fixing, choice of contour and renormalization. Gauge fixing is implemented through the Faddeev–Popov procedure, leading to the corresponding ghost contributions. We choose the integration contour to sum over complex geometries satisfying the prescribed boundary conditions with the $\mathbb{R}\times S^3$ topology.

This work primarily focuses on the regularization and renormalization of the gravitational path integral. In particular, the wavefunction defined in eq. (\ref{eq:PI_LI_g}) is confronted with issues at both ultraviolet (UV) and infrared (IR) scales. Both of them have different origins. The UV problem is intrinsic to the quantum gravity. Gravitational fluctuations generate the ultraviolet divergences in the perturbative expansion, reflecting the non-renormalizability of gravity in absence of higher-derivative terms \cite{tHooft:1974toh, Deser:1974nb, Deser:1974cz, Goroff:1985sz, Goroff:1985th, vandeVen:1991gw}. Interestingly, as we will see, the covariant boundary term required for a consistent variational principle plays an important role in regulating the early-time UV behavior and identifying the associated divergences. These divergences can then be systematically removed by introducing appropriate counterterms. In particular, for pure Einstein–Hilbert gravity without cosmological constant, the remaining one-loop divergences vanish on shell which reflects its on-shell one-loop renormalizability. 

The IR divergence, by contrast, originates at a completely different scale due to late-time deSitter asymptotics. The inflating background continuously stretches quantum fluctuations to super-Hubble scales, where their dynamics can become sensitive to the long-distance structure of spacetime. In perturbation theory, this can lead to secular growth and infrared enhancements \cite{ Akhmedov:2013vka, Akhmedov:2019cfd, Akhmedov:2024npw, Miao:2024shs, Starobinsky:1994bd}.
Moreover, it has been reported in literature that the IR divergences get sharper as the acceleration of universe slows down \cite{Konstantinidis:2016nio}. In this paper, we compute the wavefunction of a four dimensional complex universe evolving from an initial Euclidean hypersurface to a late-time Lorentzian (de Sitter) regime at the one-loop level. In particular, we study the universes that nucleate from near no-boundary and near de Sitter initial conditions.

We start by considering the spatially homogeneous and isotropic metric [which is the Friedmann–Lemaître–Robertson–Walker (FLRW) metric] in four spacetime dimensions along with the exponential metric fluctuations in the spatial part. In polar coordinates ${\bf x} = \{t_p, r, \ta, \cdots \}$, it is given by
\beq
\label{eq:frwmet}
{\rm d}s^2 = - N_c^2(t_p) {\rm d} t_p^2 
+ 
\underbrace{
a^2(t_p) \left[
\rho_{ik}\big(e^{h}\big)^{k}{}_{j}
\right]
}_{\g_{ij}} {\rm d}x^i \, {\rm d}x^j\, ,
\eeq
where $\rho_{ij}$ corresponds to the metric on $S^3$. We employ the exponential parametrization of the metric to avoid the appearance of unphysical singularities and to ensure consistency with earlier results \cite{Percacci:2015wwa}. An additional advantage of this choice is that the resulting beta functions are independent of the gauge-fixing condition, thereby providing a more robust and scheme-independent formulation.

The metric perturbations $h_{00}$ and $h_{0i}$ components have been gauge fixed to zero and will not be considered in the following. The dynamical variables appearing in the metric ansatz are lapse $N_c(t_p)$, scale factor $a(t_p)$ and fluctuation field $h_{ij}(t_p, x_m)$. The former two are only time ($t_p$) dependent while the fluctuation $h_{ij}$ is both time ($t_p$) and space ($x_m$) dependent. Imposing the proper-time gauge $N_c'(t_p)=0$, corresponding to the time-reparameterization invariance. For the tensor perturbations, we impose the transverse-traceless (TT) conditions, $\nabla^{i}h_{ij}=0$, with $\rho^{ij}h_{ij}=0$ which isolate physical degrees of freedom of graviton. With these gauge choices the lorentzian path integral (after implementing Faddeev-Popov procedure) in eq.(\ref{eq:PI_LI_g}) over the remaining fields is given by:
\beq
\label{eq:Gform_sch}
\Psi[{\rm Bd}_f, {\rm Bd}_i]
= \int_{\bf C} {\rm d}N_p
\int_{{\rm Bd}_i}^{{\rm Bd}_f} 
{\cal D} a(t_p) {\cal D} h_{ij} {\cal D} \bar{\eta}_i {\cal D} \eta^i\,\,
e^{i S_{\rm grav}[a, N_p, h_{ij}] + i S_{\rm gf} + i S_{\rm gh}/\hbar} \, ,
\eeq
where ${\bf C}$ refers to contour of integration. We take the contour to be $-\infty<N_c<+\infty$ that computes the wavefunction, instead $0< N_c<\infty$ that corresponds to propagator. Here, $S_{\rm gf}$ and $S_{\rm gh}$ are the gauge fixing and ghost action, which prevent over counting of the gauge orbits. At one-loop order, their contributions reduce to an overall numerical factor and therefore do not affect the physical conclusions of our analysis, see \cite{Mallik:2026iai} for details.

Our aim in this paper is to compute and study eq. (\ref{eq:Gform_sch}) for the gravitational theory given by eq. (\ref{eq:grav_act}), subject to the boundary conditions imposed on the scale factor $a(t_p)$ and $h_{ij}(t_p, \textbf{x})$. However, requirements of the general covariance restrict the freedom to choose the boundary conditions for the scale factor $a(t_p)$ and $h_{ij}(t_p, \textbf{x})$ independently, and the two are intimately tied to each other \cite{Brizuela:2023vmb, Ailiga:2024wdx}. It implies that if certain boundary conditions are imposed on the scale factor or its conjugate momenta, then requirements of general covariance constrains the available boundary choices for the fluctuation field $h_{ij}(t_p, \textbf{x})$. In this paper, we  explore and study this connection in more detail by fixing the conjugate momentum at the initial hypersurface, while field itself on the final boundary of background field. This choice of boundary conditions for the background field leads to an infinite family of covariant boundary terms, parameterized by a boundary parameter $(c)$ which is required for a well-defined variational principle. We focus on the simplest nontrivial choice and investigate the consequences of these boundary conditions on the boundary conditions of tensor perturbation field and associated gravitational path integral. Interestingly, as aforementioned, the parameter $c$, associated with the covariant boundary action required for consistent variational problem acts as UV regulator in the one-loop study of tensor modes.

Once the boundary conditions for the fields are found, the full gauge-fixed path integral over $a(t_p)$ and $h_{ij}(t_p,\textbf{x})$ is seen to decouple when backreaction is neglected equivalently for a special choice of perturbation field boundary conditions. However, the overall ordinary lapse integral still couples the two. The path integral over the scale factor can be done exactly in the case of Einstein-Hilbert gravity as the theory has only two time derivatives and the action is quadratic. The path integral over the fluctuation field is nontrivial as the action of the fluctuation field is no longer quadratic but will contain an infinite number of terms. As a result it is not possible to do an exact computation for the fluctuation path integral. However, up to one loop the path integral over the fluctuation field can be computed in which case one needs to retain terms up to only quadratic order in the fluctuation field in the action. The resulting quantity is one-loop corrected lapse action. However, performing the lapse integral of one-loop effective theory remains challenging due to highly nonlinear dependence of the one-loop effective action on $N_c$. 

Instead, we evaluate the one-loop wavefunction in the saddle point approximation employing Picard-Lefschetz (PL) formalism. In particular, we identify the relevant and contributing background saddle geometries, whose structure is dictated by the covariant boundary-action parameter, and subsequently analyze the one-loop behavior of tensor modes around these saddles. Interestingly, we find  the background saddle geometries to be near no-boundary and near deSitter geometries in the vanishing limit of covariant boundary term. Past studies \cite{Mallik:2026iai, Ailiga:2024wdx, Barvinsky:1992dz} about ``no-boundary" proposal shows it confronts with UV and IR divergences. In this work we aim to study these issues in the approaching limit and compare the wavefunction and results with the deSitter results. In both cases, the one-loop effective action suffers from UV divergences and requires proper renormalization and regularization. UV-divergences are carefully extracted where the covariant boundary parameter acts as a regulator and isolate these UV divergence in which we have  performed infinite summations using generalized zeta function. These UV divergences are removed by adding suitable counterterms. We find that the late-time IR behavior of the two background geometries are different. This is not unexpected, since vacuum of both geometries are different. 

\vspace{-0.5cm}
 \subsection{Outline of paper}
 \label{outline}
The rest of the paper is organized as follows:  In Sec. \ref{exp} we do the expansion of the action up to quadratic order of fluctuation field and explore the consequences of general covariance on boundary conditions of fields. In Sec. \ref{bound_act}, we compute the generic on-shell solution for the scale factor and fluctuation field with the corresponding action. Here, we study the possible boundary choices for the fluctuation field corresponding to the Neumann boundary choice imposed on scale factor at the initial hypersurface and Dirichlet boundary condition on the final hypersurface. Sec. \ref{trans_amp_absence_of_pert} computes the exact expression of wavefunction in absence of metric tensor fluctuations, and studies the dominant initial condition $\pi_i$, in the semiclassical limit. Sec. \ref{comp_sads} discusses about the complex saddle geometries associated with the each choice of $\pi_i$ and their relevance to the path integral using Picard Lefschetz formalism. In Sec. \ref{pert}, we analyse  the classical behavior of tensor fluctuation of metric around the background complex saddles. In Sec. \ref{trans_amp},  we evaluate the one-loop quantum corrected action for lapse factor which is seen to be UV divergent at the no-boundary saddles. The one loop UV divergence and contact divergence are systematically extracted from the quantum corrected lapse action for the near no-boundary saddles, where suitable counterterms are discussed in Sec. \ref{one_loop_cal_nnb}. Sec. \ref{one_loop_cal_nds} extends the similar analysis to the near de-Sitter saddles geometries which are free from any UV divergences, except for the contact divergence. The lapse integration on the finite lapse action via Picard-Lefschetz and WKB methods is done in Sec. \ref{PL_Nc_sad}. The quantum-corrected UV-finite transition amplitude is computed in Sec. \ref{subs:quant_TA}. Conclusions are presented in Sec. \ref{conc}, where we have summarized our results.

\section{Action Expansion}
\label{exp}
In this section we compute the action that governs the dynamics of gravitational perturbations $h_{ij}$ in the exponential parameterization. We substitute the metric given in eq. (\ref{eq:frwmet}) into the Einstein - Hilbert (EH) gravitational action with a positive cosmological constant mentioned in eq. (\ref{eq:grav_act})  and perform the perturbative expansion up to quadratic order in $h_{ij}$. As we are interested in spatial perturbations, it is natural to work within the Arnowitt–Deser–Misner (ADM) decomposition. In the ADM formalism, the four-dimensional Ricci scalar $R$ can be expressed in terms of the geometric quantities defined on the constant-time hypersurfaces as follows:
\bea
\label{eq:3Rgamma1}
R = {\cal R} + K_{ij} K^{ij} + K^2 + 2K^\prime/N_p \, ,
\eea
where $(^\prime)$ denotes the derivative with respect to $t_{p}$. The quantities $K_{ij}$ and $\mathcal{R}$  are the extrinsic curvature and Ricci-scalar of the 3-hypersurface, respectively. The determinant of the four-dimensional metric factorizes as: $\sqrt{-g}=N_{c}\sqrt{\g}$, where $g $ is the full metric determinant $g_{\mu\nu}$ and $\g$ denotes the determinant of induced metric $\g_{ij}$. In this ADM decomposition, the EH action becomes the following:
\bea
\label{eq:EHact_ADM}
S_{\rm grav}[g_\mn] 
= \frac{1}{16 \pi G} \int_{\cal M} {\rm d}^4x \sqrt{-g} 
\bigl[
-2 \Lam + {\cal R} + K_{ij} K^{ij} - K^2 
\bigr] 
+ \frac{2}{16 \pi G} \int_{\pt {\cal M}} {\rm d} {\bf x} \sqrt{\g} K + S_{\rm bd} \, .
\eea
Using the perturbative expansions of the geometric quantities on the spatial hypersurfaces (see appendix [\ref{gfgh1}]), one can expand the above action upto quadratic order. The zeroth-order term in $h_{ij}$ describes the background spacetime, while first-order term vanishes on background, and the second-order action then determines the dynamics of perturbation on the background spacetime. This expansion in eq.   (\ref{eq:EHact_ADM}) suffices for calculating one-loop quantum corrections within the theory. Contributions from higher-order terms become relevant only when considering multi-loop corrections, which are beyond the scope of the present work. In the following, we fix the gauge redundancy by working in the proper-time gauge $N_c^\prime =0$, implying $N_c = {\rm const}$ on background spacetime. For the metric perturbations, we impose the transverse-traceless (TT) gauge i.e $\bar{\nabla}^{i}h_{ij} = 0$ and $\gamma^{ij}h_{ij} = 0$, which projects onto the two physical graviton polarizations. Here, $\bar{\nabla}_{i}$ is the covariant derivative with respect to induced metric $\g_{ij}$. As our analysis is restricted to the one-loop level, the Faddeev–Popov ghost sector contribute as an overall numerical prefactor to the one-loop effective action (see \cite{Mallik:2026iai}).

\subsection{Boundary Action and Covariance}
\label{boundTR} 
In this section, we study the gravitational perturbations using the exponential parameterization
\beq
\label{eq:exp_split}
\g_{ij} = a^2 \rho_{ik}\big(e^{h}\big)^{k}{}_{j}
= a^2 ( \rho_{ij} + h_{ij} + \tfrac{1}{2}h_{ik}h^{k}{}_{j}
+\mathcal{O}(h^{3}) ) \, ,
\eeq
instead of the conventional linear split $\g_{ij} = a^2 (\rho_{ij} + h_{ij})$. The exponential parametrization has proven to be useful in studies of the renormalization-group (RG) flow of gravitational couplings. In particular, it has been shown to eliminate the singularity that appears in the beta functions associated with the cosmological constant, thereby leading to more stable fixed-point properties under fluctuations. This is in contrast to the conventional linear split, where such singularities emerge in the RG flow, resulting in an unstable behavior as the flow approaches the singularity \cite{Pawlowski:2018ixd, Ohta:2021bkc, Percacci:2015wwa}. This naturally motivates one to adopt the exponential parametrization over the linear parametrization. In this work, we are interested in identifying the consistent boundary conditions that can be imposed on perturbations at the hypersurfaces while preserving the general covariance and study the associated one-loop behavior. Recent studies \cite{Brizuela:2023vmb, Ailiga:2024wdx, Mallik:2026iai} reveal that general covariance puts a strong constraint on the boundary conditions of perturbation for either case - linear or exponential parameterization. In particular, a given choice of boundary condition on the background spacetime restricts the boundary choices on the perturbation field. In the following, we present this analysis for the exponential parameterization.

To begin with, consider the line element of the perturbed FLRW metric given in eq. (\ref{eq:frwmet}) in four spacetime dimensions 
\beq
\label{eq:frwmet_changed}
{\rm d}s^2 = - \frac{N_c^2}{q(t)} {\rm d} t^2 
+ \underbrace{
q(t) \left[
\rho_{ij} + h_{ij}+\frac{1}{2}h_{ik}h^{k}{}_{j}
\right]
}_{\g_{ij}} {\rm d}x^i \, {\rm d}x^j \, .
\eeq
where $h_{ij}$ is function of time and space co-ordinates i.e $h_{ij}(t, {\bf x})$. In obtaining above expression, we make the following transformation for the time-coordinate and scale factor:
\beq
\label{eq:rescale}
{\rm d} t_p = \frac{dt}{a(t)} \, ,
\hspace{5mm}
q(t) = a^2(t) \, ,
\eeq
with $0\leq t\leq1$. This transformation is beneficial as it casts the EH action in a simpler quadratic form \cite{Feldbrugge:2017kzv, Narain:2022msz} making exact computations possible. Substituting the above metric ansatz in the EH action mentioned in eq. (\ref{eq:EHact_ADM}) and utilizing the perturbative expansions of the geometric quantities (see appendix [\ref{gfgh1}]), the relevant gravitational action for one-loop computation reads as follows:
\bea
\label{eq:EHact_exp}
&&
S_{\rm grav}[g_\mn] 
= S_{\rm grav}^{(q)} + S_{\rm grav}^{(h)} = 
\frac{1}{16 \pi G}\int_{\mathcal{M}}d^4x\sqrt{\rho}\biggl[\left(6k-2\Lambda q\right)N_c-\frac{3\dot{q}^2}{2 N_c}-\frac{N_c k}{2}h_{ij}h^{ij} \notag \\
&& + \frac{N_c}{4 }h^{ij}\Box h_{ij}+\frac{q^2}{4 N_c}\dot{h}_{ij}\dot{h}^{ij}\biggr] +\frac{1}{16 \pi G}\int_{\mathcal{\partial M}}d{\bf x} \sqrt{\rho}\biggl[\frac{3\dot{q}q}{N_c}\biggr]_{0}^{1} +S_{bd} ,
\eea
where $\Box$ is laplacian with respect to $\rho_{ij}$. The action $S_{\rm grav}^{(q)}$ and $S_{\rm grav}^{(h)}$ are given by
\bea
\label{eq:Act_sq_sh_begEXP}
&&
S_{\rm grav}^{(q)}
= \frac{1}{16\pi G}\int_{\cal M} {\rm d} t {\rm d} {\bf x} \sqrt{\rho}
\biggl[
(6k -2 \Lam q )N_c - \frac{3\dot{q}^2}{2N_c} 
\biggr] +  \frac{1}{16\pi G} \int_{\pt {\cal M}} {\rm d} {\bf x} 
\sqrt{\rho} \left(\frac{3q \dot{q}}{N_c}\right) \biggr \rvert_0^1 + S_{\rm bd}^{(q)} \, ,
\notag \\
&&
S_{\rm grav}^{(h)}
= \frac{1}{16 \pi G}\int_{\mathcal{M}}d^4x\sqrt{\rho}\biggl[\frac{q^2}{4 N_c}\dot{h}_{ij}\dot{h}^{ij}-\frac{N_c k }{2}h_{ij}h^{ij}+\frac{N_c}{4 }h^{ij}\Box h_{ij}\biggr] 
     +S_{bd}^{(h)} \ ,
\eea
respectively, where we have decomposed the boundary action as $S_{\rm bd} = S_{\rm bd}^{(q)} + S_{\rm bd}^{(h)} $, with $S_{\rm bd}^{(q)}$ accounts for background field and while $S_{\rm bd}^{(h)}$ is for perturbation field. In deriving above expression, we have utilized the gauge-fixing conditions and performed integrations. The boundary $\pt{\cal{M}}$ refers to the temporal hypersurfaces of the manifold at $t=0$ and $t=1$. Certain surface contributions arising from the integration by parts vanish under the boundary condition $h_{ij}\rightarrow 0$ at spatial infinity. One can construct the boundary term $S_{\text{bd}}$  accordingly to ensure a consistent and well-defined variational principle for the background field $q$ and metric perturbations $h_{ij}$.  A general expression for $S_{\rm bd}$ which is consistent with the principle of general covariance can be written as \cite{Brizuela:2023vmb}
\beq
\label{eq:Sbd_genform}
S_{\rm bd} = \frac{1}{8\pi G} \int_{\pt {\cal M}} {\rm d} {\bf x}
\sqrt{\g} \,\, F({\cal R}, K) \biggr \rvert_0^1\, .
\eeq
Now one may proceed to perform the expansion of boundary action $S_{\rm bd}$ in $h_{ij}$ upto quadratic order as required for the one-loop computation. For the gauge-fixing choice considered here, one finds neither $K$ nor ${\cal R}$ contains a linear term in $h_{ij}$. Consequently, the Taylor expansion of boundary term $F$ around background, is given by:
\beq
F({\cal R}, K) = F({\cal R}_b, K_b) + \frac{\partial F }{\partial {\cal R}}\Big|_{{\cal R}_b, K_b} ({\cal R} - {\cal R}_{b}) + \frac{\partial F}{\partial K}\Big|_{{\cal R}_b, K_b} (K - K_{b}) \, ,
\eeq
where $K_b$ and ${\cal R}_b$ are defined on the background given by ${\cal R}_b=\bar{{\cal R}}/q = {\cal R}_{h=0}$. $\bar{{\cal R}}=6k$ is intrinsic curvature of $S^3$ , and $K_b=3\dot{q}/(2N_c \sqrt{q}) = K_{h=0}$. The expressions for ${\cal R}$ and $K$ can be read out from the appendix \ref{gfgh1}. We have neglected the higher-order terms, as they contribute only beyond quadratic order in the perturbation $h_{ij}$. The boundary action then reads as
\bea
\label{eq:Fexp}
S_{\rm bd}
= \frac{1}{8 \pi G}\int_{\partial \mathcal{M}}d^3x \sqrt{\rho}\Bigl[q^{3/2}F(\mathcal{R}_b,\mathcal{K}_b)+\sqrt{q}\left(\frac{1}{4}h^{ij}\Box h_{ij}-\frac{k}{2}h^{ij}h_{ij}\right)\partial_{{\cal R}}F(\mathcal{R}_b,\mathcal{K}_b) + \cdots\Bigr]\Bigg|_{0}^{1} \, .
\eea
The resulting boundary action can be divided into two contributions: a term that depends exclusively on the scale factor and another that incorporates the metric perturbations. The determination of these terms follows from enforcing the consistency of variational formulation involving all dynamical fields. In subsequent sections, we derive $F(\mathcal{R}_b,\mathcal{K}_b)$ demanding a well-posed variational problem for given boundary conditions on the background scale factor. We then find that general covariance further restricts the allowed boundary conditions on the perturbation fields.

\subsection{Dynamics of  Scale Factor}
\label{dynmaics_scalefactor} 
With the general boundary term mentioned in eq. (\ref{eq:Fexp}), we proceed to vary the total action with respect to dynamical fields $q(t)$ and $h_{ij}$ to obtain the respective equation of motion that governs their classical evolution. The equation of motion for the background spacetime is obtained by varying the total gravitational action $ S_{\rm grav}[g_\mn]$ given in eq.(\ref{eq:EHact_exp}), with respect to scale factor by writing $q(t) = \bar{q}(t) + \de q(t)$, while neglecting the perturbative contributions from $h_{ij}$. The variation is given by: 
\bea
\label{eq:Sexp_qvar}
\de S_{\text{grav}}^{(q)} =
\frac{1}{16 \pi G}\int_{\mathcal{M}}d^4x 
&&
\sqrt{\rho}
\Bigl[
\left(
\frac{3\ddot{\bar{q}}}{N_c}-2\Lambda N_c
\right)\delta q
\Bigr] 
\notag \\
&&
+ \frac{1}{16\pi G} \int d^3x \sqrt{\rho} \Bigl[ 
\frac{3 \bar{q }\delta\dot{q}}{N_c}
+3 \sqrt{\bar{q}}\bar{F} \delta q
+2 \bar{q}^{3/2}\delta \bar{F}
\Bigr]\Bigg|_{0}^{1} , \notag \\
\eea
with $\bar{F}\equiv F({\cal R}_b, K_b)$. With an appropriate choice of boundary term $F$, one requires
\beq
\label{eq:sby_eqn}
\frac{3 \bar{q }\delta\dot{q}}{N_c}
+3 \sqrt{\bar{q}}\bar{F} \delta q
+2 \bar{q}^{3/2}\delta \bar{F} = 0 \, ,
\eeq
at both hypersurfaces and thereby ensuring the well-posed variational problem. The corresponding surface term $ F({\cal R}_b,K_b)$ for various boundary conditions on scale factor have been derived and discussed in the past literature \cite{Brizuela:2023vmb, Ailiga:2024wdx}. The equation of motion for the scale factor is obtained from the term proportional to $\delta q$ from $\de S_{\text{grav}}^{(q)}$, yielding
\beq
\label{eq:dyn_q_eq}
\ddot{\bar{q}} - \frac{2}{3} \Lam N_c^2 =0 \, .
\eeq
We point out again that, we assume no backreaction in obtaining above expression which means perturbation field $h_{ij}$ does not affect the background $q(t)$ evolution. The equation of motion for $q(t)$ is 
easy to solve with a generic solution given by
\beq
\label{eq:qsol_gen}
\bar{q}(t) = \frac{\Lam N_c^2}{3} t^2 + b_1 t + b_2 
= \frac{\Lam N_c^2}{3} (t-r_1)(t-r_2)\, ,
\eeq
where $b_{1,2}$ are constants to be fixed based on the 
requirement of boundary conditions, while $r_ {1,2}$ are the roots of the scale factor.

\subsection{Dynamics of  Fluctuation Field}
\label{dynmaics_fluctuation field} 
Upon varying the fluctuation action $S_{\rm grav}^{(h)}$  with respect to the fluctuation field by expressing i.e $h_{ij} = \bar{h}_{ij} + \delta h_{ij}$, we get :
\bea
\label{eq:Sexp_hvar}
\de S_{\text{grav}}^{(h)} = 
&&
\frac{1}{16 \pi G}\int_{\mathcal{M}}d^4x \sqrt{\rho} \Biggl[\left(\frac{N_c}{2}\Box \bar{h}_{ij}-kN_c  \ \bar{h}_{ij}-\frac{\bar{q}\dot{\bar{q}}}{N_c}\dot{\bar{h}}_{ij}-\frac{\bar{q}^2}{2N_c}\ddot{\bar{h}}_{ij}\right)\delta h^{ij}\Biggr]
\notag \\
+ && \frac{1}{16 \pi G}\int_{\mathcal{\partial M}}d^3x \sqrt{\rho} \,
A_{ij} \de h^{ij} 
\Big|_{0}^{1}\, ,
\eea
where
\bea
\label{eq:ABexp}
&&
A_{ij} =\frac{\bar{q}^{2}}{2N_c}\dot{\bar{h}}_{ij}+2\sqrt{\bar{q}} \ \partial_{\cal R} \bar{F}\left(\frac{1}{2}\Box \bar{h}_{ij}-\bar{h}_{ij}\right) \, .
\eea
As in the case of scale factor action, the consistent variational problem requires the boundary conditions of $h_{ij}$ to obey
\beq
\label{eq:BD_h_gen}
A_{ij} \de h^{ij}  \bigr|_0^1= 0 \, .
\eeq
The condition may be realized either by fixing the boundary such that the term vanishes identically or by choosing an appropriate form of $F$. However, since $F$ is fixed by requiring a consistent variational principle for the background $\bar{q}$, the choice of boundary conditions for the fluctuation field is constrained. Once an appropriate boundary condition is chosen such that variational principle is preserved, we can get the equation of motion for the fluctuation field which is given by
\beq
\label{eq:hijEQM}
\bar{q}^2\ddot{\bar{h}}_{ij}+2\bar{q}\dot{\bar{q}}\dot{\bar{h}}_{ij}-N_c^2\Box \bar{h}_{ij} +2 k N_c^2 \bar{h}_{ij} = 0 \, .
 \, 
\eeq
We have utilized the equation of motion corresponding to $\bar{q}(t)$ in obtaining the above result. The equation of motion 
for $h_{ij}(t, {\bf x})$ is a bit more complicated to solve as it 
involves both time and space derivatives, while $\bar{q}(t)$
is given by the above solution. In order to solve it we first do
mode decomposition of the fluctuation field $h_{ij}(t, {\bf x})$. 
Respecting the background symmetry the 
field $h_{ij}(t, {\bf x})$ can be expressed in terms of tensor spherical harmonics \cite{Gerlach:1978gy} as
\bea
\label{eq:h_ij_Sph_Har}
&&
h_{ij}(t, {\bf x}) 
=\sum_{l=2}^\infty 
\sum_{n=2}^{l} \sum_{m=-n}^{n}
h^{l}_{nm}(t) (G_{ij})^l_{nm} ({\bf x}) \, , \notag \\ 
&&
D^2 G^{l}_{ij} ({\bf x}) = - [l(l+2) -2] G^{l}_{ij} ({\bf x}) \, , 
\hspace{5mm} l \geq 2 \, .
\eea
where $h^l_{nm}(t)$ can be thought of as Fourier coefficients
at any given time $t$ and 
$G^{l}_{ij} ({\bf x})$ on $S^3$ are the tensor spherical harmonic  eigenfunctions  which satisfy above relation and obeys the normalization:
\beq
\label{eq:sper_har_norm}
\int_{S^3} d\Omega_3 \sqrt{\rho} \, G_{ij}^{l}({\bf x}) G_{i'j'}^{l'}({\bf x}) = \delta_{i,i'} \,  \delta_{j,j'} \, \delta_{l,l'} \, ,
\eeq
where $\delta_{ij}$ denotes Kronecker delta function.
As the Lagrangian of the fluctuation field $h_{ij}$ mentioned 
in Eq. (\ref{eq:EHact_exp}) is seen to respect isotropy, 
it is therefore independent of $n$ and $m$ modes. For brevity, we can suppress the $n$ and $m$ indices and write $h^{l}_{nm}(t) \equiv h_l(t)$. 
Owing to the $S^3$ symmetry, there exists degeneracy in the system: 
for each $l$ there are $g_l =2 (l+3)(l-1) $
modes having the same action. The factor $2$ appearing above is due to the contributions from the two propagating polarizations of the $h_{ij}$ field in the Landau/TT gauge choice. 

This mode decomposition when utilized in the 
equation of motion for $\bar{h}_{ij}$ mentioned in 
eq. (\ref{eq:hijEQM}), it is seen that each component $\bar{h}_l(t)$
follows the following equation 
\beq
\label{eq:htl_eqm_ep}
\ddot{\bar h}_l + \frac{2 \dot{\bar{q}}}{\bar{q}} \dot{\bar h}_l
+ \frac{N_c^2}{\bar{q}^2}
l(l+2) 
\bar{h}_l = 0 \, .
\eeq
The general solution of the on-shell equation $\bar{h}_l(t)$ can be expressed in terms of special functions and given by
\bea
\label{eq:hl_sol_gen}
&&
\bar{h}_l(t) = \frac{1}{\sqrt{\bar{q}(t)}} 
\bigl[
d^{(l)}_{p} \mathbb{P}_1^{\xi_l} (\tau_t) + d^{(l)}_{q} \mathbb{P}_1^{-\xi_l} (\tau_t) 
\bigr] \, ,
\eea
with the variables $\xi_l$ and $\tau$ given by
\beq
\label{eq:xi_tau_defs}
\xi_l =  \sqrt{1-
 \frac{
36l(l+2)}{N_c^2 \Lam^2 (r_1 - r_2)^2}} \, ,
\hspace{5mm} \text{and} \hspace{5mm}
\tau_t = \frac{2t - (r_2+r_1)}{r_1 - r_2} \, .
\eeq
Here, $\mathbb{P}_1^{\xi_l}$ denotes the associated Legendre function of the first kind, whose explicit form is:
\bea
\label{Legendre_form}
\mathbb{P}_1^{\xi_l}(\tau_t)=\left(\frac{1-\tau_t}{1+\tau_t}\right)^{\xi_l/2}\left(\tau_t + \xi_l \right)\, .
\eea
Utilizing the mode decomposition, the full perturbation action can be expressed as
\beq
\label{eq:mode_hl_act}
S^{(h)}_{\rm grav}
= \frac{1}{16\pi G} \sum_{l=2}^\infty \,\,  g_l
S^{(l)}_{\rm grav}  \, ,
\eeq
where $S^{(l)}_{\rm grav} $ is the action 
corresponding to each mode and is given by
\bea
\label{eq:SL_hl_act}
S^{(l)}_{\rm grav} =  \int_0^1 {\rm d}t
\Bl[
\frac{\bar{q}^2 }{4N_c}\dot{h}_l^2
&&
- N_c\Bl\{
\frac{l(l+2) - 2 + 2k}{4}   
\Br\} h_l^2
\Br]  
\notag \\
&&
-  \Bl[\sqrt{q}\left(\frac{l(l+2) -2 + 2k}{2} \right) h_{l}^2 \partial_{{\cal R}}\bar{F} \Br]\Bigg|_{0}^{1} \, . \notag \\
\eea
%

\section{Covariant Boundary choices and On-shell action}
\label{bound_act} 

In this section, we explore the restricted choices available for the boundary condition of the perturbation $h_{ij}$ given the fixed boundary choice for the background spacetime. We first note that the conjugate momenta corresponding to field $q(t)$ is given by $\pi_q = -3\dot{q}/(2 N_c)$. This implies that $K_b=-\pi_q/\sqrt{q}$. In terms of $q$ and $\pi_q$ variables the eq. (\ref{eq:sby_eqn}) becomes as:
\beq
\label{eq:BDpart_qexp}
\biggl[
\biggl(
3 \sqrt{\bar{q}} \bar{F} 
- \frac{2\bar{\cal R}}{\sqrt{\bar{q}}} \pt_{\cal R} \bar{F} 
+ \bar{\pi}_q \pt_K \bar{F} 
\biggr) \de q
- 2 \bar{q} \biggl(
1+ \pt_K \bar{F}
\biggr) \de \pi_q 
\biggr] \Bigg|_{0}^{1}= 0 \, ,
\eeq
where as before $\bar{F} = F({\cal R}_b, {K}_b)$.  Here, we focus on exploring the boundary conditions for the fluctuation field, given that we impose Neumann boundary conditions at the initial hypersurface and Dirichlet boundary conditions at the final hypersurface on the scale factor.

\subsection{Neumann boundary condition }
\label{NBC_bg} 

If the Neumann boundary condition is imposed at the initial hypersurface it means that the conjugate momenta 
corresponding to $q(t)$ is fixed at that end point: 
$\de \pi_q =0$. In this case if $q$ is arbitrary at that end point 
then using eq. (\ref{eq:BDpart_qexp}), we have 
\beq
\label{eq:nbc_bg_q}
\biggl(
3 \sqrt{\bar{q}} \bar{F} 
- \frac{2\bar{\cal R}}{\sqrt{\bar{q}}} \pt_{\cal R} \bar{F} 
+ \bar{\pi}_q \pt_K \bar{F} 
\biggr) = 0 
\hspace{5mm}
\Rightarrow
\hspace{5mm}
2 {\cal R} \pt_{\cal R} F + K \pt_K F = 3F \, . 
\eeq
To solve this we can make an ansatz that $F({\cal R}, K)
\sim {\cal R}^n K^{2m}$. This gives $2n + 2m =3$ leading to an infinite 
set of choices for $F({\cal R}, K)$:
\beq
\label{eq:Fchoice_nbc}
F({\cal R}, K) = - c {\cal R}^{3/2-\om} K^{2\om} \, ,
\eeq 
where $c$ is a constant. For each value of $\om$ there is a choice of $F({\cal R}, K)$
giving an infinite series of available choices. All of these different $F$ will lead to the
same boundary condition for the background $q(t)$ (Neumann in this case) 
but will lead to different boundary conditions for the fluctuation field. 
This is an infinite set of covariant boundary action choices available, 
\beq
\label{eq:Sbd_nbc_cov}
S^{\rm NBC}_{\rm bd}
= -\frac{c}{8\pi G} \int_{\pt {\cal M}} {\rm d} {\bf x}
\sqrt{\g} \,\, {\cal R}^{3/2-\om} K^{2\om} 
\hspace{5mm}
\forall \om \in \ \mathbb{R} \, .
\eeq
The trivial $c=0$ situation has been explored in past literature, see \cite{Ailiga:2024wdx} for details. Here, we focus on the more general case, $c\neq0$, which implies the addition of non-trivial boundary term. This non-trivial choice could modify the saddle point structure and corresponding transition amplitude.

In this paper we will work with the 
choice $F \neq 0$. The boundary condition that is expected to
be satisfied by the fluctuation field $h_{ij}$ can be obtained
using eq. (\ref{eq:BD_h_gen}). This gives
\beq
\label{eq:hij_nbc_F0}
\Biggl[\frac{\bar{q}^{2}}{2N_c}\dot{\bar{h}}_{ij}+2\sqrt{\bar{q}} \ \partial_R \bar{F}\left(\frac{1}{2}\Box \bar{h}_{ij}-\bar{h}_{ij}\right)\Biggr]\delta h^{ij} = 0 \, .
\eeq
This gives rise to two possible scenarios: (i). having 
$h_{ij}$ is fixed (DBC), which makes the above equation satisfy identically 
and (ii). requiring 
\beq
\label{eq:hl_nbc_2ndcond}
 \frac{\bar{q}\sqrt{\bar{q}}}{2N_c}\dot{\bar{h}}_l-\partial_R \bar{F}(\mathcal{R}_b,\mathcal{K}_b)l(l+2)h_l=0
\, ,
\eeq
which is a particular case of canonical Robin boundary condition. It is worth emphasizing that the exponential parameterization allows one to impose non-trivial Dirichlet boundary condition on fluctuation field unlike in the case of linear parameterization where vanishing $h_{ij}$ is the only valid Dirichlet boundary condition.

\subsection{Dirichlet boundary condition}
\label{DBC_bg} 

Imposing Dirichlet boundary condition at the final hypersurface
imply that we fix $q$. This means $\de q=0$. 
If the momentum is arbitrary then eq. (\ref{eq:BDpart_qexp}) immediately leads to 
\beq
\label{eq:dbc_qbd_F}
1 + \pt_K \bar{F} = 0
\hspace{5mm}
\Rightarrow 
\hspace{5mm}
F = - K_b  + f({\cal R}_b)\, ,
\eeq
where $f({\cal R}_b)$ is an arbitrary function of ${\cal R}_b$
which for simplicity can be fixed to zero. Once the functional form of $F$ is known, one can use it to 
compute the boundary condition expected to be satisfied 
by the fluctuation field $h_{ij}$ for consistent variational problem. 
This is achieved by exploiting eq. (\ref{eq:BD_h_gen}). 
For the DBC case this gives,
\beq
\label{eq:DBC_hcase}
\frac{1}{2N_c} \bigl(
\bar{q}^2 \dot{\bar h}_{ij}
\bigr) \de h^{ij}  = 0 \, .
\eeq
This means we have two possible scenarios at final hypersurface as well: (i). either $h_{ij}$ is also fixed at the boundary
which implies $\de h_{ij} =0$, and (ii). we expect that the fluctuation field to satisfy the condition 
\beq
\label{eq:robinH_dbc_case}
 \bigl(
\bar{q}^2 \dot{h}_{ij}
\bigr) = 0 \, .
\eeq
The former corresponds to Dirichlet boundary conditions (DBC), in which the fluctuation field is fixed, while the latter corresponds to Neumann boundary condition, where the conjugate momentum $\pi_{h_{ij}} = q^2\dot{h}_{ij}/(2N_c) $ of the fluctuation field is fixed. Therefore, it is clearly evident that imposing NBC or DBC for the 
background scale-factor $q(t)$ restricts our available 
choices of boundary conditions for the fluctuation field $h_{ij}$. In the following, we will focus on the Neumann boundary condition at $t=0$ hypersurface and Dirichlet boundary condition at the $t=1$ hypersurface of background, while we pick the simplest boundary choices: $\de h_{ij} = 0 $ on the perturbation field on both hypersurfaces \footnote{Various other boundary conditions on fluctuation field such as: DBC-DBC, DBC-NBC, RBC-DBC or RBC-NBC are possible for the chosen background boundary conditions, we leave them for future exploration.} that aligns with general covariance and variational consistency.

\subsection{On-shell Action}
\label{onshell_pert}
Imposing NBC at the initial boundary and DBC at the final boundary on the background spacetime means
\beq
\label{eq:bg_bc}
\pi_{q}(0)= \pi_i\, , \quad \text{and} \quad q(1)=q_f \, .
\eeq
The unknown constants $b_{1,2}$ that appear in the solution to the equation of motion given in eq. (\ref{eq:dyn_q_eq}) can now be determined. This implies
\beq
\label{eq:qsol_nbc}
\bar{q}(t) = \frac{\Lam N_c^2}{3} (t^2-1) - \frac{2 N_c \pi_i}{3} (t-1) + q_f \, .
\eeq
With the corresponding boundary terms, we have the proper variational action for $q(t)$ as
\beq
\label{eq:on_act_q}
S^{(q)}_{grav} = \frac{1}{16 \pi G}\int d^4x\sqrt{\rho}\biggl[\left(6k-2\Lambda q\right)N_c-\frac{3\dot{q}^2}{2 N_c} \biggl] + \frac{1}{8 \pi G}\int d{\bf x} \sqrt{\rho} \biggl[ -\frac{3\dot{q}q}{2N_c} + c {\cal R}_b^{3/2 - \omega} K^{2\omega}  \biggr]_{t=0}\, .
\eeq
Plugging the on-shell solution of scalar factor in above equation, we obtain the expression of background on-shell action as follows
\beq
\label{eq:stot_onsh_nbc}
S_{\rm grav}^{(\bar{q})}[N_c] = \frac{V_3}{8\pi G}
\Bl[\frac{\Lam^2}{9} N_c^3 
- \frac{\Lam \pi_i}{3} N_c^2 
+ \left(3k - \Lam q_f + \frac{\pi_i^2}{3} \right) N_c
+ q_f \pi_i + c(6k)^{3/2-\omega}\left(-\pi_i\right)^{2\omega} \Br]\, .
\eeq
Here $V_3$ is the volume of $3-$sphere. Similarly, the on-shell solution for $h_l(t)$ is mentioned 
in eq. (\ref{eq:hl_sol_gen}) for the $h_{l}(0)=0$ and $h_{l}(1)=h_1^{(l)}$, with $d^{(l)}_{p}$ and $d^{(l)}_{q}$ given by
\bea
\label{eq:hl_dpdq_NBC}
d^{(l)}_{p} =  \frac{ h^{(l)}_1 \sqrt{q_f} \mathbb{P}_1^{-\xi_l}[\tau_0] }{W_l(1,0)} \, ,
\hspace{5mm}
d^{(l)}_{q} = - \frac{ h^{(l)}_1 \sqrt{q_f} \mathbb{P}_1^{\xi_l}[\tau_0] }{W_l(1,0)} \, ,
\eea
where we defined $W_l(1,0) \equiv  \mathbb{P}_1^{\xi_l}(\tau_1) \mathbb{P}_1^{-\xi_l}(\tau_0)-\mathbb{P}_1^{\xi_l}(\tau_0) \mathbb{P}_1^{-\xi_l}(\tau_1)$. The corresponding on-shell action for the perturbation mode $h_{l}$ including the contributions from the boundary term is given by
\bea
\label{eq:hh_onSH_NBC}
S^{(l)}_{\rm grav} [\bar{q}, \bar{h}_l, N_c]  
&=& 
\frac{1}{16 \pi G}\Biggl[\left(\frac{\bar{q}^2}{4 N_c}\dot{\bar{h}}_l\bar{h}_l\right)\Bigg|_1
\notag \\
&&
+\Bl\{-\frac{\bar{q}^2}{4 N_c}\dot{\bar{h}}_l\bar{h}_l+\frac{c}{2}(6k)^{1/2-\omega}(-\pi_i)^{2\omega}\left(\frac{3}{2}-\omega\right)l(l+2)\bar{h}_l^2\Br\}
\biggr \rvert_{0} \Biggr] \, ,
\notag \\
&&
= \frac{q_f \bl(h^{(l)}_1\br)^2}{(16\pi G) (12N_c)}
\Bl[ N_c\pi_i-\Lam N_c^2 + 3q_f \  D_t\bl\{ \ln W(t,0) \br\}  \br\rvert_{t=1}
\Br],
\eea
where $D_t\equiv d/dt$.

\subsection{Dominant boundary configuration in $\hbar \to 0$}
\label{trans_amp_absence_of_pert}
\begin{figure}[ht]
\centering
\begin{minipage}[t]{0.48\textwidth}
    \centering
    \includegraphics[width=\textwidth]{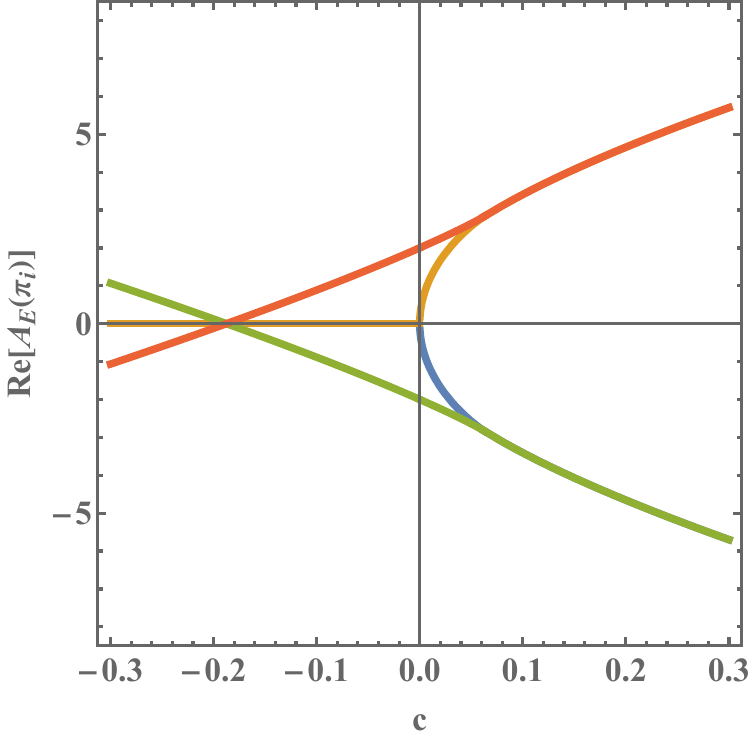}
\end{minipage}
\hspace{0.02\textwidth}
\begin{minipage}[t]{0.30\textwidth}
    \centering
    \raisebox{1cm}{
        \includegraphics[width=\textwidth]{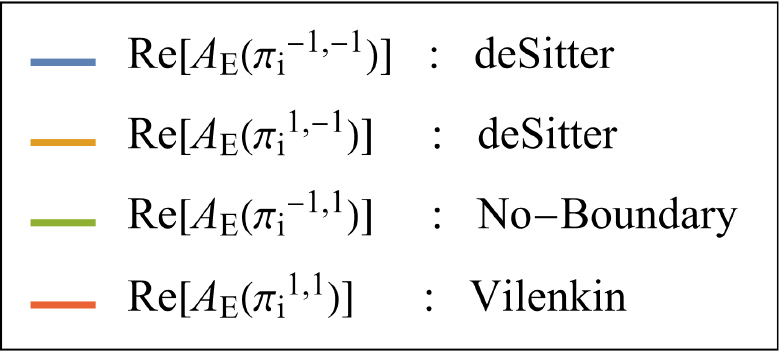}
    }
\end{minipage}
\caption{Plot of Euclidean action $\text{A}_{E}$ given in
eq. (\ref{Eucl_action_of_pi}) as a function of $c$ for each
dominant choice of $\pi_i$. For illustration, we choose $k=1$
and $\Lam=3$. The Euclidean action of $c_1=-1$ and $c_2=1$
is dominant for all values of $c\geq 0$.}
\label{fig:pi_dom}
\end{figure}

We now proceed to compute the transition amplitude between two 3-hypersurfaces in the absence of metric tensor perturbations, and subsequently identify the dominant initial condition $\pi_i$ in the semiclassical limit. Within the minisuperspace approximation, the wavefunction can be expressed as (with $V_3 = 8\pi G$):
\beq
\label{eq:TA}
\Psi[\text{Bd}_f,\text{Bd}_i]=\int_{-\infty}^{+\infty}\text{d}N_c\int_{\text{Bd}_f}^{\text{Bd}_i}\mathcal{D}q(t)\text{exp}\left(\frac{i}{\hbar}S_{\text{tot}}[q,N_c]\right) \, ,
\eeq
where $\text{Bd}_i$ and $\text{Bd}_f$ are the initial and final boundary configurations respectively, and $S_{\text{tot}}$ is the mini-superspace action, which can be calculated from the first part of eq. (\ref{eq:Act_sq_sh_begEXP}). The wavefunction after performing the functional integral over scale factor with Neumann and Dirichlet as the initial and final boundary conditions respectively, takes the form \cite{Ailiga:2023wzl}
\beq
\label{eq:Lap_int}
\Psi[\text{Bd}_f,\text{Bd}_i] = \int_{-\infty}^{+\infty}\text{d}N_c \exp\{i S_{\rm grav}^{(\bar{q})}[N_c]/\hbar \} \, .
\eeq
The on-shell action $S_{\rm grav}^{(\bar{q})}[N_c]$ is given by eq. (\ref{eq:stot_onsh_nbc}). The resultant lapse $N_c$ integral is simple one-dimensional ordinary integral, that can be done exactly following a set of transformations that have been outlined in \cite{Narain:2022msz, Ailiga:2023wzl}. The exact expression of the wavefunction is given by
\bea
\label{trans_amp_1}
\Psi[\text{Bd}_f,\text{Bd}_i] &=&
\left(\frac{3\hbar}{\Lambda^2}\right)^{1/3}\text{exp}\left(\frac{i}{\hbar}\Biggl[\frac{3}{\Lambda}\left(k\pi_i+\frac{\pi_i^3}{27}\right) + c(6k)^{3/2-\omega}\left(-\pi_i\right)^{2\omega}\Biggr]\right)
\notag \\
&&
\times \mathcal{A}i\Biggl[\left(\frac{\sqrt{3}}{\hbar \Lambda}\right)^{2/3}\left(3k-q_f\Lambda\right)\Biggr] \, .
\eea
Notice that the transition amplitude acquires an exponential prefactor that depends explicitly on the parameters $c$ and $\omega$. Our immediate step is to analyze this factor, which is solely a function of the initial boundary parameter $\pi_i$, to determine the dominant choice of initial momentum. The exponential factor is given by the following
\beq
\text{exp}\left(\frac{i}{\hbar}\Biggl[\frac{3}{\Lambda}\left(k\pi_i+\frac{\pi_i^3}{27}\right)+c(6k)^{3/2-\omega}\left(-\pi_i\right)^{2\omega}\Biggr]\right) = e^{i \text{A}(\pi_i)/\hbar} \, .
\eeq
We perform a Wick rotation of the initial canonical momentum ($\pi_i\rightarrow i y$ with $y\in \mathbb{R}$), thereby obtaining an Euclidean version of the exponential factor. The corresponding Euclidean action $\text{A}_E(y)$ is given by
\beq
\label{Eucl_action_of_pi}
\text{A}_E(y)=\frac{3}{\Lambda}\left(k y-\frac{y^3}{27}\right) - i \ c (6k)^{3/2-\omega}\left(-i y\right)^{2\omega}, \quad e^{i \text{A}(\pi_i)/\hbar}\Rightarrow e^{-\text{A}_E(y)/\hbar} \, .
\eeq
On extremization of Euclidean action  $\text{A}_E(y)$, we get $\omega =n+\frac{1}{2}, \ \text{where} \  n \in \mathbb{Z}$ from  the $  \text{Im}[\text{A}_E^{'}(y)]=0$ condition. From the condition $  \text{Re}[\text{A}_E^{'}(y)]=0$ , one gets an additional relation governing the extremized configurations, specifically given as:
\beq
\label{eq:Ext_eq}
\frac{3}{\Lambda}\left(k-\frac{y^2}{9}\right)-(-1)^n(6k)^{1-n}c(1+2n)y^{2n}=0 \, .
\eeq
Solving the extremization equation for a generic value of $n$ is challenging, as it is a polynomial equation of degree higher than four. However, for simplicity we consider the $n=-1$ special case, for which the equation becomes analytically tractable. For this choice, the extremized initial momentum choices are given by
\beq
\label{eq:dom_pi}
\pi^{(c_{1},c_{2})}_i = i \ c_1\Biggl[\frac{9k}{2}\left(1+c_2\sqrt{1-\frac{16c\Lambda}{3}}\right)\Biggr]^{1/2} \ \text{where} \ c_1,c_2 \in \{\pm1\} \, .
\eeq
Note that the initial conjugate momenta $\pi_i$ consists of two superscripts $c_1$, $c_2$ and the initial boundary parameter $c$. They specify the initial state of the Universe. For example in the case when the boundary parameter $c\to0$, the Hartle-Hawking no-boundary state correspond to $c_1=-1$ and $c_2=1$, while the Vilenkin tunneling state correspond to $c_1=+1$ and $c_2=1$. To avoid future cluttering we will refer to them as $\pi_i^{HH}$ and $\pi_i^V$ respectively. The configurations $c_1=\pm 1$ and $c_2=-1$ correspond to deSitter and we will refer to them as $\pi_i^{dS}$ wherever needed. The behavior of the associated Euclidean action $A_E$ of these extremized choices have shown in the Fig. [\ref{fig:pi_dom}].

It is clearly evident that the  weight factor of the Euclidean action $A_{E}$, associated with the root $\pi^{HH}_i$ dominates the gravitational path integral for all values of $c\geq 0$ which indicates that this branch contributes most significantly to the semiclassical amplitude. The on-shell action mentioned in eq. (\ref{eq:stot_onsh_nbc}) for these dominant choices of $\pi_{i}$ with $\omega=-1/2$ and $n=-1$ reads as follows
\beq
\label{eq:stot_nbc}
S_{\rm tot}^{(q)}[N_c] = \frac{\Lam^2}{9} N_c^3 
- \frac{\Lam \pi_i}{3} N_c^2 
+ \biggl(3k - \Lam q_f + \frac{(\pi_i)^2}{3} \biggr) N_c
+ q_f \pi_i - \frac{36ck^2}{\pi_i} \, ,
\eeq
where it is understood that $\pi_i = \{\pi_i^{HH}, \pi_i^V, \pi_i^{dS}\}$.

\section{Complex Saddles and their relevance}
\label{comp_sads}
In this section we study the saddle geometries and their relevance to the wavefunction for each choice of $\pi_{i}^{(c_1,c_2)}$ by utilizing Picard Lefschetz formalism.  In the gravitational path-integral, each value of the lapse $N_c$ gives an appropriately weighted contribution, and the dominant ones coming from the saddle points. Saddle contributions can be computed systematically from the action $\bar{S}_{\rm grav}^{(\bar{q})}[N_c]$ mentioned in eq. (\ref{eq:stot_onsh_nbc}) by the following
\beq
\label{eq:sad_eq}
{\rm d} \bar{S}^{(\bar{q})}_{\rm grav}[ N_c] / {\rm d} N_c|_{\text{saddles}} = 0 \, .
\eeq
On solving it, we get the following set of saddle points for given dominant choice of $\pi_i^{(c_1,c_2)}$
\beq
\label{sad_point}
{\rm Saddles}: \hspace{5mm}
N_{\text{sad}}^{(c_1,c_2,c_3)} = \frac{1}{\Lam} \bigl( \pi_i^{(c_1,c_2)} +  c_{3} \sqrt{3 \Lam q_f - 9k} \bigr) \, ,
\eeq
where $c_1,c_2 \ \text{and} \  c_3 = \{-1,1\}$. Here the dependence on $c_1$ and $c_2$ in the expression of saddle comes via $\pi_i$. The on-shell action evaluated at these saddle points from the eq. (\ref{eq:stot_onsh_nbc}) is given by
\beq
\label{onshell action.at sad}
S_0^{sad}=-\frac{36 c k^2}{\pi_i}+\frac{\pi_i^3}{9\Lambda} + \frac{3\pi_ik}{\Lambda}\mp \frac{6}{\Lambda}\left(\frac{q_f\Lambda}{3}-k\right)^{3/2} \, ,
\eeq
where $\pi_i = \{\pi_i^{HH}, \pi_i^V, \pi_i^{dS}\}$. The size of the initial hypersurface at each of these saddles, which is given by the value of scale-factor at $t=0$ is:
\bea  
\label{eq:ini_sc_nb}
q(0)\Big|_{\rm saddles} && = \frac{k\left(3 - c_2\sqrt{9 - 48 c \Lambda}\right)}{2 \Lambda}  \, .
\eea
Note that $q(0)$ doesn't depend on $c_1$ and $c_3$, only on $c_2$ and the boundary parameter $c$, thereby implying the saddles with same $c_2$ start with same initial size $q(0)$. 

\begin{figure}[htbp]

\begin{subfigure}
\centering
\hspace{-1cm}
\includegraphics[trim={0cm 0cm 0cm 0cm}, clip, scale=0.31]{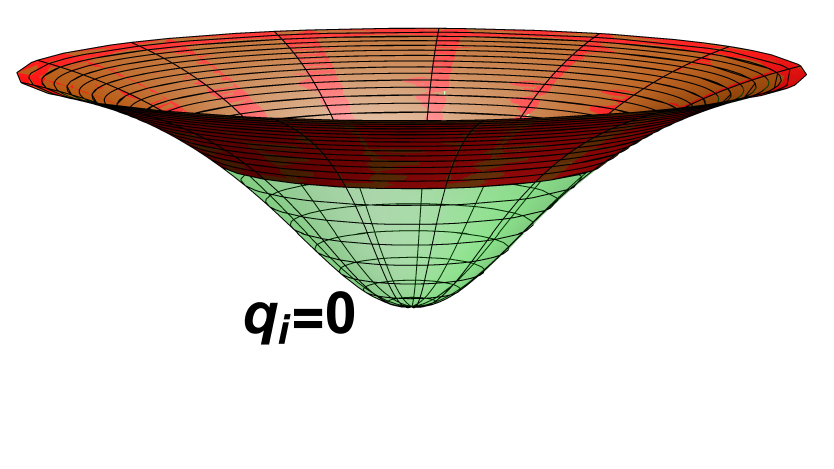}
\end{subfigure}
\hspace{-0.45cm}
\begin{subfigure}
\centering
\includegraphics[trim={0cm 0cm 0cm 0cm}, clip, scale=0.31]{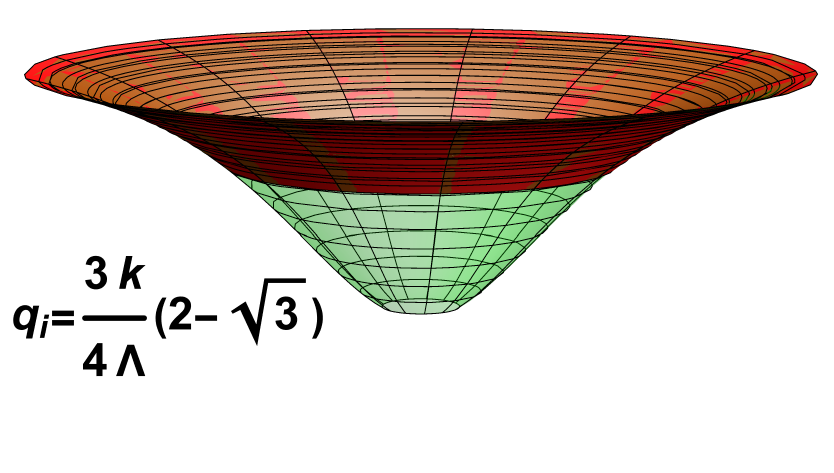}
\end{subfigure}
\hspace{-0.45cm}
\begin{subfigure}
\centering
\includegraphics[trim={0cm 0cm 0cm 0cm}, clip, scale=0.31]{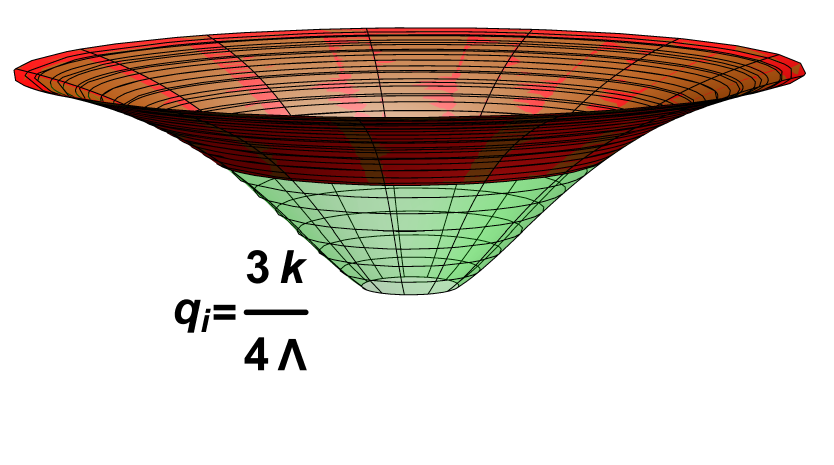}
\end{subfigure}
\hspace{-0.45cm}
\begin{subfigure}
\centering
\includegraphics[trim={0cm 0cm 0cm 0cm}, clip, scale=0.31]{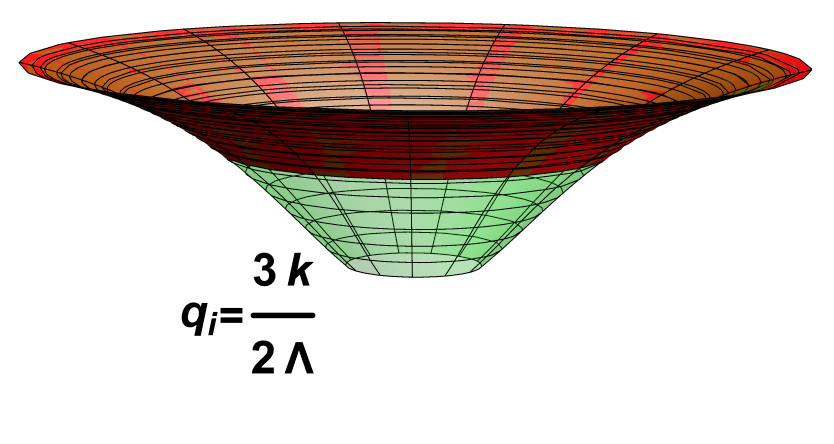}
\end{subfigure}


\begin{tikzpicture}
\draw[thick] (0,0) -- (\textwidth,0);

\foreach \x in {0.09,0.35,0.61,0.855}{
  \fill (\x\textwidth,0) circle (2.2pt);
}

\node[below] at (0.09\textwidth,-0.05) {$c=0$};
\node[below] at (0.35\textwidth,-0.05) {$c=\frac{3}{64\Lam}$};
\node[below] at (0.61\textwidth,-0.05) {$c=\frac{9}{64\Lam}$};
\node[below] at (0.855\textwidth,-0.05) {$c=\frac{3}{16\Lam}$};
\end{tikzpicture}
\caption{Illustration of spacetime geometries associated with the near no-boundary saddles $N_{\mathrm{nb}}^{(c_1,c_3)}$ for four representative values of the parameter $c$. The green region corresponds to Euclidean four sphere, while the brown region denotes the Lorentzian de Sitter $dS^4$ nucleated at the throat of radius $\sqrt{3k/\Lam}$. Only in the limit $c\rightarrow0$ one finds the no-boundary geometry corresponding to a universe that emerges smoothly from a vanishing size.  The value of $q_i$ for each choice of $c$ is indicated beneath the corresponding geometry.}
\label{fig:Near_nb_geo}
\end{figure}

\begin{figure}[htbp]

\begin{subfigure}
\centering
\hspace{-1cm}
\includegraphics[trim={0cm 0cm 0cm 0cm}, clip, scale=0.31]{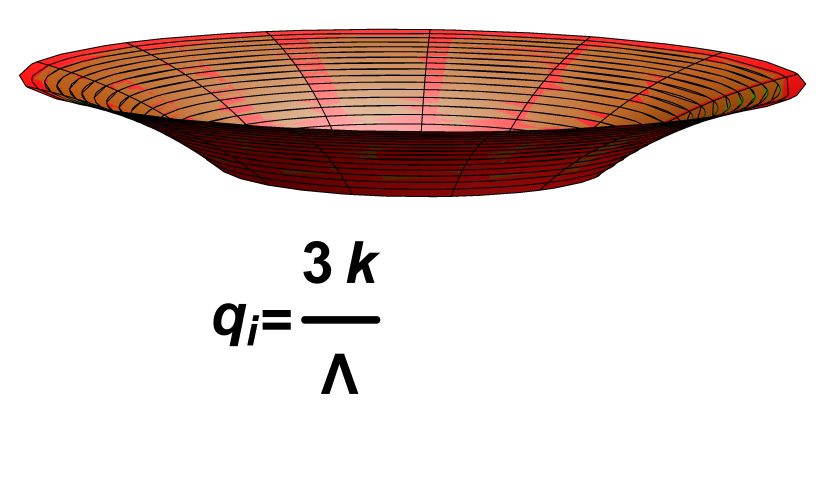}
\end{subfigure}
\hspace{-0.45cm}
\begin{subfigure}
\centering
\includegraphics[trim={0cm 0cm 0cm 0cm}, clip, scale=0.31]{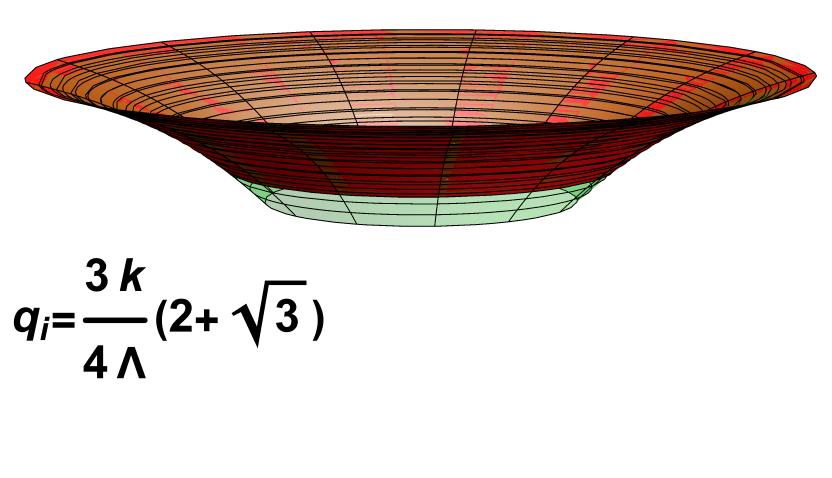}
\end{subfigure}
\hspace{-0.45cm}
\begin{subfigure}
\centering
\includegraphics[trim={0cm 0cm 0cm 0cm}, clip, scale=0.31]{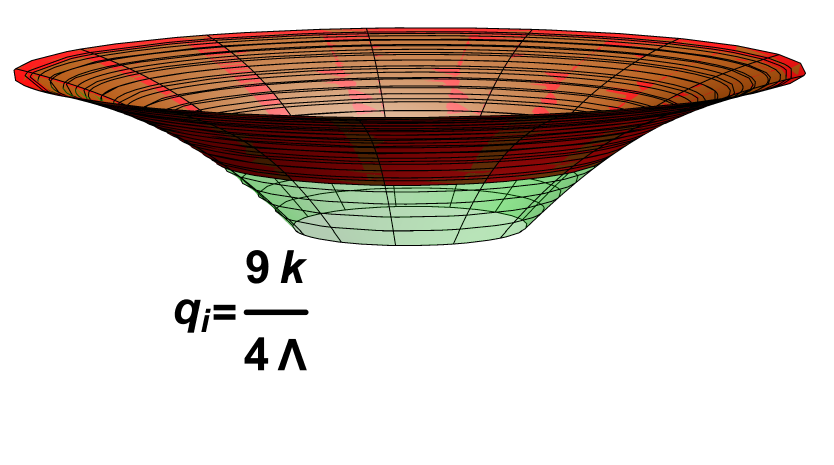}
\end{subfigure}
\hspace{-0.45cm}
\begin{subfigure}
\centering
\includegraphics[trim={0cm 0cm 0cm 0cm}, clip, scale=0.31]{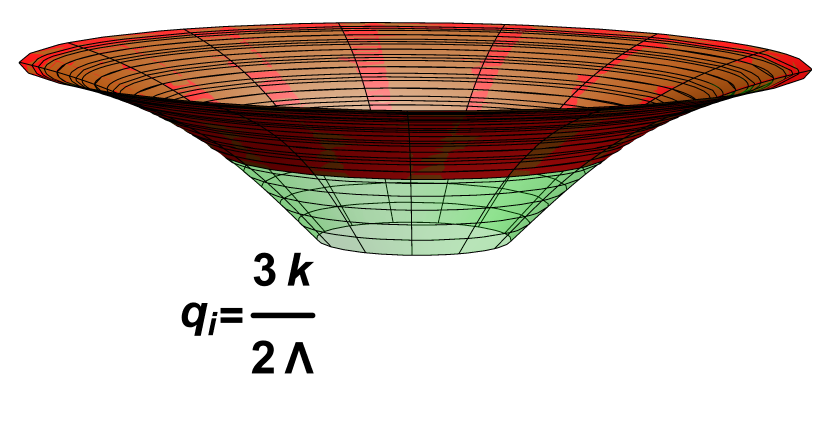}
\end{subfigure}


\begin{tikzpicture}
\draw[thick] (0,0) -- (\textwidth,0);

\foreach \x in {0.09,0.35,0.61,0.855}{
  \fill (\x\textwidth,0) circle (2.2pt);
}

\node[below] at (0.09\textwidth,-0.05) {$c=0$};
\node[below] at (0.35\textwidth,-0.05) {$c=\frac{3}{64\Lam}$};
\node[below] at (0.61\textwidth,-0.05) {$c=\frac{9}{64\Lam}$};
\node[below] at (0.855\textwidth,-0.05) {$c=\frac{3}{16\Lam}$};
\end{tikzpicture}
\caption{A figure similar to Fig. [\ref{fig:Near_nb_geo}] for near de-Sitter saddles $N_{\mathrm{dS}}^{(c_1,c_3)}$.
In the limit $c\rightarrow0$, we obtain $q_i=3k/\Lambda$ which indicates the universe emerges from de-Sitter throat with no Euclidean segment. 
As the parameter $c$ increases,the size of the initial hypersurface displaces towards $q_i=3k/(2\Lambda)$ in the Euclidean signature. }
\label{fig:Near_dS_geo}
\end{figure}

From the expression in eq. (\ref{eq:ini_sc_nb}), one may immediately realize a bound on $c$ obtained by requiring  $q(0)\big|_{\rm saddles} \geq 0$. This ensures that the initial geometry is purely real and non-singular. This condition imposes the constraint on the allowed range: $ 0 \leq c \leq \frac{3}{16\Lambda}$, such that the Universe can only start from a non-negative size, either from nothing or some finite real positive radius but not as complex or negative sized. It is worth noticing that all the initial momentum $\pi_{i}^{(c_1,c_2)}$ choices are purely imaginary in this regime. The geometries associated with these saddle points are illustrated in Figs. [\ref{fig:Near_nb_geo}] and [\ref{fig:Near_dS_geo}]. From the Fig. [\ref{fig:Near_nb_geo}], in the $c \to 0$ limit, it is evident that the initial geometry of the saddles $N_{\text{sad}}^{(c_1,1,c_3)}$ , vanishes i.e \( q(0) \to 0 \) in a smooth and regular way, yielding the universe originated from nothing. Indeed, in this limit, one finds the associated initial momentum as $\pi^{(c_1,1)} \to 3ic_{1}\sqrt{k}$. Interestingly, these correspond to the well-known complex initial conditions that give rise to the no-boundary universe. In particular, the choice $\pi_{i}^{HH} = -3i\sqrt{k}$ corresponds to Hartle and Hawking proposal, while $\pi_{i}^{V} = 3i\sqrt{k}$ corresponds to Vilenkin tunneling proposal. \footnote{A related recent study \cite{Yamada:2026lzg} proposes a small-throat prescription for the wavefunction of a closed universe, in which the universe creation viewed as the decoupling, or pinch-off, limit of a tunneling geometry connected to another universe through a small throat.}

Similarly, Fig. [\ref{fig:Near_dS_geo}] shows that the initial geometry corresponding to the saddles $N_{\text{sad}}^{(c_1,-1,c_3)}$, approaches the de Sitter throat in the limit $c \to 0$. Furthermore in this scenario the associated initial momentum vanishes, which is precisely the boundary condition characterizing the de Sitter throat geometry. For \( c \) taking values in the interval \( 0 < c \leq \tfrac{3}{16\Lambda} \), the universe instead begins from Euclidean signature with a finite, regular three-geometry. In view of the above characteristics, we refer to the saddles $N_{\text{sad}}^{(c_1,1,c_3)}$ as near no-boundary saddles $(N_{\text{nb}}^{(c_1,c_3)})$, while $N_{\text{sad}}^{(c_1,-1,c_3)}$ as near deSitter saddles $(N_{\text{dS}}^{(c_1,c_3)})$. Among the near no-boundary saddles, $N_{\rm nb}^{(-1,c_3)}$ corresponds to the Hartle–Hawking saddle point and $N_{\rm nb}^{(+1,c_3)}$ corresponds to the Vilenkin (tunneling) saddle point in the limit $c \to 0$.

Given the saddle points, one may proceed to compute the wavefunction mentioned in eq. (\ref{eq:Lap_int}) utilizing Picard Lefschetz (PL) technique \cite{Witten:2010cx, Witten:2010zr, Basar:2013eka, Tanizaki:2014xba, Lehners:2018eeo, Narain:2021bff, Ailiga:2024mmt}. According to PL theory, one analytically continue the real $N_c$ to a complex plane $N_c=N_x+i\,N_y$. With $N_c$ is being complex, the exponent $iS[N_c]$ becomes a holomorphic function of $N_c$ with real ($h$) and imaginary ($H$) part: $iS[N_c]=h(N_x,N_y)+iH(N_x,N_y)$. Each of the saddle point $N_{\sg}$ define an associated distinct Lefschetz thimble \( \mathcal{J}_\sigma \) along with its corresponding steepest-ascent contour \( \mathcal{K}_\sigma \). Collectively, these are referred to as the flow lines associated with the saddle. Picard-Lefschetz methodology then dictates the saddles which become {\it relevant} (as the steepest ascent starting from them hits the original integration contour ${\cal C}$), based on whether the corresponding steepest descent lines (also called {\it Lefschetz thimbles}) constituting the deformed integration contour or not. Schematically, this can be written as 
\begin{equation}
\label{eq:calC_def_breakup}
\mathcal{C}=\sum_\sigma n_\sigma \mathcal{J}_\sigma,
\hspace{5mm} \,n_\sigma={\rm Int} (\mathcal{C}, {\cal K}_\sg),
\end{equation}
where $\rm Int (. , .)$ counts the intersection between two curves and satisfies $\rm Int (\mathcal{J}_\sigma,\mathcal{K}_{\sigma'})=\delta_{\sigma\sigma'}$. The $n_\sigma$ is known as intersection number, that takes values $(0,\pm 1)$. A non-zero intersection number, $n_{\sg} \neq 0$ (steepest ascent intersects with the original integration contour) implies that the corresponding saddle point is relevant and contributes to the wavefunction. In contrast, saddles with zero intersection number, $n_{\sg} = 0$ saddles does not contribute (see Refs. \cite{Mahajan:2025bzo, Singhi:2025rfy, Ailiga:2025osa, Kolanowski:2026gii, Barbon:2026tri, Kolanowski:2026ylt} for recent application of PL.). 

Exploiting the property, the conservation of phase factor $H$ along the flow lines, the necessary condition to determine the relevance of a saddle point $N_\sg$ reads as: there must exist a real $N_x$ such that it satisfied 
\beq
\label{eq:rel_con}
H(N_x,0) =  H_{\sg} \quad \text{with} \quad h(N_x,0) \geq h_\sg \, ,
\eeq
where $h_\sg$ and $H_{\sg}$ are the real and imaginary part of exponent evaluated at the saddle point $N_\sg$. The number of solutions $N_x$ to the above equation determines the number of intersections between the steepest-ascent contour associated with a given saddle and the original integration contour. After carefully ignoring the intersection from the ghost solutions—i.e., trajectories that satisfy the flow equations but are not continuously connected to the saddle — the presence of an odd number of such solutions guarantees that the saddle point is relevant (modulo orientation), whereas the absence or the presence of even number of solutions indicates that the saddle does not contribute to the wavefunction. Having identified the {\it relevant} saddle points and determined the corresponding Lefschetz thimble decomposition of the original contour. The integral in eq.~(\ref{eq:Lap_int}), in the saddle-point approximation using the Picard–Lefschetz formalism is given by 
\beq
\label{eq:G_RBC_sad_appr}
\Psi[{\rm{Bd_f},\rm{Bd_i}}] \approx \sum_{\sg} n_\sg \sqrt{2\pi\hbar}
\,\, 
e^{i\pi/4} (-1)^m 
{\bar S}_{\rm grav}^{\prime\prime}[N_\sg]^{-1/2}
\exp\bl[i {\bar S}_{\rm grav}^{(q)}[N_\sg]\br] \, ,
\eeq
where $m$ takes $0$ or $1$ and $\sg$ sum over saddles.

\subsection{$q_f > 3k/\Lam$}
\label{qf>3k/Lam}
We now proceed to apply the Picard–Lefschetz (PL) formalism to our background spacetime. Plugging $N_c=N_x + i N_y $ into the on-shell action mentioned in eq. (\ref{eq:stot_onsh_nbc}) with $\omega=-1/2$ or $n=-1$, one finds the real and imaginary parts of the exponent at a generic lapse point as
\bea
\label{eq:h_H_gen_lp_qf>3k/Lam}
h[N_x,N_y] & = & \frac{1}{9}\biggl(N_y - \frac{y}{\Lam}\biggr)^3 + \frac{y^3}{9\Lam^3} + \biggl(\frac{\Lam N_x^2}{3}- q_f\biggr)(y-\Lam N_y) - 3k N_y   - \frac{36 c k^2 }{y} \, , \\
H[N_x,N_y] & = & \frac{N_x}{9} \biggl( N_x^2 \Lam^2 - 9\Lam q_f + 27k - 3\{y - \Lam N_y\}^2 \biggr) \, .
\eea
In deriving the above expressions, we have taken the initial momentum to be purely imaginary, $\pi_{i}=i y$ with $y \in \mathbb{R}$. For the universe in the lorentzian phase i.e $q_f > 3k/\Lam$,  we have at the saddle points $N_{\text{sad}}^{(c_1,c_2,c_3)}$, the above expressions simplifies to:
\bea
\label{eq:h_H_sad}
h_{\sg}  =  \frac{y^3 - 27 k y}{9\Lam} -  \frac{36 c k^2 }{y} \, , \quad \& \quad H_{\sg}  = - c_3\frac{2\left(q_f\Lambda-3k\right)^{3/2}}{\sqrt{3}\Lambda} \, ,
\eea
where $y$ at the saddles can be obtained from eq. (\ref{eq:dom_pi}). 
\begin{figure}[htbp]
\label{PL_graph1a}
\centering

\begin{minipage}[t]{0.48\textwidth}
    \centering
    \includegraphics[width=\linewidth]{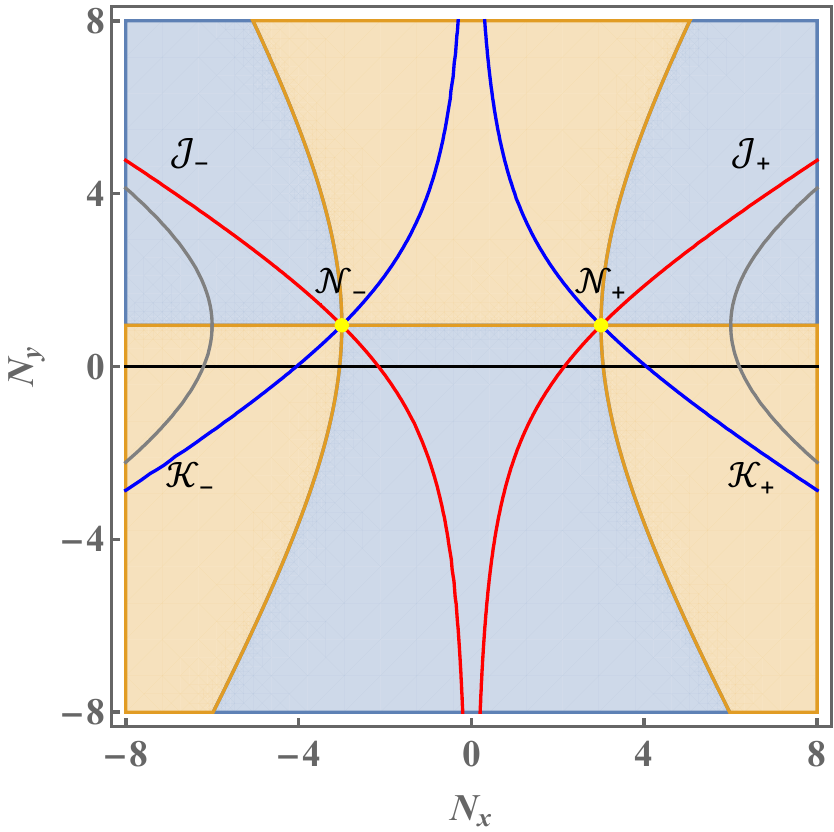}
    
    \small (a) Corresponds to $N_{\text{nb}}^{(1,c_3)}$ saddles
\end{minipage}
\hfill
\begin{minipage}[t]{0.48\textwidth}
    \centering
    \includegraphics[width=\linewidth]{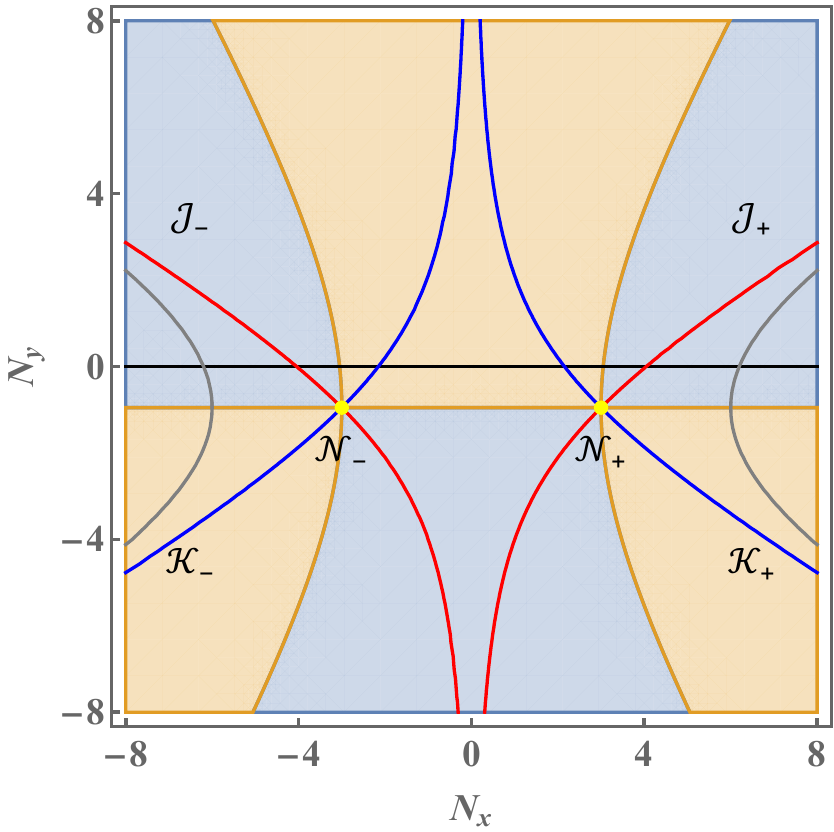}
    
    \small (b) Corresponds to $N_{\text{nb}}^{(-1,c_3)}$ saddles
\end{minipage}

\caption{The PL plot showing the relevance of near no-boundary saddles, denoted by yellow dots for the Lorentzian regime $q_f>3k/\Lam$. Similar PL analysis can be carried out for de-Sitter saddles, that reveals their relevance to wavefunction. The Lefschetz thimbles $\mathcal{J}_\pm$ shown as red curves, lie within the blue wedges, where $h<h_{\sg}$. The steepest ascent flowlines $\mathcal{K}_\pm$ are depicted by blue lines, lie within the orange wedges, where $h>h_{\sg}$. The ghost flow lines are denoted by gray lines. Here we choose parameter values: $k=1, \Lam=3, c=1/48 \ \text{and} \  q_f=10$. For notational convenience, in all subsequent PL plots, we denote $(N_{\mathrm{sad}}^{(c_1,c_2,1)})$ by $N_{+}$ and $(N_{\mathrm{sad}}^{(c_1,c_2,-1)})$ by $N_{-}$, for the respective values of $c_1$ and $c_2$. }
\label{fig:PL_near_nb}
\end{figure}
Having obtained these expressions, we solve the relevance equation $H[N_x,0]=H_\sigma$, to determine the relevance of each saddle point $N_{\sg}$. Explicitly, the relevance condition is given as
\beq
\label{cubic.eq}
\frac{\Lambda^2}{9}N_x^3 + \left(3k +\frac{\pi_i^2}{3} - q_f\Lambda\right)N_x = -c_3\frac{2\left(q_f\Lambda-3k\right)^{3/2}}{\sqrt{3}\Lambda} \, .
\eeq
As discussed before, a saddle point is said to be relevant and contributing to the wavefunction if there exists a real solution $N_x$ satisfying $h[N_x,0] > h_{\sg}$, which ensures the intersection between corresponding steepest ascent and the real line. Without solving the cubic equation, useful insights can be gained by looking at its discriminant $\Delta_{dis} $, which is given as
\beq
\label{eq:Discr_cub}
\Delta_{dis} = \frac{4\Lam^2y^2}{9}\left[\frac{y^4}{27} + \left(q_f\Lam-3k\right)^2 + y^2\left(\frac{q_f\Lam}{3} - k\right)\right], \ \text{where} \ \pi_i\rightarrow i y \, .
\eeq
One finds the discriminant is strictly positive for all saddle points when for $q_f>3k/\Lam$, implying the solutions are distinct real roots. Further analysis reveals that there always exists at least one real solution satisfying $h[N_x,0]>h_\sigma$ for all saddle points. This confirms that the steepest ascent emanating from the saddle point intersects the real line making the corresponding saddle point relevant to the lapse integral. The analysis and results remain similar for all saddles, which suggests that all eight saddle points given in eq. (\ref{sad_point}) are relevant to the wavefunction for $q_f > 3k/\Lam$. Additionally, there exists, an alternative selection criterion of saddle points, namely KSW criterion, which identifies the physically admissible saddle points contributing to the wave function. We find all the saddles satisfy the KSW criterion, see appendix \ref{KSW} for details. The PL-plots corresponding to near No-Boundary saddle are demonstrated in Fig.[\ref{fig:PL_near_nb}]. Therefore, in the saddle point approximation the wavefunction is given by:
\bea
\label{G[bdf,Bdi]}
\Psi[q_f,c] \approx 
&&
\sum_{\pi_i} \left(\frac{\hbar}{2\Lambda\sqrt{\frac{q_f\Lambda}{3}-k}}\right)^{1/2}\text{exp}\Biggl[i\frac{2\pi^2}{\hbar}\left(-\frac{36ck^2}{\pi_i} + \frac{\pi_i^3}{9\Lambda} + \frac{3k\pi_i }{\Lambda}\right)\Biggr]
\notag \\
&&
\times \ \text{cos}\Biggl[\frac{12\pi^2}{\hbar \Lambda}\left(\frac{q_f\Lambda}{3}-k\right)^{3/2}-\frac{\pi}{4}\Biggr] \, ,
\eea
where the sum is taken over $\pi_i = \{\pi_i^{HH},\pi_i^{V},\pi_i^{dS}\}$. Moreover, in the range of interest $0\leq c < 3/(16\Lam)$, one finds the saddle characterized by $\pi_i^{HH}$ 
dominates over the other, see Fig. (\ref{fig:pi_dom}). Consequently, in the $\hbar \to 0 $ limit for $q_f > 3k/\Lam$, the wavefunction can be approximated as
\bea
\label{G[bdf,Bdi]_dom}
\Psi[q_f,c] \approx 
&&
\left(\frac{\hbar}{2\Lambda\sqrt{\frac{q_f\Lambda}{3}-k}}\right)^{1/2}\text{exp}\Biggl[i\frac{2\pi^2}{\hbar}\left(-\frac{36ck^2}{\pi^{HH}_i} + \frac{(\pi^{HH}_i)^3}{9\Lambda} + \frac{3k\pi^{HH}_i }{\Lambda}\right)\Biggr]
\notag \\
&&
\times \ \text{cos}\Biggl[\frac{12\pi^2}{\hbar \Lambda}\left(\frac{q_f\Lambda}{3}-k\right)^{3/2}-\frac{\pi}{4}\Biggr] \, .
\eea
%

\subsection{$q_f < 3k/\Lam$}
\label{qf<3k/Lam}
\begin{figure}[htbp]
\centering
\subfigure[ 
\, For $N_{nb}^{(\pm 1,\pm1)}$   ]{
    \includegraphics[trim={0cm 0 0cm 0cm}, clip, scale=0.5]{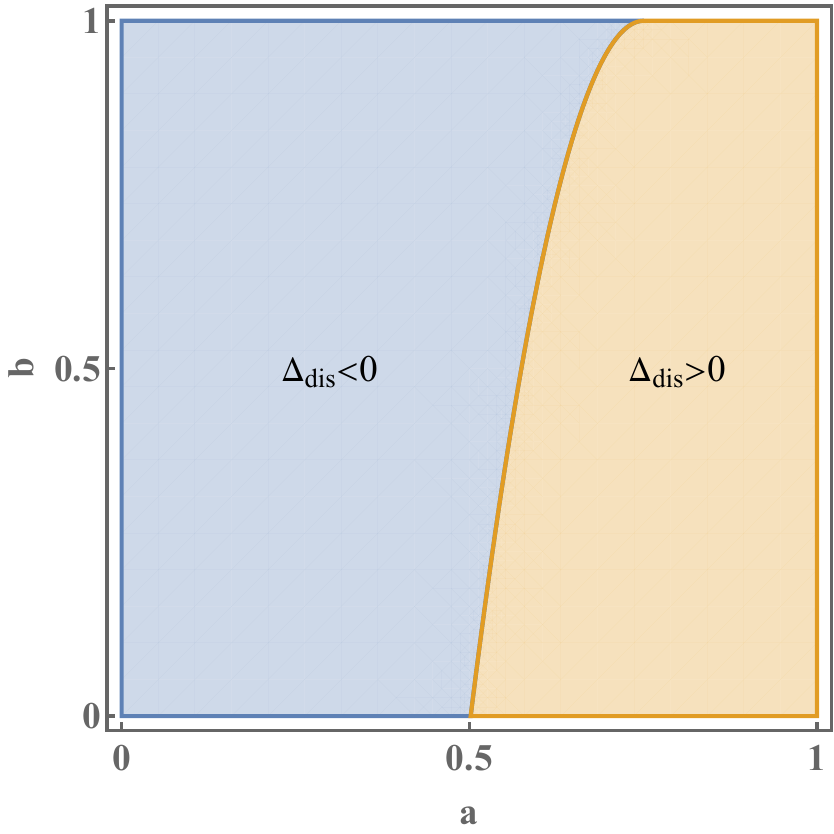}
}
\hspace{0.5cm}
\subfigure[ 
\, For $N_{nb}^{(\pm 1,\mp1)}$]{
    \includegraphics[trim={0 0 0cm 0cm}, clip,scale=0.5]{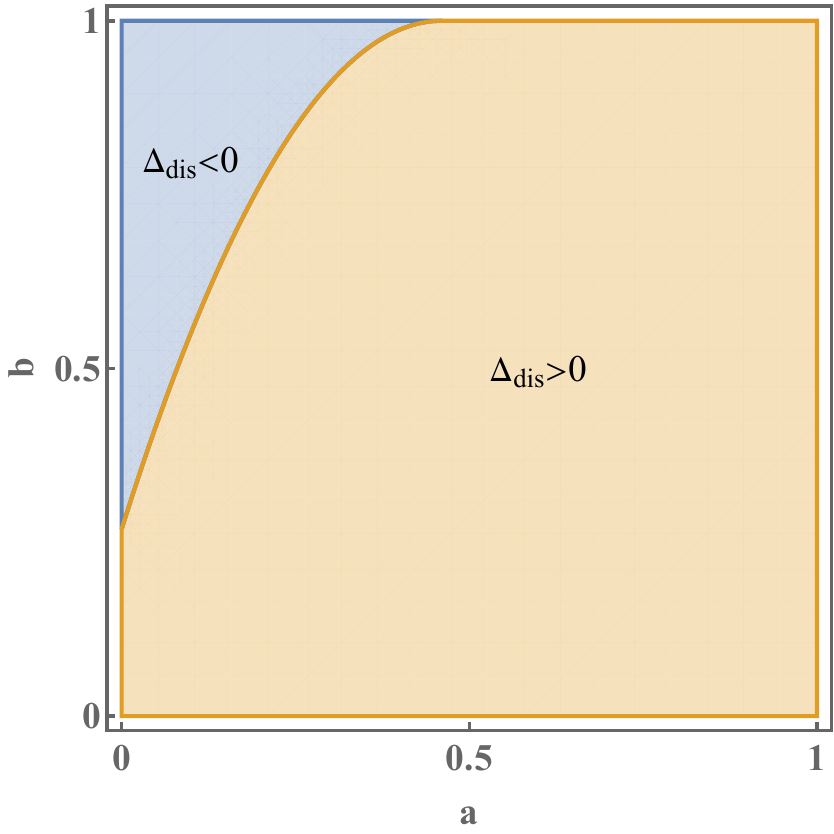}
}
\caption{ Behavior of relevance equation discriminant given in eq. (\ref{eq:dis_qf<3k/Lam})  when $q_{f}\leq3k/\Lam$ for various choice of $c_{1,2} = \pm 1$ . Here, we have rescaled the variables as $q_{f}=3k a/\Lam$ and $c=3 b/(16\Lam)$.}
\label{fig:Dis_qf_ls_qc}
\end{figure}

Similar analysis can be extended to $q_f < 3k/\Lam$, where the spacetime associated with the saddle points is Euclidean. In this case that the saddle points are purely imaginary for either choice of initial momentum $\pi_{i}^{(c_1,c_2)}$, reflecting the Euclidean nature of the corresponding geometry. The real and imaginary part of the on-shell exponent at the saddle points is given by:
\bea
\label{eq:h_H_sad_qf<3k/Lam}
h_{\sg}  =  \frac{y^3 - 27 k y}{9\Lam} -  \frac{36c k^2 }{y} -  \frac{2c_3}{9\Lam}(9 k - 3 q_f \Lam)^{3/2} \, , \quad \& \quad H_{\sg}  = 0 \, .
\eea
Unlike the $q_f > 3k/\Lam$, the $q_f < 3k/\Lam$ limit exhibits an interesting behavior. In particular, one finds $H_{\sg}=H_{\sg^{\prime}}$ with $h_{\sg} \neq h_{\sg^{'}}$ where $\sg \neq \sg^{\prime}$, indicating the presence of Stokes degeneracy. This is the kind of degeneracy, where the steepest descent  of one saddle overlaps with the steepest ascent of another. The drawback of this degeneracy is that it prevents us from unambiguously deciding the {\it relevant} saddle points, which comes from determining $n_\sigma$. Recall that in order to obtain a unique value of $n_\sigma$, one requires $\rm Int (\mathcal{J}_\sigma,\mathcal{K}_{\sigma'})=\delta_{\sigma\sigma'}$ which implies steepest descent $\mathcal{J}_\sg$ of one saddle can never intersect with the steepest ascent $\mathcal{K}_{\sg^{\prime}}$ of another saddle. Clearly, the presence of Stokes degeneracies violates this requirement and prevents one from implementing the PL method. Hence, one must lift this degeneracy to figure out the {\it relevant} saddles and the correct integration contour. Past studies \cite{Nishimura:2015loa, Tanizaki:2015pua, Tanizaki:2015gcv}, (for recent discussion see \cite{Ailiga:2025fny}) show that the cause of this degeneracy is tied up with the presence of anti-linear (AL) symmetry in the system, given by: 
\begin{equation}
\label{eq:antilin}
{\rm AL} 
\hspace{5mm}
\Rightarrow
\hspace{5mm}
N_c\rightarrow-N_c^*\, , 
\hspace{3mm}
N_x\to-N_x\, ,
\hspace{3mm}
N_y\to N_y \, ,
\end{equation}
where a point on the imaginary axis is mapped to itself and lies on the axis of symmetry. Under this symmetry, the on shell action $S^{(q)}_{\rm tot}[N_c]$ mentioned in eq. (\ref{eq:stot_nbc}) transform as, 
\begin{equation}
\label{eq:anti-linear1}
\{iS^{(q)}_{\rm tot}[N_c]\}^*=iS^{(q)}_{\rm tot}[-N_c^*]\, ,
\end{equation}
where $(\,.)^*$ implies complex conjugation. We emphasize that this symmetry exists only when $\pi_i$ is purely imaginary. Under this anti-linear symmetry, one finds the real and imaginary part of on-shell exponent satisfies:
\begin{equation}
\label{eq:handH}
h[N_x,N_y]=h[-N_x,N_y] \, ,
\hspace{5mm}
H[N_x,N_y]=-H[-N_x,N_y] \, .
\end{equation}
Along the imaginary axis $N_x=0$, we have
\begin{equation}
\label{eq:handH_im}
H[0,N_y]=-H[0,N_y] \, \implies H[0,N_y] =0  \quad \forall \, N_{y} \, .
\end{equation}
As a consequence of AL, all the purely imaginary saddles have same phase factor $H_{\sg} = 0$, giving rise to Stokes degeneracy. As pointed before, presence of such degeneracy hinders one from implementing PL mechanism, and thus urges us to lift it by alternative means. A straight forward way to lift is by breaking the AL symmetry of the system, which can be achieved in various ways. Specifically, the degeneracy can be lifted by introducing a small artificial defect, by incorporating quantum corrections \cite{Ailiga:2025fny} or by complex rotation of Newton constant $G$ or Planck constant $\hbar$ \cite{Honda:2024aro}, \cite{Tanizaki:2015gcv}. In this paper, we adopt the latter remedy. With the complex rotation of $\hbar \to \hbar e^{i\theta}$ with $|\theta| \ll 1$, the modified Morse function $h^{\theta}$ and the phase factor $H^{\theta}$ of on-shell exponent are given by:
\beq
\begin{split}
\label{eq:the_mod_h_H}
    h^{\theta}[N_x,N_y] & =  h[N_x,N_y]\cos{\theta} + H[N_x,N_y] \sin{\theta} \, ,\\
    H^{\theta}[N_x,N_y] & = H[N_x,N_y] \cos{\theta} - h[N_x,N_y]\sin{\theta} \, .
\end{split}
\eeq
It is worth emphasizing that, unlike the other approaches mentioned above, complex deformation of Planck length does not affect the saddle points. From the above equation, one finds the phase factor at the saddle point receives $\theta$ corrections, as follows:
\beq
\label{eq:H_mod_sad_qf<3k/Lam}
H^{\theta}_{\sg}  =  - h_{\sg}\sin{\theta} \, ,
\eeq
where $h_{\sg}$ is mentioned in eq. (\ref{eq:h_H_sad_qf<3k/Lam}). It is clearly evident that the lifting of Stokes degeneracy as $H^{\theta}_{\sg} \neq H^{\theta}_{\sg^{\prime}} $ as $h_{\sg} \neq h_{\sg^{\prime}}$ for $\sg \neq \sg^{\prime}$. With the degeneracy resolved, we proceed to analyze the relevance of the saddle-point contributions to the wavefunction. With the complex deformation of Planck constant, the sufficient condition for the relevance of a saddle point $N_{\sg}$ reads as: there must exist a real $N_x$ such that it satisfies 
\beq
\label{eq:rel_mod_con}
H^{\theta}(N_x,0) =  H^{\theta}_{\sg} \quad \text{with} \quad h^{\theta}(N_x,0) \geq h^{\theta}_\sg \, .
\eeq
In terms of $h$ and $H$, the condition simplifies to 
\beq
\label{eq:rel_mod_con_sim}
\begin{split}
    H(N_x,0) \cos{\theta} - \{h(N_x,0) - h_{\sg}\} \sin{\theta} & = 0 \, , \quad \text{with}\\
     \quad \{h(N_x,0)-h_{\sg}\} \cos{\theta} &\geq - H(N_x,0)\sin{\theta} \, .
\end{split}
\eeq
Utilizing the expressions given in eqs. (\ref{eq:h_H_gen_lp_qf>3k/Lam}) and (\ref{eq:h_H_sad_qf<3k/Lam}), we obtain the relevance condition as
\beq
\label{eq:rel_mod_con_exp}
 N_x \Lam \{N_x^2\Lam^2 - 3 y^2 - 9\Lam q_f + 27k\} + \{  y (y^2 - 3 N_x^2 \Lam^2  + 9 \Lam q_f  - 27k )  - 2c_3  (9k-3 q_f\Lam)^{3/2}\} \tan{\theta} = 0 \, .
\eeq
As in the case of $q_f>3k/\Lam$, we analyze the discriminant of above equation to understand the nature of roots. The discriminant of eq. (\ref{eq:rel_mod_con_exp}) is given by
\beq
\label{eq:dis_qf<3k/Lam}
\begin{split}
    \Delta_{dis}  =  \frac{4\Lam^2}{243} & \biggl[ y^3 \{y^3  + 9 y q_f  \Lam -27 ky - 2 c_3 (9k-3q_f\Lam)^{3/2} \sin^2{\theta}\}
      \\ 
    &  - (9k - 3q_f\Lam)^2 (9k-3q_f\Lam -3 y^2)\cos^2\theta \biggr] \, .
\end{split}
\eeq
Unlike for $q_f > 3k/\Lam$ case, the discriminant for $q_{f}<3k/\Lam$ can be negative, positive or zero in Fig.[\ref{fig:Dis_qf_ls_qc}]. Let us study the discriminant in more detail. One immediate feature is discriminant remains same under $\theta \to - \theta$, as a consequence nature of roots remain invariant for either choice of complex rotation $\hbar \to \hbar e^{\pm i \theta}$. It is rather challenging to treat the problem fully analytically. Instead, we present the two possible Picard–Lefschetz (PL) scenarios that collectively represent all choices of the parameters $c_{1}, c_2$ and $c_3$ choices arising under the deformation.
%
\subsubsection{Case:1}
\label{cs:1}
Consider the following parameters $k=1, \Lam=3,\theta=\pi/10, c=1/48$ and $ q_{f}=1/2$ with the initial choice of momentum as $c_1=-1$ and $c_2=1$. Fig. [\ref{fig:Dis_qf_ls_qc}] shows that the discriminant for $N_{\text{nb}}^{(-1,1)}$ is positive while the discriminant of $N_{\text{nb}}^{(-1,-1)}$ is negative. As a consequences one finds that flow lines emerging from $N_{\text{nb}}^{(-1,1)}$ saddle intersect the original integration contour at three points (cf. Fig.~\ref{fig:PL_qf_ls_qc}). Of these, two intersections correspond to the associated Lefschetz thimble, while only one corresponds to the steepest-ascent contour, which makes the $N_{\text{nb}}^{(-1,1)}$ contributing saddle to the wavefunction.  Analogously, since the discriminant associated with the $N_{\text{nb}}^{(-1,-1)}$ is negative, there is only a single intersection between its flow lines and the real axis. As shown in Fig. (\ref{fig:PL_qf_ls_qc}), this intersection occurs between the Lefschetz thimble and the original integration contour, while there is no intersection between the steepest-ascent contour of $N_{\text{nb}}^{(-1,-1)}$ with the real axis making it irrelevant according to the PL prescription. Thus, one concludes that the $N_{\text{nb}}^{(-1,1)}$ saddle as the solely contributing saddle to the wavefunction for $q_{f} \leq 3k/\Lam$.

\subsubsection{Case:2}
\label{cs:2}
\begin{figure}[htbp]
\centering
\subfigure[  Corresponds to $N_{\text{nb}}^{(-1,c_3)}$ saddles ]{
    \includegraphics[trim={0cm 0 0cm 0cm}, clip, scale=0.55]{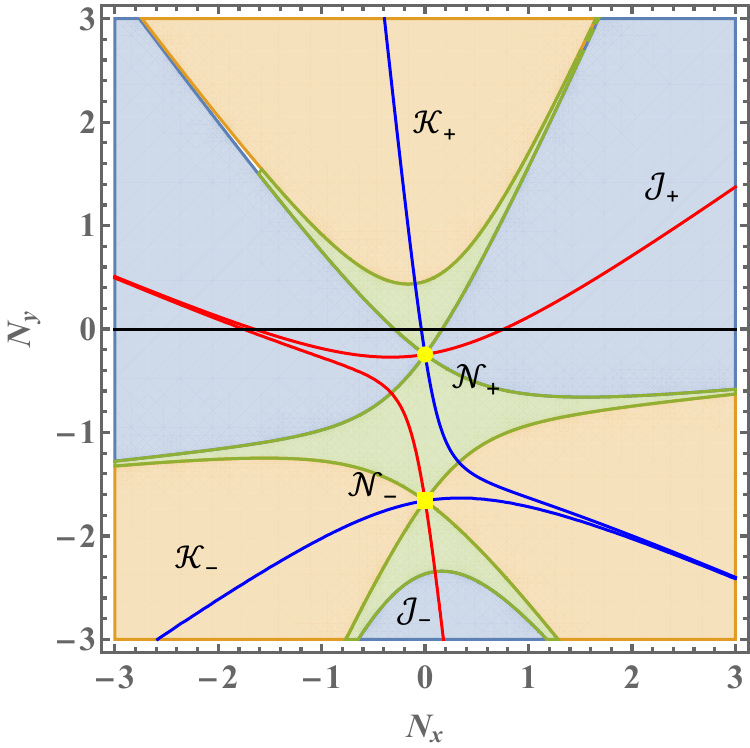}
}
\hspace{0.5cm}
\subfigure[ Corresponds to $N_{\text{nb}}^{(1,c_3)}$ saddles ]{
    \includegraphics[trim={0 0 0cm 0cm}, clip,scale=0.55]{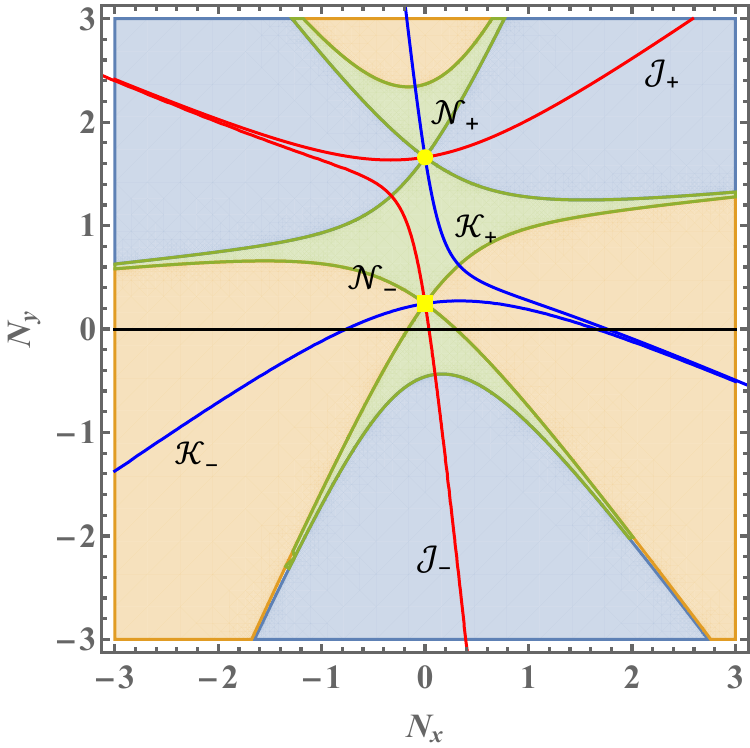}
}
\caption{The PL plot showing the relevant saddles (yellow dot) and irrelevant saddles (yellow square)
in Euclidean regimes. For demonstration, we choose parameter values as $k=1, \Lam=3, c=1/48 \ \text{and} \  q_f=1/2$. We complexify Planck’s constant, $\hbar=|\hbar|e^{i\theta} $ where we take $|\hbar|=1$ and $\theta=\pi/10$.}
\label{fig:PL_qf_ls_qc}
\end{figure}

We now consider the another case $c_1=1$ and $c_2=1$ for the same choice of parameters as in the case 1. This situation is more subtle and requires a little careful analysis to determine the relevant saddle point. From Fig. [\ref{fig:Dis_qf_ls_qc}] one may notice that the discriminant of $N_{nb}^{(1,1)}$ is negative while that of $N_{nb}^{(1,-1)}$ is positive. Thus we have a single intersection of flowline of $N_{nb}^{(1,1)}$, more precisely the steepest ascent with the real axis, making it relevant saddle (cf. Fig.[\ref{fig:PL_qf_ls_qc}]). While positive nature of discriminant indicates the steepest-ascent contour associated with the $N_{nb}^{(1,-1)}$ saddle intersects the real axis at two points (cf Fig.[\ref{fig:PL_qf_ls_qc}]). However, these intersections occur with opposite orientations. Assigning $n_{\sg} = +1$ to the flow line crossing from the negative to the positive imaginary direction (corresponding to the first intersection), and $n_\sg =-1$ to the second, the net intersection number vanishes. This effectively makes that $N_{nb}^{(1,-1)}$ irrelevant saddle despite the presence of intersections. Therefore, we conclude that $N_{nb}^{(1,1)}$ is the relevant contributing saddle.
Following the above analysis, one can generalize to arbitrary choices of parameters and finds that the system falls into one of the two cases discussed above. In all such scenarios, the only relevant contribution arises from the $N_{\rm sad}^{(c_1,c_2,1)}$ saddle for $q_f< 3k/\Lam$ for any choice of $c_{1,2}$. Thus in the saddle point approximation, we have the transition amplitude as:
\bea
\label{G[bdf,Bdi]for qf<3k/Lam}
\Psi[q_f]\approx 
&&
\sum_{\pi} \left(\frac{\hbar}{2\Lambda\sqrt{k-\frac{q_f\Lambda}{3}}}\right)^{1/2}\text{exp}\Biggl[i\frac{2\pi^2}{\hbar}\left(-\frac{36ck^2}{\pi_i} + \frac{\pi_i^3}{9\Lambda} + \frac{3k\pi_i }{\Lambda}\right)\Biggr]
\notag \\
&&
\times \ \text{exp}\left[-\frac{12\pi^2}{\hbar\Lam}\left(k-\frac{q_f\Lam}{3}\right)^{3/2}\right] \, ,
\eea
where the sum is understood over $\pi_i =\{\pi_i^{HH},\pi_i^{V},\pi_i^{dS}\}$.
\subsection{$q_f = 3k/\Lam$}
\label{qf=3k/Lam}
\begin{figure}[htbp]
\centering
\subfigure[ Corresponds to $N_{\text{nb}}^{(-1,c_3)}$ saddles  ]{
    \includegraphics[trim={0cm 0 0cm 0cm}, clip, scale=0.5]{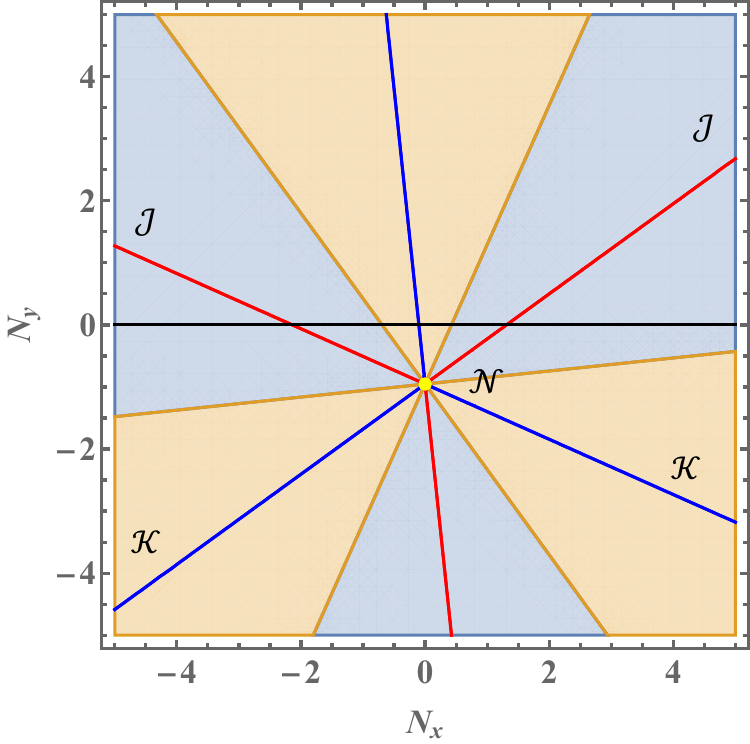}
}
\hspace{0.5cm}
\subfigure[ Corresponds to $N_{\text{nb}}^{(1,c_3)}$ saddles  ]{
    \includegraphics[trim={0 0 0cm 0cm}, clip,scale=0.5]{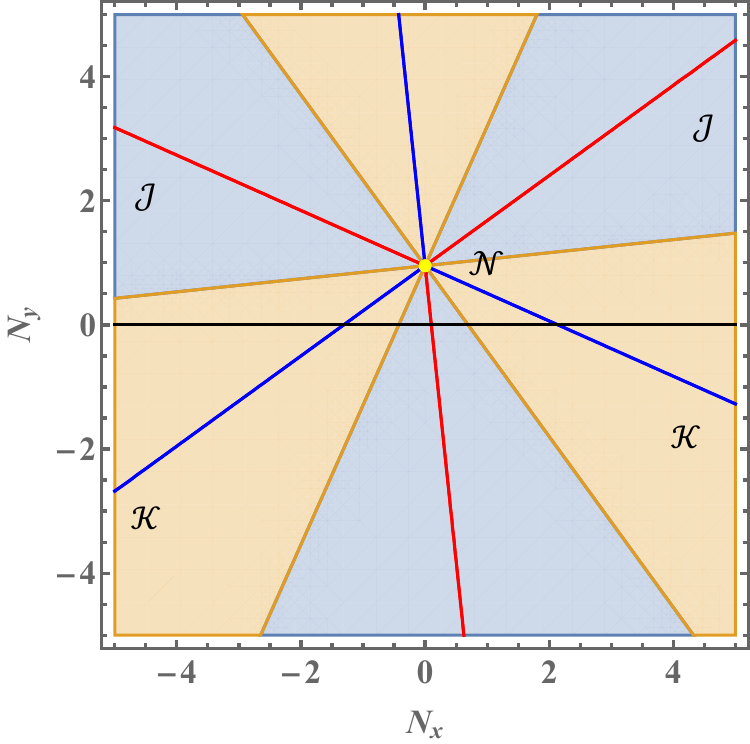}
}
\caption{ The PL plot shows the relevant saddles (yellow dot) in the transition regime from Euclidean to Lorentzian. This is degenerate case where $N_{\text{sad}}^{(c_1,c_2,1)}$ and $N_{\text{sad}}^{(c_1,c_2,-1)}$ merge.}
\label{fig:PL_qf_at_qc}
\end{figure}

This is a degenerate situation, one finds the coalesce of saddle points. In particular, $N_{\rm sad}^{(c_{1},c_2,1)}$ and $N_{\rm sad}^{(c_{1},c_2,-1)}$ merge:
\beq
\label{eq:deg_sad}
N_{\rm sad}^{(c_{1},c_2,c_3)} = \pi_{i}^{(c_1,c_2)}/\Lam \, .
\eeq
Notice that, in this case as well, the saddle turns out to be purely imaginary. Furthermore, this corresponds to the transition point between the Euclidean sphere and Lorentzian de Sitter geometry. As there is only one degenerate saddle for each choice of initial momentum, there is no Stokes degeneracy ambiguity in this case, unlike the $q_f < 3k/\Lam$. 
The relevance condition for the saddle point can be obtained by substituting $q_f=3k/\Lam$ in eq. (\ref{eq:rel_mod_con_exp}) with $\theta = 0$, which is given by:
\beq
\label{eq:rel_mod_con_exp_qf_qc}
 N_x \Lam \{N_x^2\Lam^2 - 3 y^2 \}  = 0 \, ,
\eeq
This indicates that the flow lines associated with each saddle intersect the original integration contour at three points, namely: $N_x=0$ and $N_x = \pm \sqrt{3}y/\Lam$, with at least one of these intersections corresponding to the steepest-ascent contour (see Fig \ref{fig:PL_qf_at_qc}).

According to the Picard–Lefschetz prescription, all saddles are relevant and contribute to the wavefunction. However, in this case the second derivative computed at the relevant saddles vanishes: ${\bar S}_{\rm tot}^{\prime\prime}[N_{\rm sad}]=0$. This signals a breakdown of the saddle-point approximation, indicating one needs to go beyond second order approximation. The transition amplitude at the third order is given by
\beq
\label{eq:G_sad_trd_odr_exp}
\Psi_{\rm NBC}[q_{f}=3k/\Lam] \approx \sum_{\sg} n_{\sg} \frac{(6\hbar)^{1/3}\Gamma(1/3)}{\sqrt{3}} \bl[{\bar S}_{\rm tot}^{\prime\prime\prime}[N_\sg]\br]^{-1/3} e^{i{\bar S}_{\rm tot}[N_\sg]/\hbar}\, ,
\eeq 
where $\sg$ sum over saddles. Utilizing the third derivative of the lapse action computed at the relevant saddle points ${\bar S}_{\rm tot}^{\prime\prime\prime}[N_s] = 2 \Lam^2/3 \,$, the transition amplitude in the degenerate case where $q_f=3k/\Lam$, is given by
\beq
\label{G[bdf,Bdi]for_qf_3k/Lam}
\Psi[q_f = 3k/\Lam]\approx 
\sum_{\pi_i} \frac{\Gamma(1/3)}{\sqrt{3}}\biggl(\frac{9\hbar}{\Lam^2}\biggr)^{1/3}\text{exp}\Biggl[i\frac{2\pi^2}{\hbar}\left(-\frac{36ck^2}{\pi_i} + \frac{\pi_i^3}{9\Lambda} + \frac{3k\pi_i }{\Lambda}\right)\Biggr]\, .
\eeq
where the sum is understood over $\pi_i  = \{\pi_i^{HH},\pi_i^{V},\pi_i^{dS}\}$.

\section{Perturbative stability of saddles}
\label{pert}
Let us now shift our focus to analyzing the cosmological perturbations around the above homogeneous saddle geometries. The most important ones are scalar and tensor perturbations, respectively leading to density perturbations and gravitational waves. In this article, we will focus on the tensor perturbations, as they form a part of geometry. The analysis of scalar perturbation closely goes in similar fashion. In the following section, we study the behavior of these perturbations around the relevant saddles, both at the on-shell level and one-loop corrections. Such an exploration is not new and has been done previously for the case of Dirichlet, Neumann and Robin boundary conditions around the Hartle-Hawking and Vilenkin saddles \cite{Feldbrugge:2017fcc, Vilenkin:2018oja, DiazDorronsoro:2018wro,  Matsui:2024bfn, Ailiga:2024wdx, Mallik:2026iai} and see also \cite{Mondal:2025dsc} for Horava–Lifshitz gravity. Rigorous studies have shown that tensor perturbations around Hartle–Hawking saddles are suppressed, whereas those around tunneling saddles can exhibit unsuppressed behavior \cite{Feldbrugge:2017fcc, Lehners:2023yrj}. However, with an appropriate choice of boundary conditions and a careful treatment, it has been argued that perturbations in the tunneling proposal can also remain stable, \cite{Vilenkin:2018oja}. Along the similar line, these perturbations showed stable behavior when examined on pure de Sitter background \cite{Mallik:2026iai}. In this work, we address these stability issues systematically for the different choices of $c_{1,2}$ which specify the boundary conditions on the initial momentum around the respective saddles. 

We impose trivial boundary conditions on perturbation on $h_{ij}$ which is motivated due to covariance of the gravitational system as discussed in section (\ref{bound_act}). Specifically, we choose $h_{l}(t=0) =0$ at the initial hypersurface and $h_{l}(t=1) = h_{l}^{(1)}$ at the final hypersurface. The on-shell solution is given by eq. (\ref{eq:hl_sol_gen}), where the coefficients are given by eq. (\ref{eq:hl_dpdq_NBC}). From the eq. (\ref{eq:hh_onSH_NBC}), the on-shell action of tensor perturbation mode evaluated around the saddle points for the aforementioned boundary conditions becomes
\beq
\label{eq:pert_act}
i S^{(l)}_{\rm grav}[\bar{q}, \bar{h}_l, N_{\text{sad}}^{c_1,c_2,c_3}]
= \frac{q_f \left(h_l^{(1)}\right)^2}{16 \pi G}
\Biggl[
-  \frac{ic_3}{2}\sqrt{\frac{\Lambda q_f }{3} - 1}
+ 
\frac{i\, q_f}{4 N_{\text{sad}}^{c_1,c_2,c_3}} D_t\!\left\{ \ln W(t,0) \right\}\big\rvert_{t=1}
\Biggr] \, .
\eeq
Moreover, the action can be recast in the simpler form
\beq
\label{iaction.h}
i S^{(l)}_{\rm grav}[\bar{q}, \bar{h}_l, N_{\text{sad}}^{c_1,c_2,c_3}]
= -\frac{q_f (h_l^{(1)})^2}{32 \pi G}
\Biggl[
  i c_3\sqrt{\frac{\Lambda q_f }{3}-1}
+ c_1\frac{\Lambda q_f}{6}
\left(\frac{\mathbb{P}'_{1}{}^{l+1}[\tau_1^{\text{sad}}]}
           {\mathbb{P}_{1}^{\,l+1}[\tau_1^{\text{sad}}]}\right)
\left(\frac{1+ f(c)\, g_1(q_f)}
           {1+ f(c)\, g_2(q_f)}\right) 
\Biggr]\, ,
\eeq
where we define
\bea
\label{eq:fun_Defs}
 f(c) \equiv \frac{\mathbb{P}_{1}^{\,l+1}[\tau_0^{\text{sad}}]}{\mathbb{P}_{1}^{\,-(l+1)}[\tau_0^{\text{sad}}]} , \ g_1(q_f) \equiv \frac{\mathbb{P}'_{1}{}^{-(l+1)}[\tau_1^{\text{sad}}]}{\mathbb{P}'_{1}{}^{l+1}[\tau_1^{\text{sad}}]} \ \text{and} \ g_2(q_f) \equiv \frac{\mathbb{P}_{1}^{-(l+1)}[\tau_1^{\text{sad}}]}{\mathbb{P}_{1}^{\,l+1}[\tau_1^{\text{sad}}]} \, .
\eea
Here, $(')$ denotes derivative with respect to time variable $\tau$ which is related to the time co-ordinate $t$ at a generic lapse point through:
\bea
\label{eq:tau_def}
\tau_t=\frac{i c_1}{3\sqrt{k}}\left(-\pi_i + \Lambda N_{\text{c}}t\right)  \, .
\eea
At the saddle points one finds that the $\tau_t$ coordinate becomes the following at the initial and final hypersurfaces, respectively
\beq
\label{eq:tau_0_1}
\tau_1^{\rm sad}=i c_1 c_3\sqrt{\frac{\Lambda q_f }{3k}-1} \ \text{and} \  \tau_0^{\rm sad}=-ic_1\frac{\pi_i}{3\sqrt{k}} \, .
\eeq
Interestingly, the function $f(c,\Lam)$ is independent of final scale factor $q_f$, while other functions $g_{1,2}(q_{f})$ are independent of initial parameter $c$. Further insights about the perturbation stability can be explored by utilizing the expression of $\mathbb{P}_{1}^{\,l+1}[\tau_t]$ at the saddle points
\beq
\label{eq:Leg_P_def}
\mathbb{P}_{1}^{\,l+1}[\tau_t] = \left(\frac{1-\tau_t}{1 + \tau_t}\right)^{\frac{l+1}{2}} (l + 1 + \tau_t) \, .
\eeq
Utilizing the definition mentioned, the explicit expression for the $f(c)$ function at the saddle points reads as
\beq
\label{eq:f_fun_sad}
f(c)\Big|_{c_1,c_2,c_3} = \left( \frac{6 + 6l + \sqrt{6}\sqrt{3 + c_2\sqrt{9 - 48c\Lambda}}}{6 + 6l - \sqrt{6}\sqrt{3 + c_2\sqrt{9 - 48c\Lambda}}}\right) \left(\frac{6 - \sqrt{6}\sqrt{3 + c_2\sqrt{9-48c\Lambda}}}{6 + \sqrt{6}\sqrt{3 + c_2\sqrt{9 - 48c\Lambda}}}\right)^{(l+1)} \, .
\eeq
The function $f(c)$ enters the quadratic action for perturbations and depends explicitly on the parameter $c_2$. Hence its structure is sensitive to the choice of branch in the complex lapse plane.  Although, the explicit expression of $g_{1,2}$ and the on-shell action looks cumbersome, one may gain useful insight by studying the on-shell action at small $c$ and large $q_f$ limit (classical limit). The resulting form of $ i S_{\text{grav}}^{(l)}$ associated with each contributing saddle points  enables us to determine the sign of the real part 
of the quadratic action for each perturbation mode.  A negative real part 
corresponds to a stable fluctuation sector on that saddle, while a positive 
real part indicates an instability, leading to an exponential growth of 
perturbations.

\subsection{Perturbations near NB  saddles}
Let us focus on the choice characterized by $c_1=\pm1$ and $c_2=1$, 
as this choice yields the Hartle-Hawking $(c_1=-1)$ and the Vilenkin (tunneling) $(c_1=1)$
saddle points for the corresponding initial momentum in the limit $c \to 0$. 
These saddles play a distinguished role in quantum cosmology, as they represent 
two proposals for the quantum origin of the universe, differing in their boundary conditions on the wave function of the universe.

As mentioned, we analyze the stability behavior in the small $c$ limit, in which the function $f(c)$ reduces to
\beq
\label{fequation1}
f(c)\approx \frac{2+l}{l}\biggl(\frac{c\Lambda}{3} \biggr)^{l+1} + \mathcal{O}((c\Lam)^{l+2} ) \, ,
\eeq
This reduced expression allows us to analyze the stability of scalar and 
tensor perturbations around the Hartle-Hawking and tunneling saddles in a 
transparent manner. Furthermore, evaluating the remaining function $g_{1,2}(q_f)$, together with the ratio $\mathbb{P}'_{1}{}^{l+1}[\tau_1^{\text{sad}}]/{\mathbb{P}_{1}{}^{l+1}[\tau_1^{\text{sad}}]}$ in the large $q_f$ limit, we obtain the semiclassical wave function, including the tensor perturbation contributions at the near no-boundary saddles as
\bea
\label{wavefunc1}
\Psi^{\text{nb}}
\approx 
&&
\exp\biggl\{\frac{c_1 k^{3/2} (l+1)(l+2)}{4\hbar\Lambda} \biggl[3l -6(-1)^l (l+2) \bigl(c\Lam/3\bigr)^{l+1} -\frac{ ilc_1 c_3}{l+1}
\left(\frac{3\Lam q_f}{ k}\right)^{1/2} \biggr]
\notag \\
&&
+ \mathcal{O}\left(\frac{(c\Lam)^{l+2}}{\sqrt{q_f}}\right)
\left(h_l^{(1)}\right)^2 
\biggr\} .
\eea
The stability behavior of the tensor perturbations is governed by the real part of the exponent $ i S^{(l)}_{\rm grav}$, which determines whether a given saddle leads to exponential suppression or enhancement. In this sense, the saddle-point structure directly controls the semiclassical probabilities assigned to different cosmological histories. From eq. (\ref{wavefunc1}), at the leading order in $c$, it is evident that perturbative weighting retains a Gaussian form for $c_1=-1$ and get exponential divergent as $h_l^{(1)}$ increases at $c_1=+1$. It exhibits no dependence on the final three-geometry $q_f$ at leading order.

\subsection{Perturbations near de-Sitter saddles}
We now turn to the other possible choice, characterized by $c_1=\pm1$ and $c_2=-1$.  For this choice, the initial momentum takes a degenerate value, $\pi_i=0$ at $c=0$.
These saddles represent deSitter geometry for $c\rightarrow 0$ limit, while for nonzero $c$, the corresponding geometries resemble a Euclidean extension of de Sitter spacetime at the throat. The associated geometry is illustrated in Fig [\ref{fig:Near_dS_geo}]. We examine the behavior of 
the function $f(c)$ in the small-$c$ regime in close analogy with the previous case. To leading order, we find
\beq
\label{fequation2}
f(c)\approx 1- \frac{4l(l+2)}{(1+l)}\sqrt{\frac{c\Lam}{3}}+\mathcal{O}(c\Lambda),
\eeq
In the small $c$ limit, the leading contribution to the gravitational action $ i S^{(l)}_{\rm grav}$ arises entirely from the first term, while the remaining terms are suppressed by powers of $c ^{1/2}$ and therefore provide subleading corrections. The wavefunction evaluated at these near de-Sitter saddles is given by
\bea
\label{wavefunc2}
\Psi^{\text{dS}}\Big|_{\rm odd modes} & \approx & \exp{\biggl\{\frac{c_1l^2(l+2)^2}{32\pi G} \Bigl(\frac{3k}{\Lam}\Bigr)\sqrt{\frac{c\Lam}{3}}(h_{l}^{(1)})^2 - \frac{ic_3}{64\pi G} \sqrt{\frac{\Lam q_f^3}{3k}} (h_{l}^{(1)})^2 + \mathcal{O}(q_f^{1/2},c\Lam) \biggr\}} \, , \notag \\
\Psi^{\text{dS}}\Big|_{\rm even modes} & \approx &\exp{\biggl\{\frac{9c_1 q_f}{32\pi G (l+1)^2} \Bigl(\frac{\Lam q_f}{3k}\Bigr)^2\sqrt{\frac{c\Lam}{3}}(h_{l}^{(1)})^2 - \frac{ic_3}{16\pi G} \sqrt{\frac{\Lam q_f^3}{3k}} (h_{l}^{(1)})^2+ \mathcal{O}(q_f,c\Lam)\biggr\}} \, .\notag \\
\eea
The perturbative weighting of the first term preserves a Gaussian form for 
$c_1=-1$, reflecting a well-behaved and normalizable contribution from the 
corresponding saddle point. In contrast, for $c_1=+1$ this weighting becomes 
exponentially divergent as the amplitude $h_l^{(1)}$ increases, signaling an 
instability in the perturbation sector.  Furthermore, in both cases the phase of the wave function increases 
monotonically with the final scale factor $q_f$, which is approximate classical behavior in the 
perturbation sector.

\section{Path integral over $q(t)$ and $h_{ij}$}
\label{trans_amp} 

Having formulated the consistent variational problem, obtained the corresponding boundary contributions for the allowed boundary conditions, and analyzed the classical behavior of the saddle geometries, we now turn our focus to the gravitational path integral introduced in eq. (\ref{eq:Gform_sch}). Using the metric parametrization given by eq. (\ref{eq:frwmet_changed}), the path integral can be expressed as
\beq
\label{eq:grav_path_qform}
\Psi[{\rm Bd}_f, {\rm Bd}_i]
= \int_{-\infty}^{\infty} {\rm d} N_c 
\int_{{\rm Bd}_i}^{ {\rm Bd}_f} {\cal D} q(t) {\cal D} h_{ij}(t, {\bf x})
\exp \left(\frac{i}{\hbar} S_{\rm grav}[q, h_{ij}, N_c] \right) \, ,
\eeq
where $S_{\rm grav}[q, h_{ij}, N_c]$ is the gravitational action given in
eq. (\ref{eq:EHact_exp}). The above path integral has been appropriately gauge fixed in order to eliminate the overcounting of gauge-equivalent configurations, with the corresponding Faddeev–Popov ghost contribution. In the analysis that follows, the Faddeev–Popov ghost sector will not be considered explicitly, since it is not relevant for the questions investigated in the present work, see \cite{Mallik:2026iai} for details.

The action in eq. (\ref{eq:EHact_exp}) contains quadratic contributions in the fields $q(t)$ and $h_{ij}$, together with interaction terms generating higher-order nonlinear effects. Throughout this work, we restrict ourselves to the one-loop semiclassical approximation and compute the corresponding path integral using the background field formalism \cite{Abbott:1980hw}, whereby the fields are decomposed as
\beq
\label{eq:qh_exp_bg}
q(t) = \bar{q}(t) + \ep_q Q(t) \, ,
\hspace{5mm}
h_{ij} = \bar{h}_{ij} + \ep_h H_{ij} \, ,
\eeq
where $\ep_q$ and $\ep_h$ are perturbation parameters to keep track the order of terms in the expansion. Substituting eq. (\ref{eq:qh_exp_bg}) into the gravitational action of eq. (\ref{eq:EHact_exp}) expanding up to quadratic order in $\mathcal{O}(\ep^2_{q}), \mathcal{O}(\ep_{q}\ep_{h})$, and $\mathcal{O}(\ep^2_{h})$, we obtain the following contribution: 
\bea
\label{eq:Sgrav_exp_QH}
&&
S_{\rm grav} = 
S_{\rm grav}[N_c, \bar{q}, \bar{h}_{ij}]
+ \int_{\cal M} \frac{{\rm d} t {\rm d} {\bf x} \sqrt{\rho}}{16\pi G}
\biggl[
\ep_q N_c \biggl\{
-2 \Lam Q - \frac{3 \dot{\bar q} \dot{Q}}{N_c^2}+\frac{\bar{q}Q\dot{\bar h}_{ij}\dot{\bar h}^{ij}}{2 N_c^2}\biggr\} + \ep_h
\biggl\{-N_c{\bar h}_{ij} H^{ij}
\notag \\
&&
+\frac{{\bar q}^2\dot{\bar h}_{ij}\dot{H}^{ij}}{2N_c}+\frac{N_c}{2}{\bar h}_{ij}{\bar D}^2{\bar H}^{ij}\biggr\} + \ep_q^2 \biggl\{
-\frac{3\dot{Q}^2}{2N_c}
+ \frac{Q^2 \dot{\bar h}_{ij} \dot{\bar h}^{ij}}{4N_c}
\biggl\}+ \ep_h \ep_q N_c
\biggl\{\frac{\bar{q} Q \dot{\bar h}_{ij} \dot{H}^{ij}}{N_c^2}\biggr\}
\notag \\
&&
+\ep_h^2 N_c
\biggl\{-
\frac{k}{2}H_{ij} H^{ij}
+ \frac{\bar{q}^2 \dot{H}_{ij} \dot{H}^{ij}}{4N_c^2} 
+ \frac{H_{ij}{\bar D}^2 H^{ij}}{4}  
\biggr\}
\biggr] + \cdots \, ,
\eea
where $S_{\rm grav}[N_c, \bar{q}, \bar{h}_{ij}]$ denotes the on-shell action evaluated on the classical configurations $q$ and $h_{ij}$. The terms linear in $\ep_q$ and $\ep_h$ yield the equations of motion for the classical evolution of fields $\bar{q}$ and $\bar{h}_{ij}$. In particular, the equation of motion for $\bar{q}$ receives additional contributions depending on $\bar{h}_{ij}$, beyond those already given in  eq. (\ref{eq:dyn_q_eq}). These contributions correspond to backreaction effects, wherein the dynamics of $\bar{h}_{ij}$ begin to influence the evolution of the background field $\bar{q}(t)$. Such terms arise naturally due to the nonlinear structure of the gravitational interaction. In the present analysis, however, we shall neglect these contributions and assume that the evolution of  $\bar{q}(t)$ is not sourced by $\bar{h}_{ij}$. This is consistent with working at leading order in the semiclassical approximation, where backreaction effects are ignored. This further implies that, within the present approximation, the equation of motion for $\bar{q}(t)$ remains unchanged from that given in eq. (\ref{eq:dyn_q_eq}). Moreover, in keeping with the philosophy of the background-field formalism, we assume that the quantum fluctuations $Q(t)$ and $H_{ij}(t, {\bf x})$ do not influence the classical configurations $\bar{q}(t)$ and $\bar{h}_{ij}(t, {\bf x})$. 

Under these assumptions, the fields, $\bar{q}(t)$ and $\bar{h}_{ij}(t, {\bf x})$ appearing in the second-order action satisfy eqs. (\ref{eq:dyn_q_eq}) and (\ref{eq:hijEQM}) respectively. For the boundary condition considered for the scale factor, it is observed that the choice $h_{ij}=0$ at the boundary is fully consistent with the requirements of covariance. When the fluctuation field $h_{ij}$ is required to vanish on both boundaries, 
the solution of Eq.~(\ref{eq:hijEQM}) yields the trivial background 
configuration $\bar{h}_{ij}(t,\mathbf{x})=0$. Under this condition, the 
off-diagonal terms appearing in the second variation of the action in 
eq.~(\ref{eq:Sgrav_exp_QH}) disappear, while the diagonal sector simplifies 
substantially. As a consequence, the gravitational path integral factorizes 
into independent contributions from the scale factor $q(t)$ and the 
fluctuation field $h_{ij}$. Therefore, eq.~(\ref{eq:grav_path_qform}) can be 
written as:
\bea
\label{eq:grav_path_qhform_1}
\Psi[{\rm Bd}_f, {\rm Bd}_i] \Br \rvert_{\bar{h}_{ij}(t, {\bf x})=0}
= &&
\int_{-\infty}^{\infty} {\rm d} N_c [\mathcal{G}]
\int_{{\rm Bd}_i}^{ {\rm Bd}_f} {\cal D} q(t) 
\exp\left(\frac{i}{\hbar} S^{(q)}_{\rm grav}[q, N_c] \right)
\notag \\
&& \times
\int_{{\rm Bd}_i}^{ {\rm Bd}_f} 
{\cal D} H_{ij}(t, {\bf x})
\exp \left(\frac{i}{\hbar} S^{(h)}_{\rm grav}[\bar{q}, H_{ij}, N_c] \right) \, ,
\eea
where $S^{(q)}_{\rm grav}[q, N_c]$ mentioned in eq.(\ref{eq:Act_sq_sh_begEXP}) and 
\bea
\label{eq:act_SH_begexp}
S^{(h)}_{\rm grav}[\bar{q}, H_{ij}, N_c]
=&& 
\int_{\cal M} \frac{{\rm d} t {\rm d} {\bf x} \sqrt{\rho}}{16\pi G}
N_c
\biggl[
-
\frac{k}{2}H_{ij} H^{ij}
+ \frac{\bar{q}^2 \dot{H}_{ij} \dot{H}^{ij}}{4N_c^2} 
+ \frac{H_{ij}{\bar D}^2 H^{ij}}{4}\biggr] \, .  
\eea
In the above eq.(\ref{eq:grav_path_qhform_1}), $[\mathcal{G}]$ is the ghost determinant which doesnot affect the one-loop studies; see \cite{Mallik:2026iai}.
Even though the path integrals over $q(t)$ and $h_{ij}$ decouple at the level 
of field integration, they continue to be linked through the lapse integral. 
Throughout this paper, we work within the one-loop semiclassical 
approximation, in which the background fluctuation field is assumed to vanish, 
$\bar{h}_{ij}(t,\mathbf{x})=0$, while the quantum fluctuation 
$H_{ij}(t,\mathbf{x})$ contributes through virtual propagation. In the more 
general setting where $\bar{h}_{ij}(t,\mathbf{x})$ is nonzero, the mixed 
off-diagonal terms in the quadratic action would generate additional 
contributions. The analysis of such effects is deferred to future work.

\subsection{Path integral over $q(t)$}
\label{SubS:PI_qt}

We first study the path integral over $q(t)$ appearing in the first 
line of eq.~(\ref{eq:grav_path_qhform_1}), without the lapse 
integration. The corresponding action describes the scale factor 
together with the appropriate boundary term. Although this is an 
effectively one--dimensional path integral, it still captures 
nontrivial features of gravitational dynamics. Exact results for the 
Einstein--Hilbert case with Neumann boundary 
conditions were obtained previously in 
\cite{Feldbrugge:2017kzv, Narain:2021bff, Narain:2022msz, 
Ailiga:2023wzl}. Here we simply quote the final results:
\bea
\label{eq:q_pathInt_BC}
\int_{{\rm NBC}}^{ {\rm DBC}} {\cal D} q(t) \, 
\exp\bl[\frac{i}{\hbar} S_{\rm grav}^{(q)}\br]
= \bl\{\D^{\rm NBC}_q(N_c) \br\}^{-1/2} \exp\Bl[\frac{i}{\hbar} S_{\rm grav}^{(\bar{q})}(N_c) \Br] 
= e^{i {\cal A}_{q}(N_c)/\hbar}\, ,
\eea
where 
\beq
\label{eq:DeltaN_Q}
\D_q^{\rm NBC}(N_c) = 1\, ,
\eeq
and $S_{\rm grav}^{(\bar{q})}(N_c)$ for  Neumann boundary condition is mentioned in Eq. (\ref{eq:stot_onsh_nbc}). The factor $\D_q(N_c)$ can 
be exponentiated to obtain the quantum corrected action for the lapse
$N_c$. This is given by
\beq
\label{eq:quant_Nc_act_q}
{\cal A}_{q}(N_c) = S_{\rm grav}^{(\bar{q})}(N_c) \, .
\eeq
Interestingly, in contrast to other boundary conditions (see, \cite{Feldbrugge:2017kzv, Ailiga:2023wzl, Yamada:2025rld, Mallik:2026iai}), the initial Neumann boundary condition does not generate quantum corrections arising from the scale factor sector. In the next subsection, we will compute the quantum correction coming from the $h_{ij}$. 

\subsection{Path integral over $H_{ij}$}
\label{SubS:PI_hij}

We next consider the path integral over the fluctuation field 
$H_{ij}$ appearing in the second line of 
eq.~(\ref{eq:grav_path_qhform_1}), where the action 
$S^{(h)}_{\rm grav}[\bar{q},H_{ij},N_c]$ is defined in 
eq.~(\ref{eq:act_SH_begexp}). The surface contributions vanish due to 
the imposed boundary condition that $H_{ij}$ vanishes at the 
endpoints. The path integral has already been gauge fixed, and the 
corresponding Faddeev-Popov ghost contribution has been properly 
taken into account. At one loop, the ghost 
determinant contributes only an overall numerical factor, consisting 
of divergent and finite pieces, and does not alter the 
qualitative behavior of the path integral (see ref. \cite{Mallik:2026iai}). 
We therefore suppress it in what follows.

To compute the path integral over $H_{ij}$, we employ the mode 
decomposition of eq.~(\ref{eq:h_ij_Sph_Har}) in terms of tensor spherical harmonics on $S^3$, under which the path 
integral measure becomes
\beq
\label{eq:PI_hij_mea}
{\cal D} H_{ij}  \hspace{3mm} \Rightarrow \hspace{3mm}
\prod_{l=2}^\infty \,\, \prod_{n=2}^l \,\, \prod_{m=-n}^n \,\, 
{\cal D} H^l_{nm}(t) \, ,
\eeq
with the fluctuation sector reducing to a sum over the actions of the 
individual modes.  The action associated 
with each mode, including the boundary contribution, takes the form
\beq
\label{eq:mode_hL_act}
S^{(h)}_{\rm grav}
= \frac{1}{16\pi G} \sum_{l=2}^\infty \,\, \sum_{n=2}^l \,\, \sum_{m=-n}^n
\bl(S^{(l)}_{\rm grav} \br)_{nm} \, ,
\eeq
where $\bl(S^{(l)}_{\rm grav} \br)_{nm}$ is the action 
corresponding to each mode and is given by
\bea
\label{eq:SL_hL_act}
\bl(S^{(l)}_{\rm grav} \br)_{nm} &&=  \int_0^1 {\rm d}t
\Bl[
\frac{\bar{q}^2 }{4N_c}\bl(\dot{H}^l_{mn}\br)^2
- N_c\Bl\{
\frac{l(l+2) - 2}{4}   
+ \frac{k}{2}\Br\} \bl(H^l_{mn}\br)^2
\Br]
\, .
\eea
Since $\bar{q}(t)$ is a quadratic polynomial in $t$, the action 
$\bl(S^{(l)}_{\rm grav}\br)_{nm}$ acquires a particularly simple 
structure. Furthermore, at the one-loop level, the quadratic nature 
of the action for $H^{l}_{mn}(t)$ renders the path integral over each 
mode exactly computable. Owing to the spherical symmetry of the action, 
its form is independent of the indices $n$ and $m$. Consequently, these 
indices will be suppressed in the subsequent discussion, with the 
associated degeneracy factors incorporated wherever necessary. For a 
fixed value of $l$, all modes labeled by different $n$ and $m$ yield 
identical contributions to the path integral. Therefore, it suffices to 
evaluate the path integral only over modes with distinct values of $l$, 
with each contribution weighted by the corresponding degeneracy factor 
$g_l$. This leads to 
\bea
\label{eq:h_PI_mode_degen}
\int_{{\rm Bd}_i}^{ {\rm Bd}_f} 
{\cal D} H_{ij}(t, {\bf x}) 
&&
\exp \left(\frac{i}{\hbar} S^{(h)}_{\rm grav}[\bar{q}, H_{ij}, N_c] \right) 
\notag \\
&&
= \prod_{l=2}^\infty
\biggl[
\int_{{\rm Bd}_i}^{ {\rm Bd}_f} 
{\cal D} H_l(t) \exp\left(\frac{i}{\hbar} S^{(l)}_{\rm grav}[\bar{q}, H_l, N_c] \right) 
\Br]^{g_l} 
= e^{i {\cal A}_{h}(N_c)/\hbar}\, .
\eea
Here we suppressed the mode labels $n$ and $m$, since their effects 
have already been incorporated through the degeneracy factor $g_l$. The 
action $S^{(l)}_{\rm grav}$ refers to the expression given in 
eq.~(\ref{eq:SL_hL_act}) with the indices $n$ and $m$ omitted. The 
quantities ${\rm Bd}_i$ and ${\rm Bd}_f$ denote the boundary conditions 
imposed on the fluctuation field $H_{ij}$ at the initial and final 
hypersurfaces, respectively. The background scale factor 
$\bar{q}(t)$ is the quadratic on--shell solution given in 
eq.~(\ref{eq:qsol_gen}).

Despite the fact that the path integral over $H_l(t)$ in 
eq.~(\ref{eq:h_PI_mode_degen}) is Gaussian within the one-loop 
approximation, its evaluation still requires careful treatment. In 
general, the path integral assumes the form:
\beq
\label{eq:hL_PI_stru}
\int_{{\rm Bd}_i}^{ {\rm Bd}_f} 
{\cal D} H_l(t) \exp\Bl(\frac{i}{\hbar} 
S^{(l)}_{\rm grav}[\bar{q}, H_l, N_c] \Br)
= \bl\{\D^{(l)}_h(N_c) \br\}^{-1/2}\, ,
\eeq
where the fluctuation mode $H_l$ satisfies vanishing boundary 
conditions on ${\rm Bd}_i$ and ${\rm Bd}_f$, while the corresponding 
mode action is given in eq.~(\ref{eq:SL_hL_act}). Since the path 
integral is Gaussian, it can in principle be evaluated exactly for 
each mode, leading to the expression appearing on the right-hand side 
of eq.~(\ref{eq:hL_PI_stru}). The system may be viewed as an effective 
harmonic oscillator with time-dependent mass and frequency. Owing to 
the absence of time-translational invariance, such systems do not 
possess zero-mode related subtleties. Consequently, one may employ the 
methods developed in 
\cite{Khandekar:1986ib, Sabir:1991PrJ} to compute the path integral 
exactly. Equivalent results may also be obtained through the Wronskian 
construction discussed in 
\cite{McKane:1995vp, Kleinert:1998rz, Kleinert:1999Cher}. In this work, we 
evaluate the 1-loop determinant utilizing the Gel'fand Yaglom method \cite{Ailiga:2024wdx, Matsui:2025guo, Mallik:2026iai}

However, a straightforward application of the prescriptions  
for evaluating $\D^{(l)}_h(N_c)$ does not 
necessarily ensure that the resulting functional determinant is free 
from singular behavior. Indeed, as will be demonstrated later, for 
certain values of $N_c$ and $q_f$ the determinant may diverge. It is therefore important to understand the structure and origin of the divergences 
appearing in the one-loop functional determinant. Utilizing Gel’fand–Yaglom approach, the determinant $\D^{(l)}_h(N_c)$ takes the form: 
\beq
\label{eq:Del_struct_hl}
\D^{(l)}_h(N_c) = 
\bl(-2\pi i \hbar\br)
[q_f \bar{q}(0)]^{-1/2} W_l(1,0) \mathbb{M}(\xi_{l},N_c)^{-1}\, ,
\eeq
where we have
\bea
\label{eq:M_def}
   \mathbb{M}(\xi_{l},N_{c}) && = \frac{\bar{q}(0)}{N_c}\left(\dot{\mathbb{P}}_1^{\xi_l}(\tau_0)\mathbb{P}_1^{-\xi_l}(\tau_0)
    - \dot{\mathbb{P}}_1^{-\xi_l}(\tau_0)\mathbb{P}_1^{\xi_l}(\tau_0)\right) \, ,\text{and} \notag \\
    \bar{q}(0) && = \frac{\pi_i^2 + 3\Lam q_f -(\Lam N_c - \pi^2)^2}{3\Lam} \, .
\eea
It should be highlighted that the dependence of parameters $c_{1,2,3}$ and $c$
is embedded in $\xi_l$ and the expression for $\tau_t$. 
The factor $ (-2\pi i \hbar)$ arises due to normalization and is 
standard in the path integral computations, it can be ignored for the 
purpose of this paper as it does not play role in the following. 
The quantum corrected lapse action is given by 
\bea
\label{eq:QT_Nc_act_h_divST}
{\cal A}_{h}(N_c) = \sum_{l=2}^\infty 
\Bl[
-\frac{i g_l \hbar}{4} \ln \left(
q_f \, \bar{q}(0)
\right)
+ \frac{i g_l \hbar}{2} \ln W_l(1,0)-\frac{i g_l \hbar}{2} \ln \mathbb{M}(\xi_{l},N_c) \Br]  \, .
\eea
This is the contribution of the quantum fluctuation field $H_{ij}$
at one loop to the action of the lapse $N_c$. 
The dependence on the parameters $c_{1,2,3}$ and $c$
enters implicitly through the quantity $\xi_l$, as defined in 
eq.~(\ref{eq:xi_tau_defs}). As a result, the expression for 
$W_l(1,0)$ also depends indirectly on the initial parameters. 
Since these parameters fix the initial boundary 
condition imposed on the scale factor, the fluctuation determinant 
inherits sensitivity to the choice of initial boundary data. 

\section{UV divergences}
\label{sec:one_loop_UVdiv}

It is crucial to study the $1$-loop effects of perturbation $h_{ij}$ around the background saddle geometries, which gives a better understanding of the quantum corrections and stability of the system under quantum fluctuations (quantum stability). Such a deep analysis was previously done while analysing the one-loop nature of No-boundary and deSitter wave-functions \cite{Barvinsky:1992dz, Ailiga:2024mmt, Mallik:2026iai}. In each of these studies, path-integral over metric fluctuations suffer from UV-divergences. These are regulated to arrive at a finite transition amplitude. In this section, we will focus on isolating the UV-divergences in our system at various saddles.

In the following, we investigate the one-loop determinant $\Delta^{(l)}_{h}$ given by Eq. (\ref{eq:Del_struct_hl}) in the small $c$ limit at various saddles. 

\subsection{UV divergences near No-boundary saddle}
\label{one_loop_cal_nnb}

We first focus on the computation of the one-loop determinant for the case of No-boundary saddles. 

\subsubsection{UV Divergence near no-boundary saddles: }
We wish to investigate the parameter $c$ gives any divergence or not, in the small $c$ limit. At the near no-boundary saddles $\xi_l$, $\tau_t$ and $\mathbb{M}$ from eqs. (\ref{eq:xi_tau_defs}) and (\ref{eq:M_def}) are given by 
\bea
\label{M_xi_tau_nb}
\xi_l^s=1+l, \  \ \tau_t^{nb}=\frac{i c_1}{3\sqrt{k}}\left(-\pi_i+ \Lambda N_{\text{nb}}^{(c_1,c_3)}t\right) \  \& \ \mathbb{M}(\xi_l^s, N_{nb}^{(c_1,c_3)}) = 2i\sqrt{k}l(l+1)(l+2)\, ,  
\eea
with $N_{\text{nb}}^{(c_1,c_3)} \equiv N^{(c_1,1,c_3)}_{sad}$, where $c_1=-1$ corresponds to the Hartle--Hawking proposal, while $c_1=1$ corresponds to the Vilenkin proposal. One may proceed to evaluate the one-loop determinant given by eq. (\ref{eq:Del_struct_hl}) utilizing the definition of $\mathbb{P}_1^{\xi_l}[\tau_t^{\text{nb}}]$ given in eq. (\ref{eq:Leg_P_def}). It is seen that in the small $c$ limit, at near no-boundary saddles, we obtain 
\bea
\label{eq:Del_nb}
\frac{W_l(1,0)}{\sqrt{q(0)}\mathbb{M}\left(\xi_l^s,N_{\text{nb}}\right)}\Bigg|_{\substack{  N_c \to N_{\text{nb}}}} \approx \left(c\Lambda\right)^{-(\xi_l^s + 1)/2}\mathcal{F}_1(q_f,l,\Lambda) \left[1 + \mathcal{F}_2(l) \ (c\Lambda) + \mathcal{O}(c^2)\right] \, ,
\eea
where the functions $\mathcal{F}_{1,2}$ are given by
\bea
\label{eq:F_12_defs}
\mathcal{F}_{1}(q_f, l, \Lam) & = & \left[\frac{i c_1 \  3^{\frac{1+l}{2}}\Lambda^{1/2}}{4k(l+1)(l+2)}\right]\left(\frac{1-\tau_1^{\text{nb}}}{1+\tau_1^{\text{nb}}}\right)^{\frac{1+l}{2}}\left(l+1+\tau_1^{\text{nb}}\right)\, , \\
\mathcal{F}_{2}(l) & = & \left(-l+\frac{2}{3l}-\frac{5}{3}\right), \,  \ \ \forall \, l \geq 2 \, .
\eea
Focusing till first order in $c\Lam$ in the one-loop behavior,
one can express the quantum corrected lapse action near the no-boundary saddles i.e $N_c \to N_{\rm nb}$ in the limit $c \to 0$,  as follows:
\bea
\label{Action_at_Nnb}
\mathcal{A}(N_{\rm nb}) = && S_{\text{grav}}^{({\bar{q})}}(N_{\rm nb}) - \frac{2i\hbar}{3}\text{ln}q_f - \underbrace{\frac{117 i \hbar}{80}\text{ln}(c\Lambda)}_{\text{Log-div}} - \frac{1153 i\hbar}{360}(c\Lambda)-\underbrace{2i\hbar (c\Lam)\zeta(1)}_{\rm Zeta-div} \notag \\
&&
+ \frac{i\hbar}{2} \sum_{l=2}^{\infty}g_l \ \text{ln}\left[\mathcal{F}_1(q_f,l,\Lambda)\right] + \mathcal{O}((c\Lam)^2) \, .
\eea
In above expression, we have used zeta function to regularize the infinite summation over $l$ to extract the coefficients of $\text{ln}(c\Lambda)$ and $c\Lam$ terms. In the quantum corrected action for lapse, the third term diverges logarithmically at the “no-boundary” saddles in the $c \to 0$ limit. Such log divergence arises from each mode of the field and needs to be summed up, leading to the above. In addition, the fourth term contains another divergent contribution at order $(c\Lam)$ order, arising from the $\zeta(1)$ function. The former divergence turns out to be the physical divergence, which has been studied in past literature \cite{Ailiga:2024mmt, Mallik:2026iai}. In contrast, the $\zeta(1)$-divergence could be an artifact of the perturbative treatment of $c$, and may not show up if the functional determinant is computed without treating $c$ to be small (non-perturbative in $c$). However, its true physical essence is not clear yet. 
\footnote{One might expect similar divergences to appear at higher orders in the $c\Lam$ expansion. A detailed analysis is therefore required to determine whether these divergences cancel upon resumming the perturbative series order by order in $c\Lam$. Ultimately, a non-perturbative evaluation of the full sum would be necessary to establish the true nature of this divergence, which is challenging. We leave this for future investigation.}.

\subsubsection{$l$ summation of $\text{ln}\left[\mathcal{F}_1(q_f,l,\Lambda)\right]$}
\label{subss:l-sum_nb_uvdiv}

In this section, we perform the generalized zeta regularization of the infinite $l$-sum that appears in the one-loop effective action in eq. (\ref{Action_at_Nnb}). The sum of interest is given by\footnote{The $i$ factor appearing in eq. (\ref{sum_of_F1}) is canceled by the $(-i)$ present in eq. (\ref{eq:Del_struct_hl}).}
\bea
\label{sum_of_F1}
&& \sum_{l=2}^{\infty} g_l \,\ln\left[\mathcal{F}_1(q_f,l,\Lambda)\right]
 =  \ln\left(\frac{c_1\Lambda^{1/2}}{4k}\right)\sum_{l=2}^{\infty} g_l + \frac{\ln(3)}{2}\sum_{l=2}^{\infty} g_l (1+l)
 \notag \\ 
&&
  + \frac{1}{2}\ln\left(\frac{1-\tau_1^{\text{nb}}}{1+\tau_1^{\text{nb}}}\right)\sum_{l=2}^{\infty} g_l (l+1) + \sum_{l=2}^{\infty} g_l\text{ln}\left(l+1+\tau_1^{\text{nb}}\right) -\sum_{l=2}^{\infty} g_l \text{ln}\Bigl(\frac{l+1}{l+2}\Bigr)   \, .
\eea
In the above equation, the first three terms can be easily summed using zeta functions: $\sum_{l=2}^{\infty} g_l =8/3$ and $\sum_{l=2}^{\infty} g_l (l+1)=191/60$. The infinite summation in the rest consists of three terms: one $\tau_1^{\text{nb}}$ dependent while the other is independent of any parameters of the theory. Notably, all these summation can be done by employing Hurwitz-Zeta regularization (see appendix \ref{Gen_zeta_reg} for details). Performing the regularization and collecting all the terms, we get
\bea
\label{F1_config}
&&
\frac{i\hbar}{2}\sum_{l=2}^{\infty} g_l \  \text{ln}\left[\mathcal{F}_1(q_f,l,\Lambda)\right]\bigg|_{\mathrm{reg}} = i\hbar\Biggl[\frac{191}{240}\text{ln}\left(\frac{1-\tau_1^{\text{nb}}}{1+\tau_1^{\text{nb}}}\right) - \zeta_{H}'(-2,a+2) \notag  \\
&&
+2(a-1)\zeta_{H}'(-1,a+2) -(a+1)(a-3)\zeta_{H}'(0,a+2)  + \zeta_{H}'(-2,3) +\zeta_{H}'(-2,4) \notag \\
&&
- 2\zeta_{H}'(-1,4) - \frac{529}{240}\text{ln}(3)  + \frac{2}{3}\text{ln}\left(\frac{\Lambda}{16 k^2}\right) + \frac{7}{2}\text{ln}\left(\frac{\pi}{2}\right)\Biggr] + \mathcal{O}(N_c-N_{\text{nb}}) \, ,
\eea
where we defined $a \equiv 1+\tau_1^{\text{nb}}$. This is the finite contribution computed at the near no-boundary saddle with corrections, which is summed up in $\mathcal{O}(N_c-N_{\text{nb}})$. Again, at one loop
 these $\mathcal{O}(N_c-N_{\text{nb}})$ terms do not contribute in the Picard-Lefschetz computation of the $N_c$-integration, and they become relevant beyond one loop.

\subsubsection{UV divergences for $q_f \leq 3k/\Lam$}
\label{Cont_div}
Apart from the above mentioned divergence, one may find another type of divergence that emerge from the short-distance behavior. Such divergences are known as ``contact divergence" and have been noticed before \cite{Konstantinidis:2016nio, Mallik:2026iai}. For the geometry characterized by the lapse $N_c$, the proper distance between two hypersurfaces $(t_{1}, r, \theta, \phi)$ and $(t_{2}, r, \theta, \phi)$ with the same spatial coordinates is given by $\int_{t_1}^{t_2} N_c dt/\sqrt{\bar{q}}$, where $\sqrt{\bar{q}(t)}$ is the on-shell solution given by eq. (\ref{eq:qsol_nbc}). For the coordinates that lie on the initial and final hypersurface, we have the proper distance between them as
\bea
\label{eq:pro_dis}
d(\pi_i,q_{f}) &=& \int_{0}^{1} \frac{N_c}{\sqrt{\bar{q}(t)}} dt \, , \notag \\
               &=& \sqrt{\frac{3}{\Lam}} \left[\arctanh{\biggl( \frac{\Lam N_c - \pi_i}{\sqrt{3\Lam q_f}}\biggr)} - \arctanh{\biggl( \frac{ - \pi_i}{\sqrt{3\Lam q(0)}}\biggr)} \right] \, .
\eea
One finds the proper distance vanishes $d(\pi_i,q_{f}) \to 0$, when the final hypersurface approaches the initial hypersurface. This is equivalently setting $q_{f} \to q(0) = (\pi_{i}^2 + 9k)/(3\Lam)$. As a consequence, one finds the one-loop determinant of perturbation vanishes which leads to additional UV divergence in the wavefunction. In the following, we will study the behavior of one-loop determinant in this limit at the saddle geometries. For the case of near no-boundary saddles, we find
\beq
\label{eq:del_con_div}
\Delta^{(l)}_{h}\Bigl|_{q_f-q(0) \to \ep} =  -\frac{27\Lam^3}{2\pi_i(\pi_i^2 + 9k)^2} \ep  + \mathcal{O}(\ep^2) \, ,
\eeq
obtained by expanding the determinant around $\ep$, i.e. in the limit the final hypersurface making contact with the initial hypersurface: $q_f \to q(0) + \ep$ with $\ep \to 0$. From above expression one may see the vanishing of one-loop determinant as the hypersurfaces make the contact. In the path-integral, this determinant leads to a divergence, since it appears in the path integral with the form $(\Delta^{(l)}_{h})^{-g_l/2}$. Clearly for $q_{f} \to q(0)$, $(\Delta^{(l)}_{h})^{-g_l/2} \sim (q_f - q(0))^{-g_l/2}$ is additional UV divergence apart from the UV divergence$(c\to0)$. This contact divergence need to be removed by adding  suitable counterterm to obtain the finite one-loop quantum action. The appropriate counter term after taking contribution of contact divergence from each $l-$ mode, is given by
\beq
\label{eq:ct_nnb}
\mathcal{A}_{\rm ct}^{\rm con} = -\frac{4i\hbar}{3} \ln{(q_f - q(0))} \, . 
\eeq
Along with this, there is an additional UV divergence that arises for small $q_f < 3k/\Lam$, in the limit $c \to 0$ with $q_f \to q(0)$. This limit corresponds to the final hypersurface approaching the initial hypersurface while their contact hypersurface simultaneously tends to zero. This can also be viewed as $q_f \to q(0)$ with $q(0)$ approaching the vanishing size in the $c \to 0$ limit. \footnote{Equivalently, this corresponds to taking $q_{f} \to 4kc$ with $c\to 0$ limit. Note we do not strictly consider $q_f \to 0$, as the doing so would violate causality. Instead, we approach the initial hypersurface from the causally allowed side, taking $q_f -  q(0) \to 0^{+}$ , i.e approaches the $q(0)$ from above.} The associate additional divergence can be extracted by investigating the small $q_f $ behavior of eq. (\ref{Action_at_Nnb}), after regulating the contact divergence. For $q_f<3k/\Lambda$, PL analysis reveals the solely relevant and contributing saddle is $N_{\text{nb}}^{(c_1,+1)}$. At this saddle, in the limit  $c \to 0$ with $q_f\rightarrow q(0)$, we find
\begin{align}
\label{eq:one_act_qf_sm}
\mathcal{A}(N_{nb}^{c_1,1}) + \mathcal{A}_{\rm ct}^{\rm con}(N_{nb}^{c_1,1})  &\approx  S_{\text{grav}}^{({\bar{q})}}(N_{nb}^{c_1,1}) - \underbrace{\frac{289 i\hbar}{240} \ln(q_f)}_{\text{Log-div}}  - \frac{117 i \hbar}{80}\text{ln}(c\Lambda) - \frac{1153 i\hbar}{360}(c\Lambda) \notag \\
& -2i\hbar (c\Lam)\zeta(1)+ C_{+} \, ,
\end{align}
where $C_{+}$ is some finite quantity. One may immediately identify an additional logarithmic divergence in the above limit which need to be regularized by suitable counter term in the limit $c\to 0$ and $q_f -q(0) \to 0^{+}$, namely:
\beq
A_{ct} =  \frac{289 i\hbar}{240} \ln(q_f) \,.
\eeq

\subsubsection{Regularized no-boundary lapse action}
\label{cont_log}

To ensure the finiteness of the lapse action at the near no-boundary saddles, one needs to suitably add counter terms which can overcome these UV divergences. The suitable counter term to be added is given by
\beq
\label{counterterms}
\mathcal{A}_{\text{ct}}(N_c)=\frac{117i\hbar}{80}\text{ln}(c\Lambda)  + 2i\hbar (c\Lam)\zeta(1) +i\hbar \biggl\{ \frac{289 }{240}\ln{q_f} - \frac{4}{3} \ln{(q_f - q(0))}\biggr\} \theta\biggl(\frac{3k}{\Lam}-q_f\biggr)   \, .
\eeq
These counterterms precisely cancel the divergent terms arising from the computation of the one-loop determinant at the near no-boundary saddles, ensuring the finiteness of the quantum corrected lapse action. The first two terms arise for all $q_f$ at the near no-boundary saddles. The third term cancels the UV divergence as the final hypersurface approaches the vanishing-size limit in the $c\to 0$, and the final term corresponds to contact divergence arises only for $q_f \to q(0)$. Furthermore, the $\log$ term, is the manifestation of UV divergence in the smooth vanishing limit of initial geometry $c\to 0$, which is the defining feature of no-boundary geometry. Interestingly, the boundary parameter $c$ acts as a regulator in identifying this UV divergences of the no-boundary geometry, and regularizing the divergence with appropriate counterterms.

Once the quantum corrected action for the lapse is supplemented with the counterterm,
the action no longer suffers from UV divergences, leading to a finite effective action for the
lapse ($N_c$). The finite lapse effective action is given by
\bea
\label{lapse_action}
\mathcal{A}_{\text{fin}}(N_c) &&
= \mathcal{A}_0(N_c) + \hbar \mathcal{A}_h(N_c) + \mathcal{A}_{\text{ct}}(N_c)\, , \notag \\
&&
= S_{\text{grav}}^{(\bar{q})}
- \frac{i\hbar}{2}\biggl[\frac{4}{3} \ln q_f
+ \frac{1153}{180} \, (c\Lambda)  
- \sum_{l=2}^{\infty} g_l \ln\!\left[\mathcal{F}_1(q_f,l,\Lambda)\right]\Bigg|_{\mathrm{reg}} \notag  \\
&&  + \biggl\{ \frac{289}{120} \ln{q_f}   - \frac{8}{3} \ln{(q_f - q(0))} \biggr\} \theta\left(\frac{3k}{\Lam} - q_f\right) \biggr] \, ,\notag \\ && \equiv \mathcal{A}_{\text{fin}}^{(0)}(N_c) + \hbar\mathcal{A}_{\text{fin}}^{(1)}(N_c) \, ,
\eea
where explicit expression of $\mathcal{A}_{\text{fin}}^{(0)}(N_c)+\hbar\mathcal{A}_{\text{fin}}^{(1)}(N_c)$ and $\mathcal{A}_{\text{ct}}(N_c)$ are mentioned in appendix [\ref{app:qcLapse_ex}] and eq. (\ref{counterterms}), respectively.

\subsection{
UV divergences near deSitter saddles}
\label{one_loop_cal_nds}
We now proceed to analyze the one-loop behavior of the near de-Sitter saddles, following the same approach adopted for the near no-boundary saddles. For the near dS-saddles the one-loop determinant is given by  (ignoring $2\pi\hbar$ factors)\footnote{The overall factor of $i$ is precisely canceled by the factor of $(-i)$ appearing in Eq.~(\ref{eq:Del_struct_hl}).}: 
\bea
\label{eq:Del2_ds}
\Delta_h^{(l)}=\frac{c_1}{2 \sqrt{q(0) q_f k }}\biggl[\frac{ \mathbb{P}_1^{l+1}\left( \tau_1^{\text{dS}}\right) \mathbb{P}_1^{-(l+1)}\left( \tau_0^{\text{dS}}\right) - \mathbb{P}_1^{-(l+1)}\left( \tau_1^{\text{dS}}\right) \mathbb{P}_1^{l+1}\left( \tau_0^{\text{dS}}\right)}{ l(l+1)(l+2)
}\biggr] \, ,
\eea
with $\xi_l$ and $\mathbb{M}$ reads same as in eq. (\ref{M_xi_tau_nb}), while $\tau^{dS}$ is given by
\bea
\label{tau_ds}
\tau_t^{dS}=\frac{i c_1}{3\sqrt{k}}\left(-\pi_i+ \Lambda N_{\text{dS}}^{(c_1,c_3)}t\right) \, .
\eea
For subsequent calculations in the $c \to 0$ limit, it is convenient to recast the above expression in the following form
\bea
\label{eq:Del3_ds}
\Delta_h^{(l)} = \frac{i c_1 \mathbb{P}_1^{l+1}\left( \tau_1^{\text{dS}}\right)\mathbb{P}_1^{-(l+1)} \left( \tau_0^{\text{dS}}\right) }{2\sqrt{k}\sqrt{q_f q(0)}l(l+1)(l+2)} \Biggl[1-\frac{\mathbb{P}_1^{l+1}\left( \tau_0^{\text{dS}}\right)}{\mathbb{P}_1^{-(l+1)}\left( \tau_0^{\text{dS}}\right)}\frac{\mathbb{P}_1^{-(l+1)}\left( \tau_1^{\text{dS}}\right)}{\mathbb{P}_1^{l+1}\left( \tau_1^{\text{dS}}\right)}\Biggr]\, .
\eea
Following this, the summation over the index l can be performed systematically, term by term, to obtain the quantum-corrected action. It is convenient to partition 
the resulting series into three distinct parts:
\bea
\label{eq:Del4_ds}
\frac{i\hbar}{2}\sum_{l=2}^{\infty}g_l\text{ln}\Delta_h^{(l)}= 
&&
\frac{i\hbar}{2}\sum_{l=2}^{\infty}g_l\text{ln}\Biggl[\frac{c_1\mathbb{P}_1^{l+1}\left( \tau_1^{\text{dS}}\right)}{2\sqrt{q_fk} \ l(l+1)(l+2)}\Biggr]+\frac{i\hbar}{2}\sum_{l=2}^{\infty}g_l\text{ln}\Biggl[\frac{\mathbb{P}_1^{-(l+1)}\left( \tau_0^{\text{dS}}\right)}{\sqrt{q(0)}}\Biggr]  \notag \\
&&
+\frac{i\hbar}{2}\sum_{l=2}^{\infty}g_l\text{ln}\Biggl[1-\frac{\mathbb{P}_1^{l+1}\left( \tau_0^{\text{dS}}\right)}{\mathbb{P}_1^{-(l+1)}\left( \tau_0^{\text{dS}}\right)}\frac{\mathbb{P}_1^{-(l+1)}\left( \tau_1^{\text{dS}}\right)}{\mathbb{P}_1^{l+1}\left( \tau_1^{\text{dS}}\right)}\Biggr] \, . 
\eea
The summation corresponding to the first term can be evaluated exactly 
by following the same regularization procedure employed for the 
near no-boundary case in eq.~(\ref{F1_config}). The second contribution 
contains a nontrivial dependence on $\tau_0$, which in turn depends on 
the combination $c\Lambda$. The associated infinite series can also be 
evaluated using the generalized zeta-function regularization developed 
in appendix~[\ref{Gen_zeta_reg}]. The resulting expression takes the form
\bea
\label{eq:Del5_ds}
&&
\frac{i\hbar}{2}\sum_{l=2}^{\infty}g_l\text{ln}\Biggl[\frac{\mathbb{P}_1^{-(l+1)}\left( \tau_0^{\text{dS}}\right)}{\sqrt{q(0)}}\Biggr]=
-\frac{2i\hbar}{3}\text{ln}\left(3+\sqrt{9-48c\Lam}\right)
\notag \\
&& -\frac{191i\hbar}{240}\text{ln}\left(\frac{\sqrt{6}-\sqrt{3-\sqrt{9-48c\Lam}}}{\sqrt{6}+\sqrt{3-\sqrt{9-48c\Lam}}}\right) 
+i\hbar\bigl[-\zeta_H'(-2,\Tilde{a}+2)+2(\Tilde{a}-1)\zeta_H'(-1,\Tilde{a}+2)
\notag \\
&&
-(\Tilde{a}+1)(\Tilde{a}-3)\zeta_H'(0,\Tilde{a}+2)\bigr]\, .
\eea
where $\Tilde{a}=1-\tau_0^{dS}$. The third contribution in eq.~(\ref{eq:Del4_ds}) is difficult to evaluate exactly for arbitrary finite $q_f$ and $c$, owing to its complicated functional dependence on these parameters. Nevertheless, one can extract qualitative information about its infrared (IR) behavior. In what follows, we evaluate this contribution by approximating the sum via an integral\footnote{The approximation becomes more accurate in the large $q_f$ limit.}.  Focusing on the third term, in the limit $c\to 0$ one may express it as
\bea
\label{eq:sim_cor_term}
R_c(\boldsymbol{q_f}) && = \sum_{l=2}^{\infty}g_l \text{ln}\Biggl[1-\frac{\mathbb{P}_1^{l+1}\left( \tau_0^{\text{dS}}\right)}{\mathbb{P}_1^{-(l+1)}\left( \tau_0^{\text{dS}}\right)}\frac{\mathbb{P}_1^{-(l+1)}\left( \tau_1^{\text{dS}}\right)}{\mathbb{P}_1^{l+1}\left( \tau_1^{\text{dS}}\right)}\Biggr] \, ,
\notag \\
&&
= \sum_{l=2}^{\infty}g_l \text{ln} \biggl[1 - \exp\bigl\{-2i \theta(l, x) - 4 l(l+2)\sqrt{c\Lam/3}/(l+1) \bigr\}\biggr] \, ,
\eea
where $\theta(l,x) = (l+1) \arctan(x) - \arctan(x/l+1)$ with $\tau_1=-i x$ and $x= \sqrt{\Lam q_f/3k - 1}$. From the above expression, it is evident that a small but nonzero $c$ acts as a regulator in two distinct ways. First, it improves the convergence of the mode sum. Second, it regulates the poles that would otherwise lie on the real $q_f$ axis by shifting them into the complex $q_f$ plane. Our strategy to evaluate the sum is by approximating it as integral. The sum in the limit $c \to 0$ is given by:
\beq
\label{eq:resi_l_sum}
R_c(x) = \sum_{l=2}^{\infty} g_{l} \ln{\biggl[1 - (-1)^l \exp\bigl\{2iD_{l}(x) - 4 l(l+2)\sqrt{c\Lam/3}/(l+1)\bigr\} \biggr]} \, ,
\eeq
where,
\beq
\label{eq:D_l_def}
D_{l}(x) \equiv (l+1) \arctan(1/x) - \arctan((l+1)/x) \, .
\eeq
This infinite sum $R_c(x)$ can only be computed in the large $q_f$ limit, equivalently in the IR limit. To evaluate it, we separate the even and odd modes, write $n=l/x$, do the series expansion at large $x$, while keeping $n$ fixed and pair up them. This essentially includes the contribution from the modes $l$ near $x$ i.e $l \sim \mathcal{O}(x)$. Doing the large $x$ expansion, we get the following
\beq
\label{eq:D_l_IR}
D_{l}(x) = F(n) + \frac{F'(n)}{x} + \frac{1}{x^2}\Bigl[ \frac{F''(n)}{2} -  \frac{n}{3}\Bigr] + \mathcal{O}(x^{-3}) \, ,
\eeq
where $F(n) = n - \arctan{n}$ and $(^{'})$ denotes derivative with respect to $n$. Upon approximating the sum as an integral (after pairing even and odd modes), we have at the leading order:
\beq
\label{eq:R_sum_int_app}
R(x) \approx \int_{C} \, dn \, (x^3 n^2 + 2 x^2 n - 3 x  ) \ln{\Bigl[1 - e^{4iD_{l}(x)- 8 n x\sqrt{c\Lam/3}}\Bigr]} \, . 
\eeq
The contour of integration $C = [0,\infty)$. In the limit $c\to 0$, one finds the leading order value of the sum as:
\beq
\label{eq:R_sum_val}
R_c(x) \approx b_0 x^{3} + b_1 \sqrt{c\Lam/3}  x^{4} + \mathcal{O}(x^2, c) \, ,
\eeq
with $b_0 = 0.286291-0.460498 i$ and $b_1 = -6.20075 + i 3.310824 $ \footnote{In doing the integral, one has to slightly rotate the contour into the complex plane to regulate the oscillatory integral, which is effectively complexifying cosmological constant $e^{i\ep} \Lam$ with $\ep \to 0^{+}$ \cite{Mallik:2026iai}.}. From the analysis, it is clear that, in contrast to the near no-boundary saddle geometries, the near de-Sitter geometries are free from any UV divergences, except for the contact divergence that arises in the same manner as in the near no-boundary case. This divergence is removed by the counterterm given by the last term of eq.~(\ref{counterterms}). Furthermore, the asymptotic behavior scales qualitatively similar in the limit $c\to 0$. In particular, both geometries suffer from infrared divergences, which we will discuss in detail in the later section.

\subsubsection{Regularized dS lapse action}
\label{count_dS}
As discussed, the near de-Sitter saddle geometries suffer from contact divergence due to fluctuations, similar to the near no-boundary saddles. To ensure finiteness in the one-loop effective action of near de-Sitter case, the appropriate counter term the need to be supplied is given by
\bea
\label{contact_dS}
\mathcal{A}^{con}_{\text{ct}}(N_c) =- \frac{4i\hbar}{3} \ln{(q_f - q(0))}  \, .
\eea
Furthermore, as discussed in case of near no-boundary saddles, the behavior of one-loop determinant in the limit $q_f \to q(0)$ given by eq. (\ref{eq:del_con_div}), which is true for generic $\pi_i$. Interestingly from eq. (\ref{eq:del_con_div}), one finds an additional UV divergence for near dS in the limit $c\to0$ which takes $\pi_i$ to vanishing. Thereby, upon examining the limit $q_{f} \to q(0)$ then $c\to 0$, one realizes the requirement of an additional counterterm to regulate the UV divergence that appear in the later limit. However, if we change the order of limits, we notice that $\pi_{i} \to 0$, UV-divergence manifests itself in the contact divergence which different regularization than eq. (\ref{contact_dS}). Accounting for both orders of limits, we obtain the counterterm that regulates all UV divergences in the de Sitter case:
\bea
\label{eq:contact_dS}
\mathcal{A}_{\text{ct}}(N_c) = -\frac{4i\hbar}{3} \ln{(q_f - q(0))}  + \frac{4i\hbar}{3} \ln\left[ (q_f - q(0))^{1/2} - \frac{i\pi_i c_1}{\sqrt{3\Lam}}\right] \, .
\eea
The finite lapse Effective action is given by
\bea
\label{lapse_action_dS}
&&
\mathcal{A}_{\text{fin}}(N_c)
= \mathcal{A}_0(N_c) + \hbar \mathcal{A}_h(N_c) + \mathcal{A}_{\text{ct}}(N_c)\, .
\eea
%

\subsubsection{$c$ as a regulator}
\label{c_reg}
\begin{figure}[htbp]
\centering
\hspace{1cm}
\subfigure[ \, de-Sitter.]{
    \includegraphics[trim={0cm 0 2.2cm 0cm}, clip, scale=0.58]{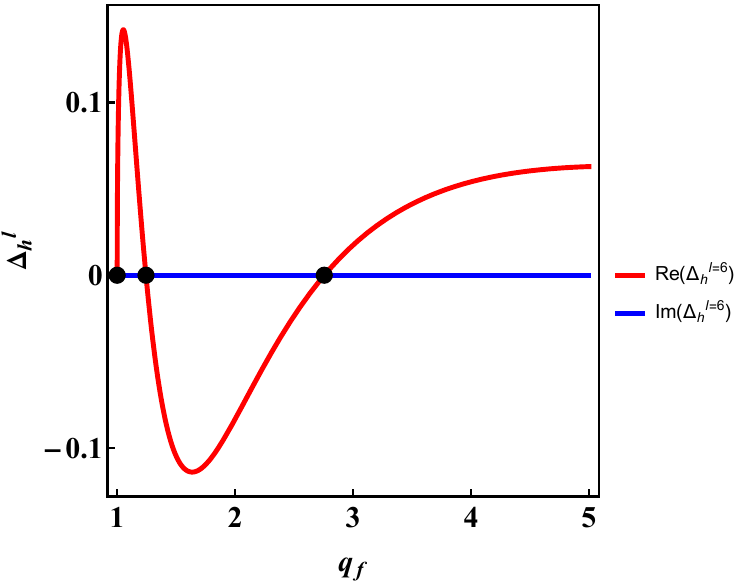}
}
\hspace{1cm}
\subfigure[\, NB]{
    \includegraphics[trim={0cm 0 2.2cm 0cm}, clip,scale=0.6]{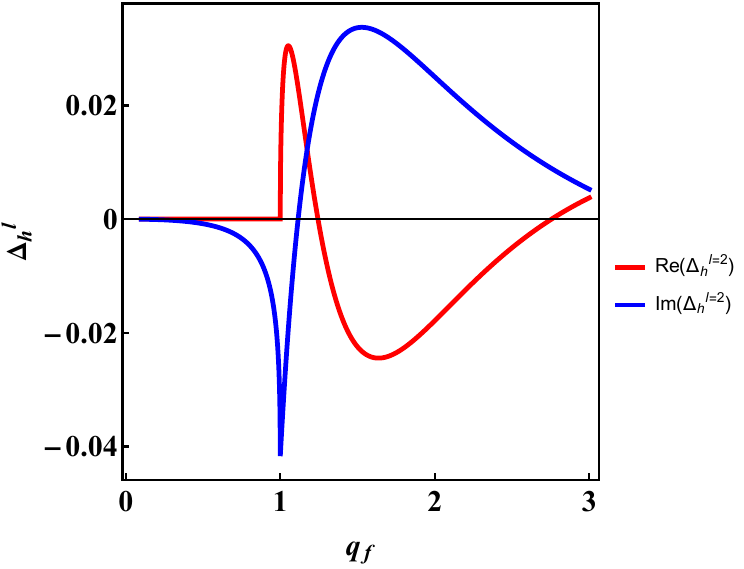}
} \\
\hspace{1cm}
\subfigure[\, Near dS ]{
    \includegraphics[trim={0cm 0 2.2cm 0cm}, clip,scale=0.58]{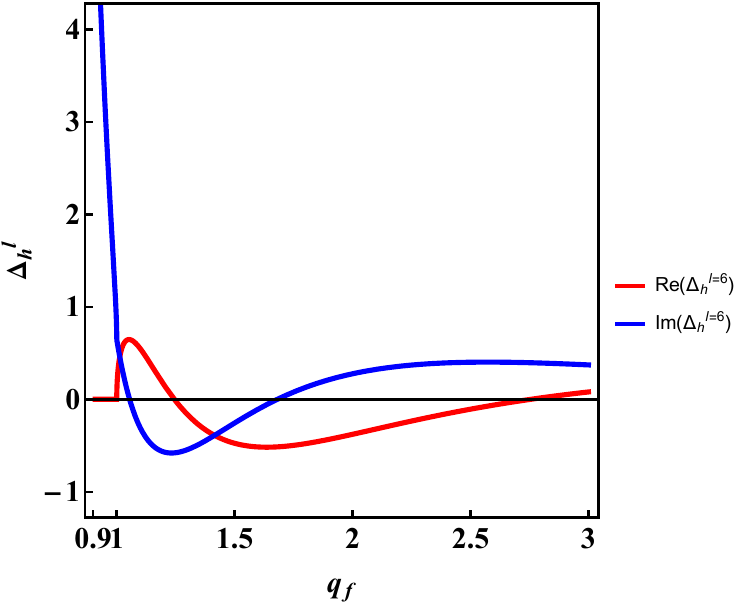}
}
\hspace{1cm}
\subfigure[\, dS poles]{
    \includegraphics[trim={0cm 0 0cm 0cm}, clip,scale=0.58]{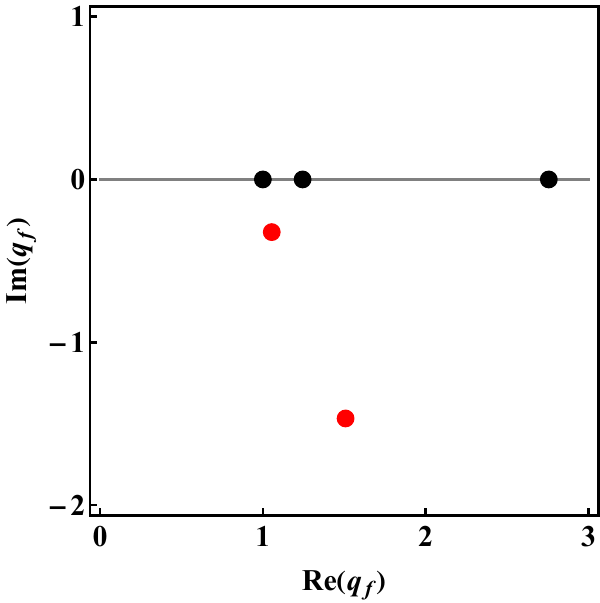}
}
\caption{Behavior of $l=6$ tensor mode one-loop determinant. The red and blue curves denote the real and imaginary parts of the determinant, respectively, while the black and red dots indicate the values of $q_f$ at which the one-loop determinant vanishes. For illustration we take parameters as: $k=1$ and $\Lam=3$. a) Pure deSitter background i.e $c=0$ with $c_{1,3}=1$ and $c_2=-1$. b) No-boundary background i.e $c=0$ with $c_{1,3}=1$ and $c_2=1$. c) Near deSitter background i.e $c=1/48$ with $c_{1,3}=1$ and $c_2=-1$. d) Poles of the one-loop determinant of the $l=6$ tensor mode in the complex $q_f$ plane with $\Lam =3, c=1/48$, $c_{1,3}=1$ and $c_2=-1$: black and red dots correspond to the pure de Sitter and near de Sitter backgrounds, respectively.}
\label{fig:1_loop_det_l6}
\end{figure}

In this section, we will outline the regulator role played by $c$ in case of de-Sitter spacetime. From sec. (\ref{one_loop_cal_nnb}), it  is clearly evident that the boundary parameter $c$ acts as a UV regulator, allowing one to identify and systematically extract the corresponding UV divergences associated with the no-boundary geometry. One can also understand the regulator character of $c$ from the perspective of the near de Sitter saddles. The non-zero $c$ regulates the one-loop determinant of the $l$-th tensor mode on dS background from developing poles along the real $q_f$ axis with $q_f>3k/\Lambda$, where the one-loop determinant vanishes when $c=0$. To be precise, for $c=0$ the background becomes purely deSitter for $c_2=-1$. On this, the $l=2m+1$ mode determinant modes posses $m$ poles, while  $l=2m$ mode determinant posses $m-1$ poles on the real and positive $q_f$ axis in addition to contact divergence $q_f \to q(0)$, where $m \in \mathbb{N}$. This renders the one-loop analysis problematic, in contrast to the no-boundary case, where the determinant is free of singularities along the real $q_f$ axis; cf. [\ref{fig:1_loop_det_l6}]. Interestingly, the non-vanishing $c$ extends the throat of dS to Euclidean geometry, making a near deSitter geometry. In this scenarios, one realizes the problem of singularities is circumvented, as the poles have now shifted to complex $q_f$ from the real axis which make the one-loop determinant singularity free \footnote{The contact divergence which is also a manifestation of pole on real $q_f$ axis, cannot be regulated by the boundary parameter $c$. Instead, it must be regulated by introducing an appropriate counterterm, as discussed in Sec. [\ref{count_dS}].}, see figure [\ref{fig:1_loop_det_l6}].  This illustrates the non-trivial regulating behavior of $c$ in case of near de-Sitter saddles.
%
\section{Regularized lapse action at saddles}
\label{PL_Nc_sad}

The finite lapse effective action allows one to carry forward with the computation of
the lapse-integral via Picard-Lefschetz methods. This leads to a finite wavefunction,
including the quantum corrections. The finite 1-loop wavefunction that we will focus our
attention on is given by

\beq
\label{eq:grav_path_NC_form_fin}
\Psi_{\rm fin}[{\rm Bd}_f, {\rm Bd}_i]
= \int_{-\infty}^{\infty} {\rm d} N_c \,\,
\exp \bl\{i {\cal A}_{\rm fin}(N_c)/\hbar \br\} \, ,
\eeq
where ${\cal A}_{\rm fin}(N_c)$ is finite lapse action defined in eq. (\ref{lapse_action}). This is computed by making use of Picard-Lefschetz and WKB methods. In the saddle-point approximation, one needs to compute the saddles of the one-loop finite lapse action ${\cal A}_{\rm fin}(N_c)$. Ignoring the contribution from $h_{ij}$ to the saddles, for the case of NBC, there will be two complex conjugate saddles as shown in eq. (\ref{sad_point}) for each choice of dominant $\pi_i$. All of these saddles are seen to be relevant as the steepest ascent emanating from them hits the original integration contour [\ref{PL_graph1a}]. The original integration contour can then be deformed to lie along the thimbles passing through these saddles, along which the integral becomes convergent.

When $h_{ij}$ contributions are incorporated, these saddles, along with the set of steepest ascent/descent lines, will be modified. The background saddles $N_{\sg}^{(q)}$ get corrections. But to leading order, the qualitative picture remains the same in the sense the relevance of the saddle remains unaffected. In particular, on inserting the expansion of generic saddle point $N_\sg = N_{\sg}^{(q)} + N_{\sg}^{(h)} + \hbar \, \mathcal{N}^{(h)}_{\sg} + \cdots$ into ${\cal A}_{\rm fin}(N_c)$, it is seen we get the following expansion:
\bea
\label{eq:ANc_sad_exp_1st}
{\cal A}_{\rm fin}(N_\sg)
&& = {\cal A}_{\rm fin}^{(0)}(N_\sg^{(q)})
+ \hbar {\cal A}_{\rm fin}^{(1)}(N_\sg^{(q)}) 
+ \frac{1}{2}(N_c - N_\sg)^2 \bl[
{\cal A}_{0, {\rm fin}}^{\prime\prime} (N_\sg^{(q)})
\notag \\
&&
+ \hbar {\cal A}_{0, {\rm fin}}^{\prime\prime\prime} (N_\sg^{(q)})\bl(
 {\cal N}_\sg^{(h)}
\br)
+ \hbar {\cal A}_{1, {\rm fin}}^{\prime\prime} (N_\sg^{(q)})\br]
 + \cdots \, ,
\eea
where $N_{\sg}^{(q)}$ and $N_{\sg}^{(h)}$ denote the classical contributions to the saddle points arising from the scale factor and the perturbations, respectively, while $\mathcal{N}_{\sg}^{(h)}$ encodes the leading quantum correction arising from the perturbations. The $(\cdots)$ denotes higher order contributions. Here $({}^\prime)$, denotes the derivative with respect to lapse $N_c$. In the saddle-point approximation and retaining terms relevant at one loop, one obtains the expression of 1-loop wavefunction as:
\beq
\label{eq:LDordI_fin}
\Psi_{\rm fin}[{\rm Bd}_f, {\rm Bd}_i]
= 
\sum_\sg \frac{e^{i\theta_\sg}}{\sqrt{\bl\vert {\cal A}_{0, {\rm fin}}^{\prime\prime} (N_\sg^{(q)}) \br\rvert}}
\times 
\exp\Bl[ 
\frac{i}{\hbar} {\cal A}_{\rm fin}(N_\sg^{(q)})
+ \cdots
\Br] \, ,
\eeq
where $\sg$ sums over relevant saddle points. This is schematically the expression of transition amplitude in saddle-point approximation up to one loop. In the following we will make use of this to compute the wavefunction for various values of the scale factor.

\subsection{$q_f>3k/\Lam$}
\label{PL_Nc_sad_qfLG} 

According to the Picard-Lefschetz theory, for $q_f > 3k/\Lam$, all the corrected saddles are relevant as detailed in Sec. [\ref{comp_sads}].  Under the saddle-point approximation, the transition amplitude is computed using the Picard-Lefschetz method given as \cite{Narain:2022msz, DiTucci:2019dji}. Making use of the expression mentioned in eq. (\ref{eq:LDordI_fin}) we get
\begin{equation}
\label{eq:Gtrans_qf>3/lam_gen}
\begin{split}
\Psi_{\rm fin}[\rm Bd_f,\rm Bd_i] & \approx \sum_{c_{1,2,3} =\pm 1} \frac{e^{i\theta_{\rm sad}^{c_3} }e^{i\mathcal{A}_{\rm fin}(N_{\text{sad}}^{(c_1,c_2,c_3)})/\hbar} } 
{
\sqrt{\bl\vert {\cal A}_{0,{\rm fin}}^{\prime\prime} (N_{\text{sad}}^{(c_1,c_2,c_3)}) \br\rvert}} \, .
\end{split}
\end{equation}
Here, $\mathcal{A}_{\rm fin}(N_{\text{sad}}^{(c_1,c_2,c_3)}) $ is the one-loop finite lapse action (obtained after adding the suitable counterterm) evaluated at the background saddle points $N_{\text{sad}}^{(c_1,c_2,c_3)}$ and $\theta^{c_3}_{\sg}$ for the corresponding saddles given in eq. (\ref{eq:theta_s_qf>3/lam}). The one-loop corrected {\it finite} lapse action at the relevant saddles for $q_f>3k/\Lam$ can be obtained from Eq. (\ref{lapse_action}). In the $c \to 0$ limit, the complete expression for the quantum corrected lapse action $\mathcal{A}_{\rm fin}(N_c)$ 
for $N_c\rightarrow N_{\text{nb}}^{(c_1,c_3)}$ at relevant saddles for various boundary choices is written in eq.(\ref{gen_config_lapse_act}). Expressing the one-loop action as $ \mathcal{A}_{\rm fin}(N_{sad}^{c_1,c_2,c_3}) = \mathcal{A}_{\rm fin}^{\rm real}(N_{sad}^{c_1,c_2,c_3}) + i \mathcal{A}_{\rm fin}^{\rm imag}(N_{sad}^{(c_1,c_2,c_3)})$. For generic $q_f$, there expressions are involved and are given in appendix [\ref{app:qcLapse_ex}]. However, in the large $q_f$ limit, they admit simpler form 
\bea
\label{eq:Afin_ReIm_LG_qf}
&&
\mathcal{A}_{\rm fin}^{\rm real}(N_{\rm nb}^{c_1,c_3})
\Big\rvert_{\rm one\mbox{-}loop}
\approx
c_1\, c_3\, \hbar
\left[
\frac{1}{18}\left(11-3\ln \bold{q_f}\right) \bold{q_f}^{3/2} - \frac{7}{4}\bold{q_f}^{1/2}\ln{\bold{q_f}}
+\cdots
\right]
\notag \\
&&
\mathcal{A}_{\rm fin}^{\rm imag}(N_{\rm nb}^{c_1,c_3})
\Big\rvert_{\rm one\mbox{-}loop}
\approx
\hbar
\left[
-\frac{1153}{360}(c\Lambda)
-\frac{\pi}{6}\bold{q_f}^{3/2} - \frac{7\pi}{4}\bold{q_f}^{1/2}
+\cdots
\right]\, ,
\eea
for near no-boundary saddles, while for near dS saddles it becomes 
\begin{equation}
\label{eq:Afin_ReIm_LG_qf_nds}
   \begin{split}
        \mathcal{A}_{\rm fin}^{\rm real}(N_{\rm dS}^{c_1,c_3}) \Br\rvert_{\rm one-loop} \approx &
c_1\, c_3\, \hbar
\left[
\frac{1}{18}\left(11 - 9 \Im(b_0)  -3 \ln \bold{q_f}\right) \bold{q_f}^{3/2} -  \frac{\Im(b_1)}{2} \sqrt{\frac{c\Lam}{3}} \bold{q_f}^{2}
+\cdots
\right]  \, ,\\
         \mathcal{A}_{\rm fin}^{\rm imag}(N_{\rm dS}^{c_1, c_3})\Br\rvert_{\rm one-loop} \approx &  \hbar
\left[
-\Bigl(\frac{\pi}{6} - \frac{\Re (b_0)}{2}\Bigr) \bold{q_f}^{3/2} 
+ \frac{\Re(b_1)}{2} \sqrt{\frac{c\Lam}{3}} \bold{q_f}^{2} + \cdots
\right]  \, ,
   \end{split} 
\end{equation}
where $\Re$ and $\Im$ denotes the real and imaginary parts, respectively and  we have defined $\bold{q_f}=\Lam q_f/3k$. This concludes our analysis, demonstrating that the infrared (IR) behavior of de Sitter (dS) geometry differs from that of the no-boundary saddle. Furthermore, the coupling between the initial parameter \(c\) and the IR parameter \(\boldsymbol{q}_f\) provides exciting evidence for the UV dependence of IR physics, highlighting the ultraviolet (UV) sensitivity of the infrared (IR) behavior. This is not unexpected, as the late-time dynamics depend on the choice of vacuum state, and different vacuum states can consequently lead to distinct IR behaviors. Interestingly, the late-time behavior at both  saddle points scales identically. The leading contribution to the real part scales as $(\ln \bold{q_f})  \bold{q_f}^{3/2}$, while the imaginary part scale as $\bold{q_f}^{3/2}$. 

It is crucial to notice the  appearance of a new $c$-dependent IR-divergent term in the dS-wavefunction which doesn't show up in the No-boundary wavefunction. This term indicates that the perturbative expansion done in $c$ in the computation of functional determinant remains valid only for $c \ll 1/\Lam^2 q_f$. This condition ensures that the late-time asymptotic behavior, and hence the IR scaling, remains unchanged. In our analysis, this requirement is satisfied as we treat $c$ as very small perturbative parameter from the beginning. It is crucial to note that the terms proportional to $\hbar$, indicate a growing behavior for large $q_f$. This is a typical characteristic of the de Sitter geometry where infrared divergences are known to show up in quantum corrections and indicate a secular growth of terms \cite{Akhmedov:2013vka, Akhmedov:2019cfd, Akhmedov:2024npw, Miao:2024shs, Mallik:2026iai, Ailiga:2024wdx}.

\subsection{$q_f<3k/\Lam$}
\label{PL_Nc_sad_qf_sm} 

 In this case for $q_f<3k/\Lam$, we have the relevant saddle as $(N_{\rm sad}^{(c_1,c_2, 1)})$, unlike the $q_f > 3k/\Lam$ case. The corresponding transition amplitude, in this situation, can be computed using the expression mentioned in eq. (\ref{eq:LDordI_fin}). This is given by \cite{Narain:2022msz,DiTucci:2019dji}
\begin{equation}
\label{eq:Gfin_qf_less_3/lam}
\Psi_{\rm fin}[\rm Bd_f,\rm Bd_i] \approx \sum_{c_{1,2} = {\pm 1}} \frac{e^{i\theta_{s}^+ }e^{i\mathcal{A}_{\rm fin}(N_{sad}^{c_1,c_2, 1})/\hbar} } {
\sqrt{\bl\vert {\cal A}_{0,{\rm fin}}^{\prime\prime} (N_{sad}^{c_1,c_2, 1}) \br\rvert}} \, .
\end{equation}
where the complete expression for $\mathcal{A}_{\rm fin}(N_{\rm sad}^{(c_1,c_2, 1)})$ for $q_f<3k/\Lam$ can be read from eq. (\ref{lapse_action}) Although, one finds that the one-loop finite action acquires a simplified structure in the $q_f \to q(0)$ limit. In this limit, the finite lapse action at the relevant saddle $(N_{\rm sad}^{(c_1,c_2, 1)})$ reads as: 
\bea
\label{A_fin_small_qf}
\mathcal{A}_{\rm fin}(N_{\rm nb}^{c_1, 1})\bigg|_{c\rightarrow 0} &\approx& S^{(q)}\bigg|_{c\rightarrow 0}+\frac{4 i\hbar}{3} \ln{\biggl(\frac{ c_1}{2\sqrt{k}}\biggl)}  + \frac{117 i\hbar}{80} \ln \left(\frac{\Lam}{4 k} \right)\, ,\\
\mathcal{A}_{\rm fin}(N_{\rm dS}^{c_1,1})\bigg|_{c\rightarrow 0} &\approx& S^{(q)}\bigg|_{c\rightarrow 0}+\frac{4 i\hbar}{3} \ln{\biggl[\frac{ c_1}{2k^2}\left(\frac{\Lam}{3}\right)^{3/2}\biggr]}  \, .
\eea
The above result follows after adding the required counterterm to the 
expression in eq.~(\ref{eq:QT_Nc_act_h_divST}). An important feature is 
that the quantum-corrected lapse action is purely imaginary. We also note that for $\bold{q_f}<1$, at the relevant  saddles, geometry looks like a sphere ($S^4$). Our finding is in agreement with the previous conclusion where the authors have shown the one-loop correction to the gravity effective action on a sphere is real \cite{Polchinski:1988ua, Mazur:1989ch}.

\subsection{$q_f=3k/\Lam$}
\label{PL_Nc_sad_qf_eq} 
This is a degenerate situation, where one finds the break down of saddle point approximation and one need to proceed to third order contributions to evaluate the wavefunction in the semi classical limit. The wave function is given by
\bea
\label{q:Gfin_qf_eq_3/lam}
\Psi[\text{Bd}_i,\text{Bd}_f]\approx 
\sum_{c_{1,2} =\pm1} \frac{(6\hbar)^{1/3}\Gamma(1/3)}{\sqrt{3}} \bl[{\cal A}_{0,{\rm fin}}^{\prime\prime\prime} (N_{sad}^{(c_1,c_2, 1)})\br]^{-1/3} e^{i{\cal A}_{{\rm fin}}[N_{sad}^{(c_1,c_2, 1)}]/\hbar}\, \, .
\eea
The special case of $q_f=3k/\Lam$, is discussed in eq. (\ref{app:subs_qf_eq_lam}) in Appendix [\ref{app:qcLapse_ex}], where the exact expression of ${\cal A}_{\rm fin}$ is given.

\section{UV Regularized Wavefunction and IR Issues}
\label{subs:quant_TA}

After computing the quantum corrected lapse action for various $q_f$ regimes and extracting the UV divergences, the next natural step is to evaluate the wavefunction of universe, where quantum fluctuations from the fields have been incorporated. It is worthwhile to investigate the wavefunction with vanishing perturbations ($h_{ij} =0$) at the two hypersurfaces while accounting for the contribution through their virtual fluctuations. $h_{ij}$ contributes solely through its virtual propagation at one loop only in this case. Loop contributions are mainly expected to provide subleading corrections to the leading-order result in a conventional flat-spacetime QFT setting. Conversely, this expectation does not necessarily hold in de Sitter spacetime, where quantum corrections can undergo secular growth in the deep infrared \cite{ Akhmedov:2013vka, Akhmedov:2019cfd, Akhmedov:2024npw, Miao:2024shs, Starobinsky:1994bd}. This may indicate a breakdown of the perturbation theory. While there is no clear agreement over how this can be resolved, several proposals debating on it exist. Some of these involve the resummation of the secularly growing terms \cite{Baumgart:2019clc, Honda:2023unh, Cespedes:2023aal, Starobinsky:1994bd, Huenupi:2024ztu, Kamenshchik:2024ybm}, incorporating an IR-cutoff \cite{Huenupi:2024ksc} and regulating the theory \cite{Xue:2011hm}, or incorporating nonlocality \cite{Narain:2018rif} which also acts as a kind of IR regulator.
Latest studies (Refs.\cite{Kamenshchik:2024ybm, Kamenshchik:2025ses}) show that correlation functions of a free scalar field in de Sitter spacetime remain infrared finite in the massive case, while the massless case exhibits secular growth. 

In the present study of computing the wavefunction of the Universe by analyzing the gravitational path integral directly in the Lorentzian signature, we get a similar growth coming from the quantum fluctuations of the metric perturbations $h_{ij}$ for near no-boundary and near dS-saddle geometries. In the deep-infrared regime, the secularly growing contributions become dominant over the leading-order behavior which is consistent with the findings of previous studies \cite{Akhmedov:2013vka, Akhmedov:2019cfd, Akhmedov:2024npw, Miao:2024shs}. Although such infrared behavior is expected to some extent for quantum metric fluctuations in de Sitter spacetime, its implications have been investigated using the Picard--Lefschetz formalism in the context of Lorentzian quantum cosmology \cite{Ailiga:2024wdx, Mallik:2026iai}.

 The metric fluctuations disappears ($h_{ij}=0$) at the two boundary hypersurfaces which implies that each mode of the $h_{ij}$-field in the mode decomposition vanishes: $h_1^{(l)}=0$. The metric fluctuations only propagate virtually in the loop and disappear at the boundary hypersurfaces in this framework. For such a situation, the correction to the orientation of thimbles goes to zero, i.e., $\delta \theta_{\rm sad}=0$.

\subsection{$q_f>3k/\Lam$}
\label{subs:GTR_qf>3/Lam_LG} 

For $q_f>\frac{3k}{\Lambda}$, we have $\theta_{\rm sad}^\pm$ given by the following: 
\begin{equation}
\label{eq:theta_s_qf>3/lam}
    \theta_{\rm sad}^{+}(x)=\frac{m\pi}{2}-\frac{\pi}{4},\hspace{5mm} m\in \mathbb{Z}_{odd}
\end{equation}
with $0\leq\theta_{\rm sad}^+(x)\leq\pi/2$, $\theta_{\rm sad}^-(x)=-\theta_{\rm sad}^+(x)$. As a result, the wavefunction in eq. (\ref{eq:Gtrans_qf>3/lam_gen}) can be approximated as following
\begin{equation}
\label{eq:Gtrans_qf>3/Lam}
\Psi_{\rm fin}[q_f>3k/\Lambda] 
\approx 
\sum_{c_{1,2}=\pm 1} \frac{e^{i\theta_{\rm sad}^+(x)}\exp\left(i\mathcal{A}_{\rm fin}(N_{\rm sad}^{(c_1,c_2, 1)})/\hbar\right)}
{
\sqrt{\bl\vert {\cal A}_{0,{\rm fin}}^{\prime\prime} (N_{\rm sad}^{(c_1,c_2,1)}) \br\rvert}}
+\frac{e^{-i\theta_{\rm sad}^+(x)}\exp\left(i\mathcal{A}_{\rm fin}(N_{sad}^{(c_1,c_2,-1)})/\hbar\right)}
{
\sqrt{\bl\vert {\cal A}_{0,{\rm fin}}^{\prime\prime} (N_{\rm sad}^{(c_1,c_2,-1)}) \br\rvert}} \, .
\end{equation}
Interestingly, the one-loop effective action exhibits an appealing feature,
\begin{equation}
\label{identity}
Re[i\mathcal{A}_{\rm fin}(N_{\rm sad}^{(c_1,c_2,1)})] = Re[i\mathcal{A}_{\rm fin}(N_{\rm sad}^{(c_1,c_2,-1)})]\, , \,
 \& \,
Im[i\mathcal{A}_{\rm fin}(N_{\rm sad}^{(c_1,c_2,1)})] = -Im[i\mathcal{A}_{\rm fin}(N_{\rm sad}^{(c_1,c_2,-1)})] \, .
\end{equation}
Exploiting this identity, the wavefunction can be expressed as the following:
\bea
\label{amplitude}
\Psi_{\rm fin}[q_f>3k/\Lambda] && 
\approx 
\sum_{c_{1,2}=\pm 1} \frac{2\exp\left(-Im[\mathcal{A}_{\rm fin}(N_{\rm sad}^{(c_1,c_2, 1)})]/\hbar\right)}
{\bl\{(Re[{\cal A}_{0,{\rm fin}}^{\prime\prime} (N_{\rm sad}^{(c_1,c_2, 1)})])^2
+(Im[{\cal A}_{0,{\rm fin}}^{\prime\prime} (N_{\rm sad}^{(c_1,c_2, 1)})])^2 \br\}^{1/4}}
\notag \\
&&
\times 
\cos\left(Re[\mathcal{A}_{\rm fin}(N_{\rm sad}^{(c_1,c_2, 1)})]/\hbar+\theta_{\rm sad}^+(x)\right) \, .
\eea
Within the allowed parameter region $0 \leq c \leq 3/(16\Lam)$, as stated the saddle characterized by $c_1=-1$ dominates over other relevant saddles associated with both the no-boundary and de Sitter geometry, c.f Fig [\ref{fig:pi_dom}]. This is the one-loop finite transition amplitude for the case of $q_f>3k/\Lam$. The quantum corrected finite lapse action at the saddle point for $q_f>3k/\Lam$ is mentioned in eq. (\ref{gen_config_lapse_act}) in Appendix [\ref{app:qcLapse_ex}]. The one-loop graviton contribution enters at two places in eq. (\ref{amplitude}): one as an weight factor given by $\exp\bigl(-Im[{\cal A}_{\rm fin}^{(1)}(N_{sad}^{(c_1,c_2, 1)})]\bigr)$ and other inside the cosine function as a phase factor. In the early stages of the Universe, when $q_f$ is not very large, this exponential gives a suppression, but as the Universe grows in size, this term starts to grow exponentially.

We revisit the wavefunction expression mentioned in eq. (\ref{amplitude}) and can see that  it contains for generic $q_f$. Although, the asymptotic limits provide further information about the behavior of the system. For large $q_f \to \infty$, we can more simplify the expression. The denominator occuring in eq. (\ref{amplitude}) can be evaluated as follows
\bea
\label{oneloopampltitude}
\bl\{(Re[{\cal A}_{0,{\rm fin}}^{\prime\prime} (N_{\rm sad}^{(c_1,c_2, 1)})])^2
&&
+(Im[{\cal A}_{0,{\rm fin}}^{\prime\prime} (N_{\rm sad}^{(c_1,c_2, 1)})])^2 \br\}^{1/4} 
\notag \\
&&
=
\left[\frac{\Lambda  V_3}{4 \pi  G }\right]^{1/2}\left(\frac{3}{\Lambda  q_f-3k}\right)^{-1/4}
\, .
\eea
This is relevant only for $q_f>3k/\Lam$. When this is combined with the large $q_f$ behavior of ${\cal A}_{\rm fin}(N_c)$ using eq. (\ref{eq:Afin_ReIm_LG_qf}), it will lead to the large $q_f$ behavior of the finite transition amplitude at one loop. We can compute the simplified expression for the transition amplitude in the large $q_f$ limit and is given by, 
\bea
\label{eq:Gtr_qf>3k/lam_LG_nb}
&&
\Psi_{\rm fin}^{\text{nb}}[q_f>3k/\Lambda,\pi_i]\Biggr \rvert_{q_f\to \infty}\sim \,\frac{1}{\mathcal{F}_\infty}\times \exp\bigg[\frac{3\sqrt{2}V_3}{8\pi G\hbar\Lambda}\left(1+\sqrt{1-\frac{16c\Lam}{3}}\right)^{1/2}\left(2-\sqrt{1-\frac{16c\Lam}{3}}\right)  \, \notag \\
&&
+ \frac{1153}{360}(c\Lam)+\frac{\pi}{6}\bold{q_f}^{3/2} + \frac{7\pi}{4}\bold{q_f}^{1/2} + \cdots+\mathcal{O}(\hbar)\bigg]\, \notag \\
&&
\times \cos\left[\frac{3V_3}{4\pi G\hbar\Lambda}\bold{q_f}^{3/2}- \frac{1}{18}(11-3\ln \bold{q_f})\bold{q_f}^{3/2} + \frac{7}{4}\bold{q_f}^{1/2}\ln \bold{q_f}+\cdots+\mathcal{O}(\hbar)-\bar{\theta}_{\rm sad}^+\right] \, ,
\eea
By contrast, the transition amplitude associated with the de Sitter geometry takes the form
\bea
\label{eq:Gtr_qf>3k/lam_LG_dS}
&&
\Psi_{\rm fin}^{\text{dS}}[q_f>3k/\Lambda,\pi_i]\Biggr \rvert_{q_f\to \infty}\sim \,\frac{1}{\mathcal{F}_\infty}\times \exp\bigg[\frac{3\sqrt{2}V_3}{8\pi G\hbar\Lambda}\left(1-\sqrt{1-\frac{16c\Lam}{3}}\right)^{1/2}\left(2+\sqrt{1-\frac{16c\Lam}{3}}\right) \, \notag \\
&&
+\Bigl(\frac{\pi}{6} - \frac{\Re (b_0)}{2}\Bigr)\bold{q_f}^{3/2} - \frac{\Re(b_1)}{2} \sqrt{\frac{c\Lam}{3}} \bold{q_f}^{2} + \cdots+\mathcal{O}(\hbar)\bigg]\, \\
&&
\times \cos\left[\frac{3V_3}{4\pi G\hbar\Lambda}\bold{q_f}^{3/2}-\frac{1}{18}\left(11 - 9 \Im(b_0)  - 3 \ln \bold{q_f}\right) \bold{q_f}^{3/2} + \frac{\Im(b_1)}{2} \sqrt{\frac{c\Lam}{3}} \bold{q_f}^{2}+\cdots+\mathcal{O}(\hbar)-\bar{\theta}_{\rm sad}^+\right] \, ,\notag
\eea

where $\mathcal{F}_\infty$ is the contribution coming from the denominator in eq. (\ref{oneloopampltitude}):
\begin{equation}
\label{eq:calF_form_LGqf}
    \begin{split}
        \mathcal{F}_\infty &\sim \bold{q_f}^{1/4}\hspace{5mm},
    \end{split}
\end{equation}
and $\bar{\theta}_\sg^+=\pi/4$ in the large $\bold{q_f}$ limit. The IR growth observed in the graviton sector from eq. (\ref{eq:Afin_ReIm_LG_qf}) is consistent with the above growing behavior in the exponential. For generic  $q_f>3k/\Lam$, the real and imaginary parts in $\mathcal{A}_{\rm fin}(N_{\rm nb}^{(c_1,1)})$ mix in a nontrivial manner, forbidding one to obtain such a simplified analytic expression. $\mathcal{A}_{\rm fin}  (N_{\rm nb}^{(c_1,1)})$ can be computed from the Appendix [\ref{app:qcLapse_ex}] for generic $q_f$, in which the one loop wavefunction can only be analyzed numerically. It is worth noting that for large $q_f$, the amplitude shows a growing behavior and dominates over the leading $1/\hbar$ contribution which arises because of the quantum fluctuations of the $h_{ij}$ field. 
\begin{figure}[htbp]
\label{graviton_pic_nb}
\centering

\begin{minipage}[t]{0.48\textwidth}
    \centering
    \includegraphics[width=\linewidth]{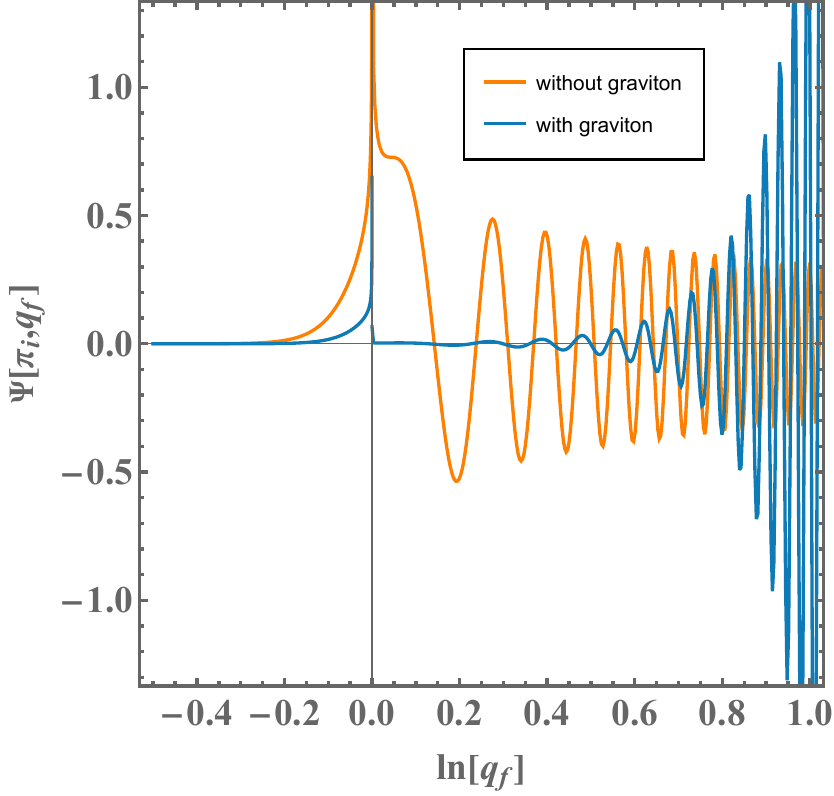}
    
    \small (a) Near no-boundary
\end{minipage}
\hfill
\begin{minipage}[t]{0.48\textwidth}
    \centering
    \includegraphics[width=\linewidth]{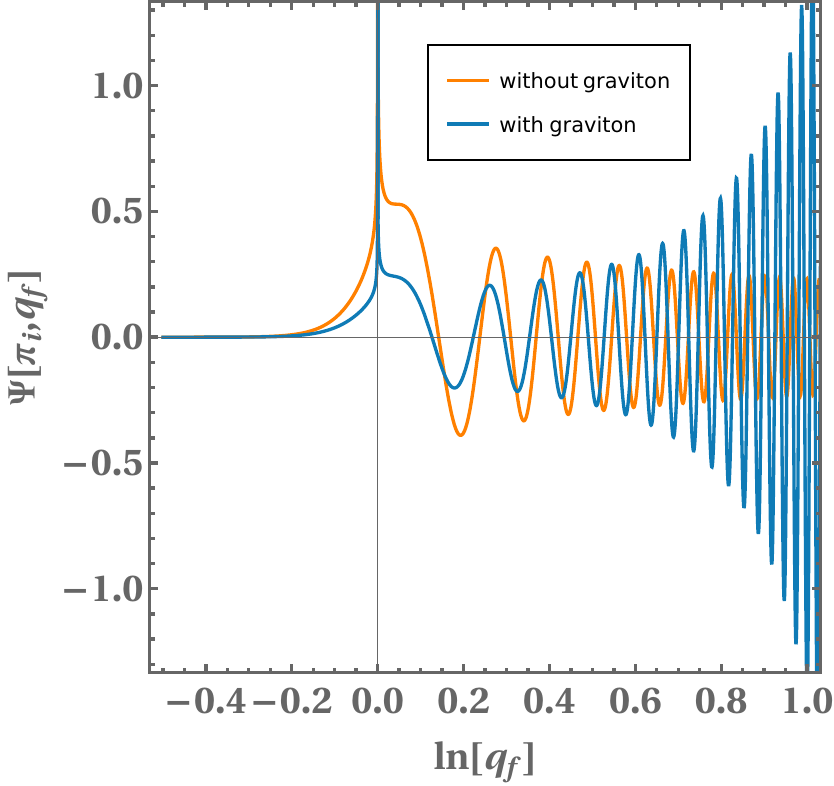}
    
    \small (b) Near-dS
\end{minipage}

\caption{This plot demonstrates the behavior of finite one-loop wavefunction evaluated around near no-boundary and near de-Sitter backgrounds as the universe expands, with the initial condition Neumann boundary condition with respective choices, and final boundary condition being Dirichlet. The orange curve depicts the wavefunction in the absence of any metric fluctuations $h_{ij}$, while blue denotes the one-loop finite regularized wavefunction in presence of virtual gravitons. For the case without gravitons, one finds as the Universe expands, the amplitude decays
monotonically for $q_f > 3k/\Lam$ for both the backgrounds. While in presence of virtual gravitons, the regularized wavefunction shows monotonically growing behavior for $q_f > 3k/\Lam$, illustrating the secular growth and late-time infrared divergence. We have taken $\hbar=1, \Lam=3, k=1, V_3=2\pi^2, G=1$ and $c=1/48$ for the plot. To bring out the features elegantly, we have scaled down the $y-$ axis by $10^3$ and took the $x-$axis in log-scale}
\label{graviton_pic}
\end{figure}
This infrared characteristic is illustrated in Fig.[\ref{graviton_pic}], where the wavefunction with and without graviton contribution is compared for both near NB and near dS  case. The one loop wavefunction with graviton for both case grows in a comparable manner, though with a distinct rate while the later, without graviton figures decays as the Universe expands. This infrared divergent behavior is similar to  QFT on de Sitter, which by now is known to experience infrared divergences and secular growth of quantum corrected terms \cite{Akhmedov:2013vka, Akhmedov:2019cfd, Akhmedov:2024npw, Miao:2024shs}. The leading growth term of $\sim q_f^{3/2}$ was also noticed in Euclidean quantum cosmology studies in \cite{Barvinsky:1992dz}.

\subsection{$q_f<3k/\Lam$}
\label{subs:GTR_qf<3k/Lam_LG} 

For $q_f<3k/\Lambda$, the wavefunction can be approximated by $N_{\rm sad}^{(-1,c_2, 1)}$ saddle at the one loop, as it is the dominant saddles among the relevant saddles for both near NB and dS geometry dictated by Picard Lefschetz formalism. The wavefunction in the Euclidean regime for $q_f < 3k/\Lam$ is given by ($\theta_{\rm sad}^+=0$ for all $q_f$, and $\Lambda$) 
\bea
\label{eq:GTfin_qf<3/lam}
&&
\Psi_{\rm fin}[q_f<3k/\Lambda]
\approx
\sum_{c_{2}=\pm 1} \frac{\exp\left(i\mathcal{A}_{\rm fin}(N_{\rm sad}^{(-1,c_2, 1)})/\hbar\right)}
{
\sqrt{\bl\vert {\cal A}_{0,{\rm fin}}^{\prime\prime} (N_{\rm sad}^{(-1,c_2, 1)}) \br\rvert}} \, ,
\\
&&
\approx 
\exp\left(-Im[{\cal A}_{\rm fin}(N_{\rm sad}^{(-1,c_2, 1)})]/\hbar\right)
\bl\{(Re[{\cal A}_{0,{\rm fin}}^{\prime\prime} (N_{\rm sad}^{(-1,c_2, 1)})^2
+(Im[{\cal A}_{0,{\rm fin}}^{\prime\prime} (N_{\rm sad}^{(-1,c_2, 1)})^2 \br\}^{-1/4} \notag\, .
\eea
As $\mathcal{A}_{\rm fin}(N_{\rm sad}^{(-1,c_2, 1)})$ is purely imaginary for $q_f<3k/\Lambda$, leading to a real amplitude. We can get the expression for $\mathcal{A}_{\rm fin}(N_{\rm sad}^{(-1,c_2, 1)})$ from eq. (\ref{A_fin_small_qf}) . The last factor in the last line of eq. (\ref{eq:GTfin_qf<3/lam}) is given by
\begin{equation}
\label{eq:WKB_fac_qf<3/lam}
\Bl\{(Re[{\cal A}_{0,{\rm fin}}^{\prime\prime} (N_{\rm sad}^{(-1,c_2, 1)})])^2
+(Im[{\cal A}_{0,{\rm fin}}^{\prime\prime} (N_{\rm sad}^{(-1,c_2, 1)})])^2 \Br\}^{1/4}
= \left[\frac{\Lambda V_3}{4\pi G}\right]^{1/2}\left(\frac{3}{3k-q_f\Lambda}\right)^{-1/4} \, .
\end{equation}
Although the finite wavefunction for generic $q_f$ is not illuminating, the expression in the small $q_f$  and $c$ limit: $q_f -q(0) \to 0^{+}$ and $c\to 0$, given by: 
\begin{align}
\label{eq:GTr_fin_qf<3/lam_small_nb}
&&
\Psi_{\rm fin}^{\text{nb}}[q_f<3k/\Lambda,\pi_i] \approx 
\frac{1}{\mathcal{F}_0}\times \exp \Biggl[
-\frac{4}{3} \ln{\biggl(\frac{c_1}{2\sqrt{k}}\biggl)}  - \frac{117}{80} \ln \left(\frac{\Lam}{4 k}\right)\Biggr] \, .
\end{align}
The corresponding transition amplitude for the de Sitter geometry is given by
\begin{align}
\label{eq:GTr_fin_qf<3/lam_small_dS}
&&
\Psi_{\rm fin}^{\text{dS}}[q_f<3k/\Lambda,\pi_i]\approx 
\frac{1}{\mathcal{F}_0}\times \exp \Biggl(
-\frac{4}{3} \ln{\biggl[\frac{ c_1}{2k^2}\left(\frac{\Lam}{3}\right)^{3/2}\biggr]}  \Biggr) \, .
\end{align}
and $\mathcal{F}_0$ is given by: 
\begin{equation}
\label{eq:calF_qf<3/lam_small}
\begin{split}
   \mathcal{F}_0\Bigl|_{\rm nb} &=  \left[1+\mathcal{O}(c\Lam)\right], \\
   \mathcal{F}_0\Bigl|_{\rm dS} &=  \left[(4k c \Lam/3)^{1/4} + \mathcal{O}(c\Lam)^{5/4}\right].
\end{split}
\end{equation}
It is worth noting that in the limit $\bold{q_f}\rightarrow q(0)$ limit, the leading contribution which is constant and independent of $\bold{q_f}$, comes from the quantum fluctuations of the fields.

\section{Conclusions and Outlook}
\label{conc}
In this work, we investigate the gravitational path integral of a four dimensional complex universe evolving from an initial Euclidean hypersurface to a late-time Lorentzian (de Sitter) regime at the one-loop level. Past studies \cite{Feldbrugge:2017kzv, DiTucci:2019dji, DiTucci:2019bui, Narain:2021bff} have shown that boundary conditions play a significant role in dictating the relevance and behavior of saddle points in the path integral. Motivated by our previous findings \cite{Ailiga:2024wdx}, here, we aim to analyse the nature of wave-function of Universe for various initial boundary choice implemented covariantly to one-loop, and investigate the role of boundary choices in dictating the behaviour of UV and IR physics. 

By restricting to class of geometries respecting spatial homogeneity and isotropy, the full gravitational path-integral reduces to a path-integral over lapse and scale-factor. This allows one to compute the wave-function of the Universe transiting from one hypersurface to another. The wave-function, however, is sensitive to the boundary choices imposed at the hypersurfaces, which in our case is implemented either on scale-factor or its conjugate momenta or a combination of two. Typically, these are imposed in canonical manner where there is no guarantee of covariance being respected. In this paper we go beyond canonical implementation of boundary choices by imposing them in a covariant manner. This was first discussed in \cite{Brizuela:2023vmb, Ailiga:2024wdx}, where it was also seen that demand of covariance leads to an infinite family of action, all leading to same canonical boundary condition. Such covariance also implies that metric perturbations living on the hypersurfaces can't be arbitrarily chosen, irrespective of the way they are parameterized \cite{Ailiga:2024wdx, Mallik:2026iai}.

In this paper we seek answer to the question: what is the implication of such an infinite family of covariant boundary action on the wave-function of Universe to one-loop? To address this, we focus on a particular scenario where Neumann boundary condition is imposed on initial hypersurface while Dirichlet is imposed on the final hypersurface. In this case, the infinite family of covariant boundary action is characterized by two parameters: $c$ and $\omega$, where $c=0$ corresponds to the standard scenario of having no boundary action. The interest to pursue investigations in to the case of nonzero $c$ is not purely academic, but also has surprising physical implications.  

Ignoring perturbations on each of the hypersurfaces, it is seen that the wavefunction for the transition from initial to final hypersurfaces can be computed exactly, where Neumann BC is imposed on initial hypersurface while Dirichlet BC on the final one. This wavefunction cleanly factorizes in two pieces: one entirely dependent on initial boundary parameters and other entirely on the final boundary parameters. The part dependent on the initial boundary parameters also depend on $c$ and $\omega$, beside having dependence on initial conjugate momenta $\pi_i$. A search for initial boundary choices giving dominant contribution in this wavefunction relates $\pi_i$ to $c$ and $\omega$. This is found via extremization procedure, which also shows that $\omega$ is quantized and is given by $\omega=(2n+1)/2$, where $n$ is integer. It is realized from eq.(\ref{eq:Ext_eq}) that $n=0$ and $n=1$ case is not expected to lead to substantially too different scenario than already investigated in \cite{Ailiga:2025fny}. The first non-trivial situation arises for $n=-1$ and $n=2$. In both the case eq. (\ref{eq:Ext_eq}) becomes quadratic in $y^2$ thereby giving four possibilities of initial momenta, however, their dominance in path-integral depends on $c$. In the paper we focus only on the case $n=-1$ in detail and study its implications. This itself is sufficient to convey the ideas. Out of the four branches of initial momenta $\pi_i$, one correspond to the Hartle-Hawking no-boundary Universe, one correspond to the Vilenkin-tunneling, while other two correspond to deSitter. This can be inferred by taking $c\to0$ limit and realizing that these four values of initial momenta $\pi_i$ gives: $-3i\sqrt{k}$ (Hartle-Hawking), $+3i\sqrt{k}$ (tunneling), and zero for other two indicating deSitter. 

For each value of $\pi_i(c)$, there are two saddle-geometries in the complex-$N_c$ plane. A Picard-Lefschetz analysis shows that each of these is {\it relevant}, which has also been analytically demonstrated. An interesting thing is noticed, that for each of these saddles, none starts with a vanishing size unless $c=0$. Only in the case when $c\to 0$, two of the initial momenta leads to No-boundary saddle geometries starting from vanishing size, while the others are Lorentzian dS. For nonzero $c$, all saddle geometries starts with non-vanishing size. Two of them however, carries a Euclidean section, these are precisely the ones which become No-boundary or Vilenkin geometries in the limit $c\to0$. This is unlike anything seen before, where the coupling constant appearing in the covariant boundary action dictates the initial size of Universe. The requirements of positivity of the scale-factor constrain the boundary parameter $c$ to obey $0\leq c\leq 3/ (16\Lambda)$. In this domain of $c$, all the saddles are seen to be KSW (Kontsevich–Segal–Witten) allowed, though only the saddle which reduces to Hartle-Hawking choice in limit $c\to0$ dominates in the path-integral over others in the allowed domain. 

It is worth evaluating the perturbative stability of these $c$-dependent saddles, which we do by analysing the behaviour of the tensor perturbations. These tensor perturbations are defined in exponential parameterization as fluctuations over the metric of spatial hypersurface. Due to restrictions imposed by general covariance, the boundary choices available for these perturbations are limited. We choose the simplest allowed boundary choice: vanishing perturbation at the boundaries. The perturbations are then decomposed into spherical harmonics over $S^3$ to analyze the nature of each mode. The behaviour of each mode is determined for these boundary choice in terms of associated Legendre functions. In general the behaviour of each mode depends non-trivially on $c$, which is still hard to analyse even in the allowed domain $0\leq c\leq 3/ (16\Lambda)$. However, progress can be made in the $c\to0$ limit, which is also the scenario where chance of Harle-Hawking type or Vilenkin-type Universe emerges. As expected, the $c$-dependent saddles which goes to Harle-Hawking saddles in the limit $c\to0$ are stable, as various modes will be Gaussian-suppressed in the gravitational path-integral. The $c$-dependent saddles which goes to Vilenkin-tunneling saddles in the limit $c\to0$ are unstable, as various modes will behave like inverse Gaussian in the gravitational path-integral. The $c$-dependent saddles that go to deSitter, one branch is stable under perturbation while the other is not.  

After having analyzed the classical stability of the saddles under metric perturbations, we proceed to compute the one-loop wave-function where the contribution to the path-integral is understood to be coming from the {\it relevant} saddles. To reach there, the gravitational path-integral over the scale-factor $q(t)$ and metric-fluctuation $h_{ij}$ is carefully and systematically performed after gauge-fixing and considering the effects of Faddeev-Popov ghost fields. As our fluctuation field $h_{ij}$ lives on the co-moving hypersurface, the gauge-fixing implemented leads to ghost field action whose contribution in the one-loop path-integral is a numerical factor. This is beneficial as then one can safely ignore ghost contributions in the one-loop study. This wouldn't have been possible if the metric perturbations weren't comoving, and were instead defined as fluctuation over the hypersurfaces. For this reason we prefer to work with former decomposition in present framework. The path-integral over the scale-factor and $h_{ij}$ is performed taking inspiration from background field formalism. The preferred choice vanishing perturbations at the boundaries (which also is compatible with the requirements of covariance), get rids of the mixing terms between $q(t)$ and $h_{ij}$ in the action. 

The path-integral over the scalar factor $q(t)$ is doable exactly for the choice of boundary condition focused in this paper. The path-integral over metric-fluctuations needs mode-decomposition beside the usage of background-field methods. The full path-integral reduces to path-integral over each mode which is computed via Gelfand-Yaglom method \cite{Ailiga:2024mmt, Ailiga:2023wzl, Ailiga:2024wdx, Matsui:2025guo}. This is the one-loop contribution coming from fluctuation $h_{ij}$, which in general is function of lapse-$N_c$. The one-loop contribution is analysed at the saddles to extract the UV divergences. Elimination of which allows one to compute the finite wavefunction. The standard source of these divergences are the singularities arising due to vanishing scale-factor at the boundaries, contact-divergences arising when the two boundary hypersurfaces are close to each other and divergence arising due to working perturbatively in the boundary parameter $c$ (which we believe that last one is an artifact of the perturbation and may go away if analysis is done non-perturbatively). These divergences agree with our previous studies \cite{Ailiga:2024wdx, Mallik:2026iai}. However, the $c$-dependent divergence is new. Also, and most importantly, we note the interesting role played by the boundary parameter-$c$ which appears as the coefficient of the covariant boundary action. It acts as a UV-regulator in extracting the UV divergences. As a result, one doesn't need to introduce an external regulator by hand. The requirements of having covariant boundary action itself interestingly assist in achieving UV-regulation at one-loop. Whether this feature persists at higher-loops is not known yet to us. This is something which is new and hasn't been witnessed before to the best of our knowledge. 

For the deSitter branch of saddles, the $c$-parameter offers an additional interesting benefit of slightly perturbing the vanishing initial conjugate momenta. This leads to moving of the poles of one-loop determinant away lapse-$N_c$ contour of integration, thereby allowing us to perform the lapse-$N_c$ integration cleanly without encountering these spurious singularities. These spurious singularities were also seen in \cite{Mallik:2026iai}, where they moved away by complexifying the cosmological constant $\Lam$. Our current analysis shows, that such complexification of $\Lam$ is no longer needed as the boundary parameter $c$ itself offers a ``cure'' to the problem. We therefore realize that boundary conditions when imposed in covariant form helps in not only UV-regulation but also in moving the spurious poles away from integration contour. 

Having eradicated the UV-divergences and spurious poles, we proceed to the computation of the UV-renormalized wavefunction, and analyze its infrared behaviour. This we compute for both for the no-boundary and deSitter saddles carefully. For No-boundary saddles we could do the computation exactly, and the results agrees with our past work in the limit $c\to0$ \cite{Ailiga:2024wdx}. For the deSitter case, we had to be extra vigilant due to the structure of the infinite summation over the modes whose arguments consist of transcendental functions. It is seen that a part of this infinite series can be computed exactly, while the rest we are able to compute only in the IR limit. The IR behvaiour of the No-boundary and DeSitter wavefunction are structurally same but differ in numerical prefactors in the limit $c\to0$. However, for nonzero (though small) $c$, there is a new $c$-dependent IR-divergent term in the dS-wavefunction which doesn't show up in the No-boundary wavefunction. This term is strongly sensitive to UV boundary choices as it depends on $c$. The appearance of this term also indicates that the perturbative expansion done in $c$ in the computation of functional determinant remains valid only for $c \ll 1/\Lam^2 q_f$. Overall, it is seen that the IR-behaviour of wavefunction is sensitive to the UV boundary choices, which is the main result of the paper. The other main result is the role of parameter $c$ (which enters via initial boundary requirements) in regulating UV-divergences and shifting the spurious poles away real $q_f$ axis. 

We conclude the paper by realizing that although we do come across some interesting surprises like the IR behaviour of the wavefunction being sensitive to UV boundary choices, the study is conducted where certain assumptions were made. We considered vanishing metric fluctuations at the boundaries, which also helps in eradicating the mixing terms. This boundary choice for metric fluctuations although is compatible with covariance, in our study we also noticed other non-trivial boundary choices which fall under Robin class of boundary conditions. These have not been considered in this paper due to complications associated with them. As a result it is hard to speculate about their implications. Furthermore, one-loop determinant is computed under small $c$ approximation, which as can be seen from the IR behaviour of wavefunction remains valid as long as $c \ll 1/(\Lam^2 q_f)$, hinting to go beyond the perturbative treatment, which is beyond the scope of this work as of now. In this paper we haven't used resummation methods to deal with the IR divergences which we hope to return to sometime in future. It will be interesting to compare the findings in this paper with the work done in \cite{Abdalla:2026mxn, Lehners:2026tej} for closed Universe, where the author investigates the normalization of the wavefunction of closed Universe. This will be considered in our future work. 

\bigskip
\centerline{\bf Acknowledgements} 
\label{ack}
We are grateful to Justin David for illuminating discussions 
on Zeta Regularization. We would like to thank Shubhashis Mallik for discussions related to IR behavior of de Sitter. GN would like to thank Chethan Krishnan for discussion on boundary choices. We thank Alexander Yu. Kamenshchik for valuable insights on zero modes and IR divergence issues. We appreciate the ongoing “Gauge-Gravity by Ghats” seminar series which helped us in addressing several issues during the course of this work. GN also acknowledges the startup support from IISc leading to workstation purchase where some of the simulations
have been performed. 

\bigskip
\textbf{Statement on AI use}: The manuscript was written and verified by the authors, who take full responsibility for any errors. We acknowledge the use of ChatGPT and Claude for correcting grammatical typos and literature survey .

\appendix

\section{Quadratic Expansion of Curvature Invariants}
\label{gfgh1} 

In this appendix, we present the expansions of the relevant geometric quantities up to quadratic order in the perturbation field $h_{ij}$. The Christoffel connection $\Theta^{k}{}_{ij}$, extrinsic curvature $K_{ij}$ and its trace $K$ associated with the induced $\g_{ij}$ metric are given by
\bea
\label{eq:chris_gam}
&&
\Theta^{k}{}_{ij}= 2^{-1} \g^{km} \left[
\pt_i \g_{mj} + \pt_j \g_{mi} - \pt_m \g_{ij}
\right]\, , 
\hspace{5mm}
K_{ij} = \frac{\g_{ij}^\prime}{2N_p} \, ,
\notag \\
&&
K = \g^{ij} K_{ij} = \frac{(\ln \sqrt{\g})^\prime}{N_p} \, .
\eea
The Ricci-tensor corresponding to $\g_{ij}$ then reads as
\bea
\label{eq:3Rgamma}
&&
{\cal R}_{ij} = \pt_m \Theta^{m}{}_{ij}- \pt_i \Theta^{m}{}_{mj}
+ \Theta^{m}{}_{mn} \Theta^{n}{}_{ij} - \Theta^{m}{}_{in} \Theta^{n}{}_{mj} \, ,
\eea
where $({}^\prime)$ denotes the derivative with respect to $t_p$
and ${\cal R} = \g^{ij} \,\, {\cal R}_{ij}$ is the Ricci-scalar 
of the three hypersurface. Expanding the metric perturbation using the exponential parametrization,
\beq
\g_{ij} = a^{2}\rho_{ik}\big(e^{h}\big)^{k}{}_{j} \, ,
\eeq
we have the various quantities entering the gravitational action upto quadratic order as follows:
\bea
\label{eq:exp1}
&&
\g^{ij} = a^{-2} \left[
\rho^{ij} - h^{ij} + \frac{1}{2}h^{ik} h_{k}{}^{j} + \cdots
\right] \, ,
\notag \\
&&
\sqrt{\g} = a^{3} \sqrt{\rho} \left[
1 + \frac{h}{2}+\frac{h^2}{8} +\cdots
\right] \, ,
\notag \\
&&
\Theta^{i}{}_{jk} =\underbrace{\frac{1}{2}\Bigl[\partial_{k}\rho^{i}{}_{j}+\partial_{j}\rho_{k}{}^{i}- \partial^{i}\rho_{jk}\Bigr]}_{\bar{\Theta}^{i}{}_{jk}} +\underbrace{\frac{1}{2}\Bigl[\bar{D}_{j}h^{i}{}_{k}+\bar{D}_{k}h_{j}{}^{i}-\bar{D}^{i}h_{jk}\Bigr]}_{\Theta^{(1)i}{}_{jk}} \nonumber \\
    &&  + \underbrace{\frac{1}{4}\Bigl[-\bar{D}^{i}h_{jm}h^{m}{}_{k}+\bar{D}_{k}h^{i}{}_{m}h^{m}{}_{j}+\bar{D}_{j}h_{km}h^{mi}\Bigr]-\frac{h^{il}}{2}\Bigl[-\bar{D}_{l}h_{jk}+\bar{D}_{k}h_{lj}+\bar{D}_{j}h_{kl}\Bigr]}_{\Theta^{(2)i}{}_{jk}} + \cdots \, . \notag \\
\eea
where $(\cdots)$ denotes terms cubic and higher order terms in $h_{ij}$, which we neglect. Here, $\rho_{ij}$ is the background metric, $h_{ij}$ is the perturbation and the operator $\bar{D}_i$ denotes the covariant derivative with the metric $\rho_{ij}$. The indices of $h_{ij}$ are raised or lowered using $\rho_{ij}$ and the trace $h$ is defined $h = \rho^{ij} h_{ij}$ . 
We proceed to expand the other geometric quantities mentioned in Eqs. (\ref{eq:chris_gam}) and (\ref{eq:3Rgamma}) by making use of the exponential parameterization. This gives
\bea
\label{eq:K_exp}
K &=& \frac{3a^\prime}{a N_p} 
+ \frac{h^\prime}{2 N_p} +\cdots \ ,
\notag \\
K_{ij} K^{ij} &=& \frac{1}{4N_p^2} \left[
\frac{12 a^{\prime 2}}{a^2} + \frac{4 a^\prime h^\prime}{a}
+ h^\prime_{im} h^{\prime im} \cdots
\right] \, ,
\notag \\
{\cal R} &=& a^{-2} \biggl[
\bar{\cal R} 
+ \left(\bar{D}_i \bar{D}_j h^{ij} - \bar{\Box} h - \bar{\cal R}_{ij} h^{ij}\right)
+ \rho^{ij} \left(\bar{D}_k \Theta^{(2)k}{}_{ij}
- \bar{D}_i \Theta^{(2)k}{}_{kj} \right) 
\notag \\
&&
+ \bar{\cal R}_{ijkl} h^{ik} h^{jl} 
- h^{ij} \bar{D}_i \bar{D}_k h^k{}_j 
+ \frac{1}{2} h^{ij} \bar{\Box} h_{ij} 
+\frac{1}{2} h^{ij}\bar{D}_i\bar{D}_j h 
+\frac{1}{2}\bar{D}^ih\bar{D}^jh_{ij}
-\frac{1}{4}\bar{D}^ih\bar{D}_ih
\notag \\
&&
+\frac{1}{4}\bar{D}^kh_{ij}\bar{D}_kh^{ij}
-\frac{1}{2}\bar{D}^kh^{ij}\bar{D}_ih_{jk}
-\frac{\bar{\cal R}}{6}h^{ij}h_{ij}
\biggr]
\, , \quad \bar{\Box} \equiv \bar{D}^2 \, ,
\eea
where $\bar{\cal R}_{ijmn}$, $\bar{\cal R}_{ij}$, $\bar{\cal R}$ 
are the Riemann tensor, Ricci tensor and Ricci scalar corresponding to the metric 
$\rho_{ij}$ respectively.

\section{KSW - Criterion } 
\label{KSW}
In this appendix we outline the compatibility of background saddles with the KSW allowability criterion \cite{Lehners:2021mah, Jonas:2022uqb, Hertog:2023vot, Ailiga:2025fny}. Consider the background D($\geq 2$) spacetimes $g_{\mu\nu}$ on a manifold $\mathcal{M}$ on which we define the Euclidean path integral of a $p$-form real gauge field with field strength being a $q=p+1$-form denoted by $F$. According to the KSW-criterion, the path-integral for the $p$-form field is convergent on the given background metric if and only if the metric satisfies the following condition
\bea
\label{KSW_cri}
\begin{split}
\mathcal{I}_q[A]& =\frac{1}{2q!}\int_M d^Dx\sqrt{det\,g}\,g^{\mu_1 \nu_1}\cdots g^{\mu_q \nu_q} F_{\mu_1 \mu_2...\mu_q} F_{\nu_1 \nu_2...\nu_q} \, ,\\
g_{\mu\nu} & \, \text{is allowable iff} \,  Re\left(\sqrt{det\,g} g^{\mu_1 \nu_1}\cdots g^{\mu_q \nu_q}F_{\mu_1 \mu_2...\mu_q} F_{\nu_1 \nu_2...\nu_q}\right) > 0 \, ,
\end{split}
\eea
for all $q \in \{0, ... , D\}$, where $\mathcal{I}_q[A]$ is the corresponding Euclidean action. For the metrics that are diagonal in a real basis, with diagonal elements $\lambda_{\mu}$ at a spacetime point $x \in \mathcal{M}$, i.e., $g_{\mu\nu} = \lambda_{\mu}(x)\delta_{\mu\nu}$, the criterion simplifies to the point-wise requirement of
\beq
\label{eq:KSW_criterion}
\Sigma \equiv \sum_{\mu=0}^{D-1} |\arg \, \lam_{\mu}(x)| < \pi \quad \forall x \in \mathcal{M}\, ,
\eeq
where $\arg (z) \in (-\pi,\pi]$ with $z$ being any complex variable. The KSW criterion has undergone extensive investigation, yielding several important physical implications, see \cite{Lehners:2021mah, Jonas:2022uqb, Lehners:2023pcn, Hertog:2023vot,  Ailiga:2025fny, Lehners:2026tej, Ailiga:2025osa, BenettiGenolini:2026raa, Krishna:2026rma, Barbon:2026tri}. 
Here, we utilize the extremal curve technique that has been discussed in \cite{Hertog:2023vot, Ailiga:2025fny}, to determine whether the saddle points are allowable or not. 

The Extremal curve method offers a strong and conclusive classification of allowable and disallowable geometries under the KSW criterion. For the time contour $\tau_{p}(N_c, u) \equiv i t(N_c,u)$ that runs in the complex plane, with the real parameter $u$, the metric reads as
\beq
\label{eq:frwmet_changed_comp_t}
{\rm d}s^2 = \biggl(\frac{{\rm d}\tau_{p}}{{\rm d}u}\biggr)^2 {\rm d} u^2 
+ q(\tau_{p}(u)) dx^2 \, .
\eeq
The KSW function reads as
\beq
\label{eq:KSW_def}
\Sigma(N_c, \tau_{p}(u)) = \left|\arg\left(\tau_{p}'(N_c, u)^2\right)\right| + 3\left|\arg q(N_c, \tau_{p})\right|\, ,
\eeq
where $(^{\prime})$ denotes derivative with respect to $u$ parameter. We define a generic geometry $N_c \in \mathbb{C}$ as allowable, if and only if there exists a curve $\tau_{p}(u): 0 \rightarrow \nu$ in the complex plane, such that the corresponding on-shell solution $\bar{q}(N_c,\tau_{p})$ satisfies the KSW criterion along the curve, namely:
\beq
\label{eq:KSW_Ext_curv_cond}
\left|\arg\left(\tau_{p}'(N_c, u)^2\right)\right| + 3\left|\arg \bar{q}(N_c, \tau_{p})\right| < \pi \, .
\eeq
Our strategy to verify the KSW criterion is based on the construction of an ``extremal curve" $\tau_{e}$ that saturates the inequality in eq. (\ref{eq:KSW_Ext_curv_cond}): 
\beq
\label{eq:ext_curve_1}
\left|\arg(\tau_{e}'(N_c, u))^2\right| +  3\left|\arg \bar{q}(N_c, \tau_{e}(u))\right| = \pi \, .
\eeq
Upon complexifying the Euclidean time as $\tau_{p} = \tau_{x} + i \tau_{y}$, the extremal equation becomes
\beq
\label{eq:ext_curve_2}
2 \left|\tan^{-1}\biggl(\frac{d\tau_{y}}{d\tau_{x}}\biggr)\right| +  3\left|\arg \bar{q}(N_c, \tau_{x}, \tau_{y})\right| = \pi \, ,
\eeq
which simplifies to
\beq
\label{eq:ext_curve_3}
\frac{d\tau_{y}}{d\tau_{x}}  = \pm \tan{\biggl(\frac{\pi}{2} - \frac{3}{2} \left|\arg \bar{q}(N_c, \tau_{x}, \tau_{y})\right|\biggr)} \, ,
\eeq
with the initial conditions as $\tau_{y}(\tau_{x} = 0) = 0$
\footnote{For numerical implementation, we start at $\tau_{y} = l \sin{\theta_{0}}$ when $\tau_{x} = l \cos{\theta_{0}}$ with $l \rightarrow 0$, where the initial angle $\theta_{0} = n\pi \pm (\pi -  |\arg \bar{q}(N_c,0,0)|)/2$ with $n = \{0,1\}$.}. 
One obtains two extremal solutions in the complex physical time plane to the eq. (\ref{eq:ext_curve_3}), with the stated initial conditions. Then, it follows from the Petrovitsch's theorem \cite{Jonas:2022uqb, michel1901maniere}, the curves everywhere satisfying eq. (\ref{eq:KSW_Ext_curv_cond}) and starting from $\tau_{y}=\tau_x=0$ are constrained to remain between these two extremal curves. This suggests that an allowable curve — one that respects the KSW criterion — may exist between these extremal boundaries. Such a curve can be constructed by continuously deforming the right-hand side of eq. (\ref{eq:ext_curve_1}) and solving it. Following this, we classify a generic point $N_c$ as allowable, if the corresponding $\nu$ obeys \cite{Hertog:2023vot}: $|\tau_{y}|> |Im(\nu)|$ at $\tau_x = Re(\nu)$ or $|\tau_x| < |Re(\nu)|$ at $\tau_{y}= Im(\nu)$. We now implement the extremal curve test to the complex lapse points.

\begin{figure}[htbp]
\centering
\subfigure[  ]{
    \includegraphics[trim={0cm 0 1.3cm 0cm}, clip, scale=0.5]{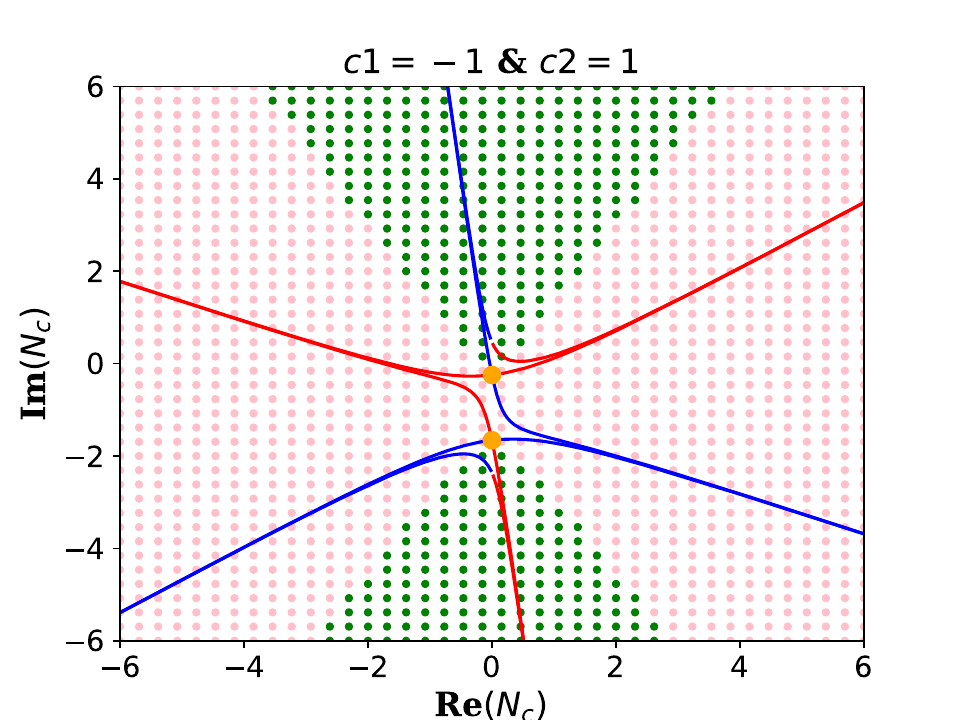}
}
\subfigure[ ]{
    \includegraphics[trim={0 0 1.3cm 0cm}, clip,scale=0.5]{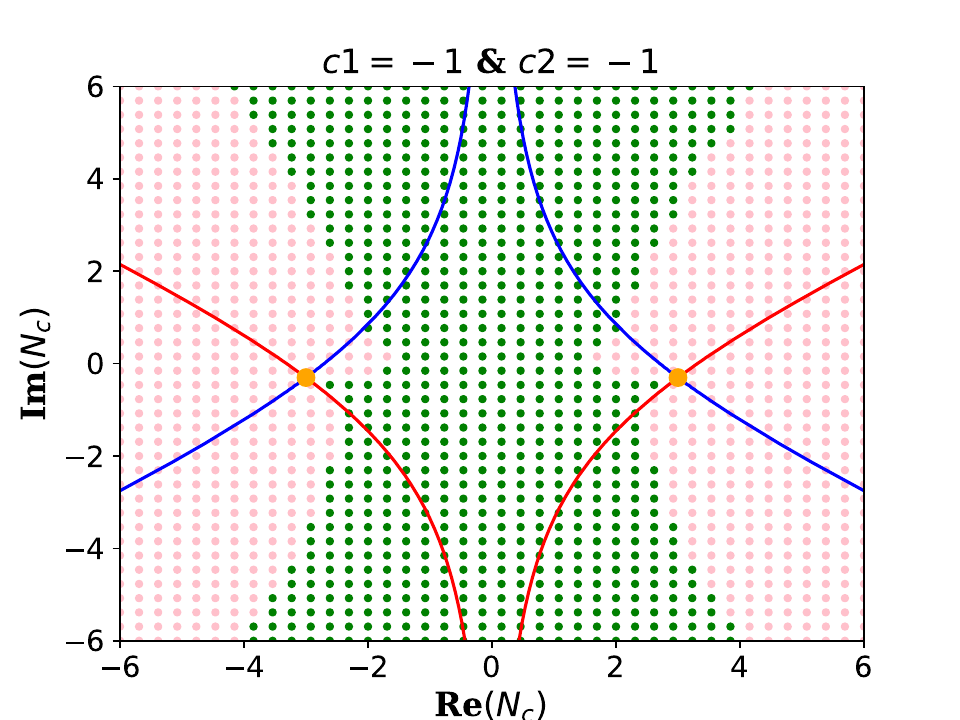}
}
\subfigure[ ]{
    \includegraphics[trim={0 0 1.3cm 0cm}, clip,scale=0.5]{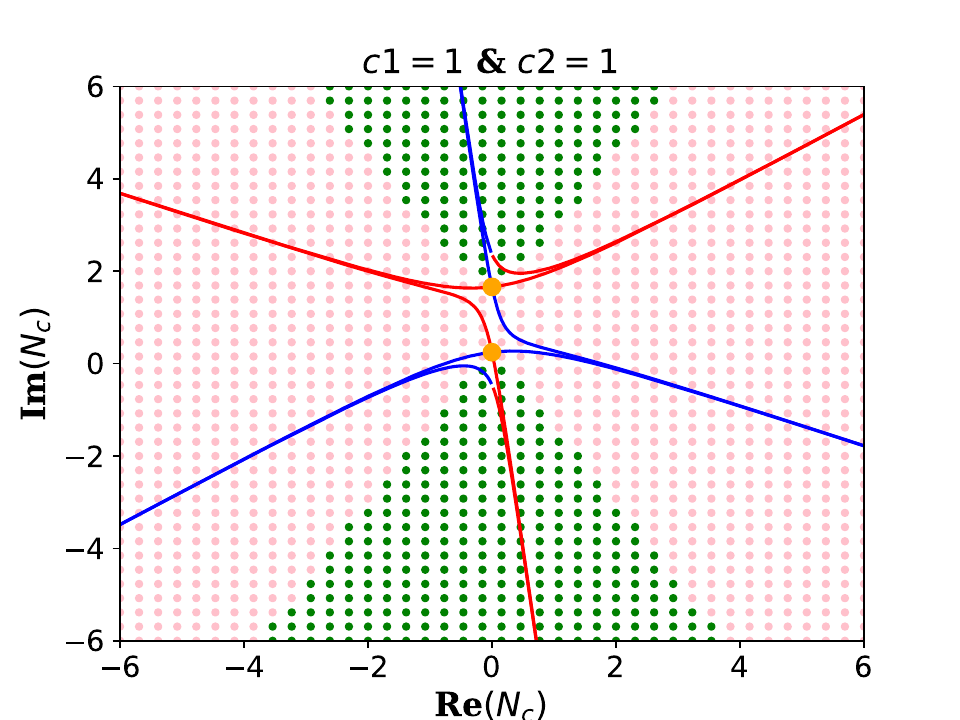}
}
\subfigure[]{
    \includegraphics[trim={0 0 1.3cm 0cm}, clip,scale=0.5]{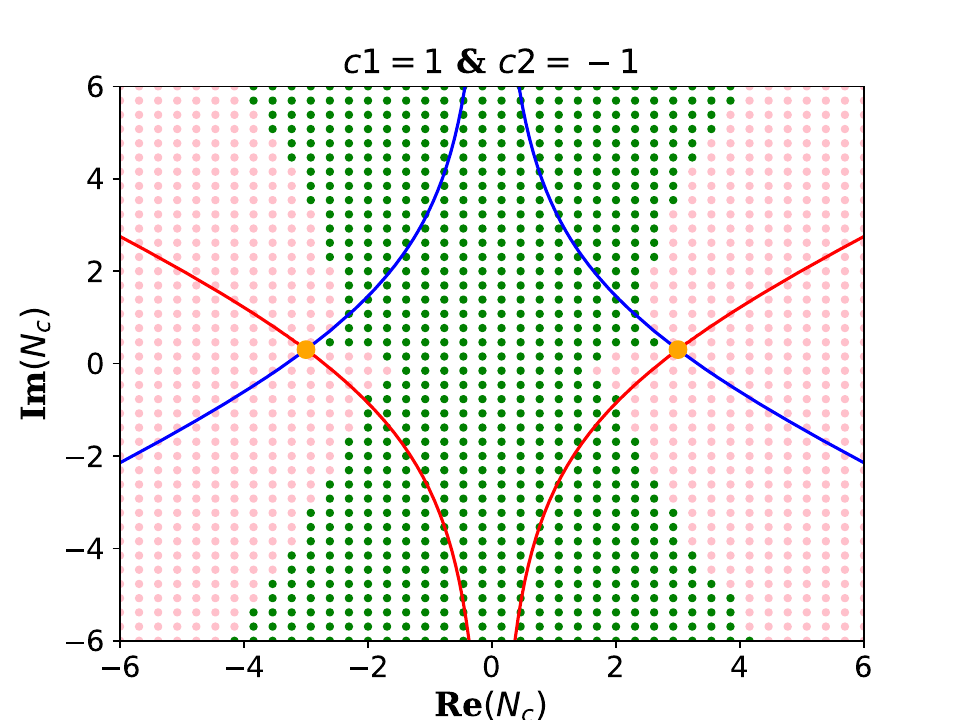}
}
\caption{ Illustration of the numerical results obtained from the KSW analysis. The region marked by pink dots corresponds to configurations for which no complex-time path exists connecting the boundary data while satisfying the KSW bound, and is therefore excluded. In contrast, the light-green region denotes configurations admitting at least one such path and hence satisfies the KSW criterion. The blue lines represents the steepest
ascents and the red lines corresponds to the steepest descents of one-loop exponent. (a) and (c) show the numerical KSW analysis for the saddle points 
$N_{\rm nb}^{(-1,c_3)}$ and 
$N_{\rm nb}^{(1,c_3)}$ respectively, in the regime 
$q_f <3k/\Lam$. Similarly, (b) and (d) depict the KSW analysis for
$N_{\rm dS}^{(-1,c_3)}$ and
$N_{\rm dS}^{(1,c_3)}$ respectively, with
$q_f > 3k/\Lambda$. For this
numerical illustration we set the parameters: $k=1, \Lam=3, $ and $c=1/16$.}
\label{fig:KSW_}
\end{figure}
The numerical implementation of the KSW criterion indicates that both the near no-boundary and near de Sitter saddle points are admissible, see fig. (\ref{fig:KSW_}). Consequently, all of these saddle configurations contribute to the gravitational path integral.


\section{Generalized zeta regularization of infinite series}
\label{Gen_zeta_reg}

In this appendix, we analyze the infinite series appearing in the 
gravitational and ghost parts of the effective action and compute 
their finite zeta--regularized contributions. For real parameters 
$x$ and $y$, the relevant series has the form
\bea
\label{Sum_series}
S(a)=\sum_{l=2}^{\infty}(l+x)(l+y)\text{log}(l+a).
\eea
where $\mathrm{Re}[a]>-2$. The above infinite series is divergent and 
contains both logarithmic and polynomial ultraviolet divergences. The 
complete divergence structure may be analyzed using the integral 
representation
$
\log(y)=\int_0^\infty \frac{ds}{s}\left(e^{-s}-e^{-sy}\right),
$
from which the divergent behavior arises in the small-$s$ region 
\cite{Ailiga:2024wdx}. We now describe a regularization prescription that 
extracts the finite renormalized contribution of the series by 
systematically separating the divergent terms. Using the identity
$
\ln(a)=-\left.\frac{\partial}{\partial s}(a^{-s})\right|_{s\to0^+},
$
we can rewrite the renormalized form of Eq.~(\ref{Sum_series}) as
\bea
\label{renorm_sum}
S^{\text{ren}}(a)=-\sum_{l=2}^{\infty}(l+x)(l+y)\frac{\partial}{\partial s}\left[(l+a)^{-s}\right]\Bigg|_{s=0^+},
\eea
Interchanging the summation and differentiation, we get
\bea
\label{renorm_form}
S^{\text{ren}}(a)=-\frac{\partial}{\partial s}\sum_{l=2}^{\infty}\frac{(l+x)(l+y)}{(l+a)^{s}}\Bigg|_{s=0^+},
\eea
To evaluate it, we rewrite each term in the summation as
\bea
\label{frac_form}
\frac{(l+x)(l+y)}{(l+a)^{s}}=\frac{1}{(l+a)^{s-2}}+\frac{x+y-2a}{(l+a)^{s-1}}+\frac{(a-x)(a-y)}{(l+a)^{s}},
\eea
The individual terms appearing in the above expression may be summed 
straightforwardly using the Hurwitz zeta function $\zeta_H(s,a)$.
\bea
\label{Hurwitz_zeta}
\zeta_H(s,a)=\sum_{l=0}^{\infty}\frac{1}{(l+a)^{s}}\, .
\eea
Upon evaluating the infinite sums and taking the derivative at 
$s=0^+$, the finite Hurwitz-zeta regularized contribution is found to be
\bea
\label{renorm_final}
S^{\text{ren}}(a)=-\zeta_H'(-2,a+2)+(2a-x-y)\zeta_H'(-1,a+2)-(a-x)(a-y)\zeta_H'(0,a+2)\, .
\eea
where prime denotes derivative with respect to $s$ variable. 
%
\section{Quantum Corrected lapse action: exact expressions}
\label{app:qcLapse_ex}

In this appendix, we write down the exact expression for the quantum-corrected transition amplitude for different final sizes of the Universe ($q_f$). This consists of three cases:$q_f=3k/\Lam$, $q_f<3k/\Lam$ and $q_f>3k/\Lam$. Quantum corrected lapse action generally is given below respectively.

\subsection{$q_f>3k/\Lambda$}
\label{app:subs_qf>lam}
For the case of $q_f>3k/\Lam$ the quantum corrected lapse action at the near no-boundary saddle is given by the following:
\bea
\label{gen_config_lapse_act}
&&
\mathcal{A}_{\rm fin}(N_{\rm nb}^{(c_1,c_3)})=-\frac{V_3}{36\pi\text{G}\Lam}\left[-\frac{27 ic_1k^{3/2}}{\sqrt{2}}\sqrt{1+s}\left(2-s\right)+c_3\left(-9k+3q_f\Lam\right)^{3/2}\right] \notag \\
&&
-\frac{1153}{360}i\hbar(c\Lam)- \frac{2}{3}i\hbar \ \text{ln}q_f - i\hbar\Biggl[\frac{191}{240}\text{ln}\left(\frac{1-\tau_1^{\text{nb}}}{1+\tau_1^{\text{nb}}}\right) - \zeta_{H}'(-2,a+2) \notag \\
&&
+2(a-1)\zeta_{H}'(-1,a+2)-(a+1)(a-3)\zeta_{H}'(0,a+2)  + \zeta_{H}'(-2,3)+\zeta_{H}'(-2,4)   \\
&&
- 2\zeta_{H}'(-1,4)- \frac{529}{240}\text{ln}(3) + \frac{2}{3}\text{ln}\left(\frac{\Lambda}{16 k^2}\right) + \frac{7}{2}\text{ln}\left(\frac{\pi}{2}\right)\Biggr]+\mathcal{O}(N_c-N_{\text{nb}}), \notag \\
&&
\, \, \text{where}, \  a\equiv1+\tau_1^{\text{nb}}\, \text{and} \, \, s\equiv\sqrt{1-\frac{16c\Lam}{3}} \notag.
\eea
%
\subsection{$q_f=3k/\Lambda$} 
\label{app:subs_qf_eq_lam}
In the case of $q_f=3k/\Lam$, the two no-boundary saddles merge, leading to a degenerate situation. In such a scenario, only one saddle contributes and the expression for the finite lapse function will be
\bea
\label{eq:finite_A_qf_equals_3k_by_Lam}
&&
\mathcal{A}_{\rm fin}(N_{nb}^{(c_1,c_3)})=\frac{V_3}{8\pi\text{G}\Lam}\left[i c_1 3\sqrt{2}k^{3/2}\sqrt{1+s}\left(2-s\right)\right]-\frac{1153}{240}i\hbar(c\Lam) -\frac{191}{240}i\hbar \  \text{ln}\left(\frac{3k}{\Lambda}\right)\notag \\
&&
-\frac{4}{3}i\hbar \ \text{ln}\left[\frac{1}{2\Lambda}\left(1+\sqrt{1-\frac{16c\Lam}{3}}\right)\right]+i\hbar\Biggl[4\zeta_{H}'(0,3)+\zeta_{H}'(-2,4)- 2\zeta_{H}'(-1,4)  \\
&&
- \frac{529}{240}\text{ln}(3) + \frac{2}{3}\text{ln}\left(\frac{\Lambda}{16 k^2}\right) + \frac{7}{2}\text{ln}\left(\frac{\pi}{2}\right)\Biggr]+\mathcal{O}(N_c-N_{\text{nb}}) \notag.
\eea
%
\subsection{$q_f< 3k/\Lambda$}
\label{app:subs_qf<lam}

For the case of $q_f<3k/\Lam$, only one saddle $N_{\rm sad}^{(c_1,c_2, 1)}$ is relevant. In the limit $q_f - q(0) \rightarrow 0^{+} $, one needs to add the counterterm mentioned in eq. (\ref{eq:ct_nnb}) to cancel the contact divergence appearing in the eq. (\ref{eq:QT_Nc_act_h_divST}). By doing that, we get

\beq
\label{eq:A_fin_at_qf_less_3k_by_Lam}
\mathcal{A}_{\rm fin}\bigl(N_{nb}^{(c_1,1)}\bigr)\Biggr\rvert_{\substack{q_f- q(0)\to 0^{+} \\ c \to 0}} \approx
\frac{4 i\hbar}{3} \ln\left(\frac{c_1}{2\sqrt{k}}\right) + \frac{117 i\hbar}{80} \ln\left(\frac{\Lam}{4 k}\right),
\quad \text{where} \ S_{\text{grav}}^{(\bar{q})}\Biggr\rvert_{\substack{q_f- q(0)\to 0^{+} \\ c \to 0}}=0.
\eeq



%

\begin{thebibliography}{99}
\bibitem{Gibbons:1976ue}
G.~W.~Gibbons and S.~W.~Hawking,
``Action Integrals and Partition Functions in Quantum Gravity,''
Phys. Rev. D \textbf{15} (1977), 2752-2756
doi:10.1103/PhysRevD.15.2752

\bibitem{Brown:1992bq}
J.~D.~Brown and J.~W.~York, Jr.,
``The Microcanonical functional integral. 1. The Gravitational field,''
Phys. Rev. D \textbf{47} (1993), 1420-1431
doi:10.1103/PhysRevD.47.1420
[arXiv:gr-qc/9209014 [gr-qc]].

\bibitem{Krishnan:2016mcj}
C.~Krishnan and A.~Raju,
``A Neumann Boundary Term for Gravity,''
Mod. Phys. Lett. A \textbf{32} (2017) no.14, 1750077
doi:10.1142/S0217732317500778
[arXiv:1605.01603 [hep-th]].

\bibitem{Krishnan:2017bte}
C.~Krishnan, S.~Maheshwari and P.~N.~Bala Subramanian,
``Robin Gravity,''
J. Phys. Conf. Ser. \textbf{883} (2017) no.1, 012011
doi:10.1088/1742-6596/883/1/012011
[arXiv:1702.01429 [gr-qc]].

\bibitem{Krishnan:2016tqj}
C.~Krishnan, K.~V.~P.~Kumar and A.~Raju,
``An alternative path integral for quantum gravity,''
JHEP \textbf{10} (2016), 043
doi:10.1007/JHEP10(2016)043
[arXiv:1609.04719 [hep-th]].

\bibitem{Hartle:1983ai}
J.~B.~Hartle and S.~W.~Hawking,
``Wave Function of the Universe,''
Phys. Rev. D \textbf{28} (1983), 2960-2975
doi:10.1103/PhysRevD.28.2960

\bibitem{Vilenkin:1987kf}
A.~Vilenkin,
``Quantum Cosmology and the Initial State of the Universe,''
Phys. Rev. D \textbf{37} (1988), 888
doi:10.1103/PhysRevD.37.888

\bibitem{Betzios:2024oli}
P.~Betzios and O.~Papadoulaki,
``Inflationary Cosmology from Anti-de Sitter Wormholes,''
Phys. Rev. Lett. \textbf{133} (2024) no.2, 021501
doi:10.1103/PhysRevLett.133.021501
[arXiv:2403.17046 [hep-th]].

\bibitem{DiTucci:2019bui}
A.~Di Tucci, J.~L.~Lehners and L.~Sberna,
``No-boundary prescriptions in Lorentzian quantum cosmology,''
Phys. Rev. D \textbf{100} (2019) no.12, 123543
doi:10.1103/PhysRevD.100.123543
[arXiv:1911.06701 [hep-th]].

\bibitem{Lehners:2018eeo}
J.~L.~Lehners,
``No smooth beginning for spacetime,''
Int. J. Mod. Phys. D \textbf{28} (2018) no.02, 1930005
doi:10.1142/9789811258251{\_}0005

\bibitem{Hawking:1982dh}
S.~W.~Hawking and D.~N.~Page,
``Thermodynamics of Black Holes in anti-De Sitter Space,''
Commun. Math. Phys. \textbf{87} (1983), 577
doi:10.1007/BF01208266

\bibitem{Saad:2018bqo}
P.~Saad, S.~H.~Shenker and D.~Stanford,
``A semiclassical ramp in SYK and in gravity,''
[arXiv:1806.06840 [hep-th]].

\bibitem{Saad:2019lba}
P.~Saad, S.~H.~Shenker and D.~Stanford,
``JT gravity as a matrix integral,''
[arXiv:1903.11115 [hep-th]].

\bibitem{Benini:2015eyy}
F.~Benini, K.~Hristov and A.~Zaffaroni,
``Black hole microstates in AdS$_{4}$ from supersymmetric localization,''
JHEP \textbf{05} (2016), 054
doi:10.1007/JHEP05(2016)054
[arXiv:1511.04085 [hep-th]].

\bibitem{tHooft:1974toh}
G.~'t Hooft and M.~J.~G.~Veltman,
``One-loop divergencies in the theory of gravitation,''
Ann. Inst. H. Poincare Phys. Theor. A \textbf{20} (1974) no.1, 69-94
doi:10.1142/9789814539395{\_}0001

\bibitem{Deser:1974nb}
S.~Deser, H.~S.~Tsao and P.~van Nieuwenhuizen,
``Nonrenormalizability of Einstein Yang-Mills Interactions at the One Loop Level,''
Phys. Lett. B \textbf{50} (1974), 491-493
doi:10.1016/0370-2693(74)90268-8

\bibitem{Deser:1974cz}
S.~Deser and P.~van Nieuwenhuizen,
``One Loop Divergences of Quantized Einstein-Maxwell Fields,''
Phys. Rev. D \textbf{10} (1974), 401
doi:10.1103/PhysRevD.10.401

\bibitem{Goroff:1985sz}
M.~H.~Goroff and A.~Sagnotti,
``QUANTUM GRAVITY AT TWO LOOPS,''
Phys. Lett. B \textbf{160} (1985), 81-86
doi:10.1016/0370-2693(85)91470-4

\bibitem{Goroff:1985th}
M.~H.~Goroff and A.~Sagnotti,
``The Ultraviolet Behavior of Einstein Gravity,''
Nucl. Phys. B \textbf{266} (1986), 709-736
doi:10.1016/0550-3213(86)90193-8

\bibitem{vandeVen:1991gw}
A.~E.~M.~van de Ven,
``Two loop quantum gravity,''
Nucl. Phys. B \textbf{378} (1992), 309-366
doi:10.1016/0550-3213(92)90011-Y

\bibitem{Akhmedov:2013vka}
E.~T.~Akhmedov,
``Lecture notes on interacting quantum fields in de Sitter space,''
Int. J. Mod. Phys. D \textbf{23} (2014), 1430001
doi:10.1142/S0218271814300018
[arXiv:1309.2557 [hep-th]].

\bibitem{Akhmedov:2019cfd}
E.~T.~Akhmedov, U.~Moschella and F.~K.~Popov,
``Characters of different secular effects in various patches of de Sitter space,''
Phys. Rev. D \textbf{99} (2019) no.8, 086009
doi:10.1103/PhysRevD.99.086009
[arXiv:1901.07293 [hep-th]].

\bibitem{Akhmedov:2024npw}
E.~T.~Akhmedov, V.~I.~Lapushkin and D.~I.~Sadekov,
``Light fields in various patches of de~Sitter spacetime,''
Phys. Rev. D \textbf{111} (2025) no.12, 125015
doi:10.1103/k65j-1jn4
[arXiv:2411.11106 [hep-th]].

\bibitem{Miao:2024shs}
S.~P.~Miao, N.~C.~Tsamis and R.~P.~Woodard,
``Leading Logarithm Quantum Gravity,''
Universe \textbf{11} (2025) no.7, 223
doi:10.3390/universe11070223
[arXiv:2409.12003 [gr-qc]].

\bibitem{Starobinsky:1994bd}
A.~A.~Starobinsky and J.~Yokoyama,
``Equilibrium state of a selfinteracting scalar field in the De Sitter background,''
Phys. Rev. D \textbf{50} (1994), 6357-6368
doi:10.1103/PhysRevD.50.6357
[arXiv:astro-ph/9407016 [astro-ph]].

\bibitem{Konstantinidis:2016nio}
G.~Konstantinidis, R.~Mahajan and E.~Shaghoulian,
``Late-time Structure of the Bunch-Davies FRW Wavefunction,''
JHEP \textbf{10} (2016), 103
doi:10.1007/JHEP10(2016)103
[arXiv:1608.06163 [hep-th]].

\bibitem{Percacci:2015wwa}
R.~Percacci and G.~P.~Vacca,
``Search of scaling solutions in scalar-tensor gravity,''
Eur. Phys. J. C \textbf{75} (2015) no.5, 188
doi:10.1140/epjc/s10052-015-3410-0
[arXiv:1501.00888 [hep-th]].

\bibitem{Mallik:2026iai}
S.~Mallik and G.~Narain,
``IR behaviour of one-loop complex $\mathbb{R}\times S^3$ saddles,''
[arXiv:2604.21736 [hep-th]].

\bibitem{Brizuela:2023vmb}
D.~Brizuela and M.~de Cesare,
``Generalized boundary conditions in closed cosmologies,''
Phys. Rev. D \textbf{107} (2023) no.10, 104054
doi:10.1103/PhysRevD.107.104054
[arXiv:2303.04007 [gr-qc]].

\bibitem{Ailiga:2024wdx}
M.~Ailiga, S.~Mallik and G.~Narain,
``Boundary choices and one-loop complex gravitational path integral,''
Phys. Rev. D \textbf{111} (2025) no.12, 123538
doi:10.1103/ytxr-4x1d
[arXiv:2410.19724 [gr-qc]].

\bibitem{Barvinsky:1992dz}
A.~O.~Barvinsky, A.~Y.~Kamenshchik and I.~P.~Karmazin,
``One loop quantum cosmology: Zeta function technique for the Hartle-Hawking wave function of the universe,''
Annals Phys. \textbf{219} (1992), 201-242
doi:10.1016/0003-4916(92)90347-O

\bibitem{Pawlowski:2018ixd}
J.~M.~Pawlowski, M.~Reichert, C.~Wetterich and M.~Yamada,
``Higgs scalar potential in asymptotically safe quantum gravity,''
Phys. Rev. D \textbf{99} (2019) no.8, 086010
doi:10.1103/PhysRevD.99.086010
[arXiv:1811.11706 [hep-th]].

\bibitem{Ohta:2021bkc}
N.~Ohta and M.~Yamada,
``Higgs scalar potential coupled to gravity in the exponential parametrization in arbitrary gauge,''
Phys. Rev. D \textbf{105} (2022) no.2, 026013
doi:10.1103/PhysRevD.105.026013
[arXiv:2110.08594 [hep-th]].

\bibitem{Feldbrugge:2017kzv}
J.~Feldbrugge, J.~L.~Lehners and N.~Turok,
``Lorentzian Quantum Cosmology,''
Phys. Rev. D \textbf{95} (2017) no.10, 103508
doi:10.1103/PhysRevD.95.103508
[arXiv:1703.02076 [hep-th]].

\bibitem{Narain:2022msz}
G.~Narain,
``Surprises in Lorentzian path-integral of Gauss-Bonnet gravity,''
JHEP \textbf{04} (2022), 153
doi:10.1007/JHEP04(2022)153
[arXiv:2203.05475 [gr-qc]].

\bibitem{Gerlach:1978gy}
U.~H.~Gerlach and U.~K.~Sengupta,
``Homogeneous Collapsing Star: Tensor and Vector Harmonics for Matter and Field Asymmetries,''
Phys. Rev. D \textbf{18} (1978), 1773-1784
doi:10.1103/PhysRevD.18.1773

\bibitem{Ailiga:2023wzl}
M.~Ailiga, S.~Mallik and G.~Narain,
``Lorentzian Robin Universe,''
JHEP \textbf{01} (2024), 124
doi:10.1007/JHEP01(2024)124
[arXiv:2308.01310 [gr-qc]].

\bibitem{Yamada:2026lzg}
M.~Yamada,
``A Small-Throat Boundary Condition for the Tunneling Wave Function of the Universe,''
[arXiv:2607.08815 [gr-qc]].

\bibitem{Witten:2010cx}
E.~Witten,
``Analytic Continuation Of Chern-Simons Theory,''
AMS/IP Stud. Adv. Math. \textbf{50} (2011), 347-446
[arXiv:1001.2933 [hep-th]].

\bibitem{Witten:2010zr}
E.~Witten,
``A New Look At The Path Integral Of Quantum Mechanics,''
[arXiv:1009.6032 [hep-th]].

\bibitem{Basar:2013eka}
G.~Basar, G.~V.~Dunne and M.~Unsal,
``Resurgence theory, ghost-instantons, and analytic continuation of path integrals,''
JHEP \textbf{10} (2013), 041
doi:10.1007/JHEP10(2013)041
[arXiv:1308.1108 [hep-th]].

\bibitem{Tanizaki:2014xba}
Y.~Tanizaki and T.~Koike,
``Real-time Feynman path integral with Picard{\textendash}Lefschetz theory and its applications to quantum tunneling,''
Annals Phys. \textbf{351} (2014), 250-274
doi:10.1016/j.aop.2014.09.003
[arXiv:1406.2386 [math-ph]].

\bibitem{Narain:2021bff}
G.~Narain,
``On Gauss-bonnet gravity and boundary conditions in Lorentzian path-integral quantization,''
JHEP \textbf{05} (2021), 273
doi:10.1007/JHEP05(2021)273
[arXiv:2101.04644 [gr-qc]].

\bibitem{Ailiga:2024mmt}
M.~Ailiga, S.~Mallik and G.~Narain,
``Lorentzian Robin Universe of Gauss-Bonnet Gravity,''
Gen. Rel. Grav. \textbf{57} (2025) no.2, 29
doi:10.1007/s10714-025-03369-2
[arXiv:2407.16692 [gr-qc]].

\bibitem{Mahajan:2025bzo}
R.~Mahajan and K.~Singhi,
``A brief note on complex AdS-Schwarzschild black holes,''
JHEP \textbf{11} (2025), 164
doi:10.1007/JHEP11(2025)164
[arXiv:2509.08883 [hep-th]].

\bibitem{Singhi:2025rfy}
K.~Singhi,
``Complex Kerr-AdS black holes,''
JHEP \textbf{02} (2026), 138
doi:10.1007/JHEP02(2026)138
[arXiv:2510.01313 [hep-th]].

\bibitem{Ailiga:2025osa}
M.~Ailiga, S.~Mallik and G.~Narain,
``Complex saddles of charged-AdS gravitational partition function,''
JHEP \textbf{02} (2026), 054
doi:10.1007/JHEP02(2026)054
[arXiv:2510.25396 [hep-th]].

\bibitem{Kolanowski:2026gii}
M.~Kolanowski and D.~Marolf,
``How to tame your (black hole) saddles: Lessons from the Lorentzian Gravitational Path Integral,''
[arXiv:2603.24681 [hep-th]].

\bibitem{Barbon:2026tri}
J.~L.~F.~Barb{\'o}n and E.~Velasco-Aja,
``AdS Black Holes Are Short-Lived inside the Spectral Form Factor,''
[arXiv:2607.21704 [hep-th]].

\bibitem{Kolanowski:2026ylt}
M.~Kolanowski, D.~Marolf, Z.~Y.~Wang and W.~Zheng,
``Superconformal indices and black hole saddles,''
[arXiv:2608.07660 [hep-th]].

\bibitem{Nishimura:2015loa}
H.~Nishimura, M.~C.~Ogilvie and K.~Pangeni,
``Complex saddle points in finite-density QCD,''
PoS \textbf{CPOD2014} (2015), 068
doi:10.22323/1.217.0068
[arXiv:1503.00060 [hep-ph]].

\bibitem{Tanizaki:2015pua}
Y.~Tanizaki, H.~Nishimura and K.~Kashiwa,
``Evading the sign problem in the mean-field approximation through Lefschetz-thimble path integral,''
Phys. Rev. D \textbf{91} (2015) no.10, 101701
doi:10.1103/PhysRevD.91.101701
[arXiv:1504.02979 [hep-th]].

\bibitem{Tanizaki:2015gcv}
Y.~Tanizaki, H.~Nishimura and K.~Kashiwa,
``Lefschetz-thimble path integral for solving the mean-field sign problem,''
PoS \textbf{LATTICE2015} (2016), 282
doi:10.22323/1.251.0282

\bibitem{Ailiga:2025fny}
M.~Ailiga, S.~Mallik and G.~Narain,
``Resolving degeneracies in complex {\ensuremath{\mathbb{R}}} {\texttimes} S$^{3}$ and {\ensuremath{\theta}}-KSW,''
JHEP \textbf{02} (2026), 249
doi:10.1007/JHEP02(2026)249
[arXiv:2507.10537 [hep-th]].

\bibitem{Honda:2024aro}
M.~Honda, H.~Matsui, K.~Okabayashi and T.~Terada,
``Resurgence in Lorentzian quantum cosmology: No-boundary saddles and resummation of quantum gravity corrections around tunneling saddle points,''
Phys. Rev. D \textbf{110} (2024) no.8, 8
doi:10.1103/PhysRevD.110.083508
[arXiv:2402.09981 [gr-qc]].

\bibitem{Feldbrugge:2017fcc}
J.~Feldbrugge, J.~L.~Lehners and N.~Turok,
``No smooth beginning for spacetime,''
Phys. Rev. Lett. \textbf{119} (2017) no.17, 171301
doi:10.1103/PhysRevLett.119.171301
[arXiv:1705.00192 [hep-th]].

\bibitem{Vilenkin:2018oja}
A.~Vilenkin and M.~Yamada,
``Tunneling wave function of the universe II: the backreaction problem,''
Phys. Rev. D \textbf{99} (2019) no.6, 066010
doi:10.1103/PhysRevD.99.066010
[arXiv:1812.08084 [gr-qc]].

\bibitem{DiazDorronsoro:2018wro}
J.~Diaz Dorronsoro, J.~J.~Halliwell, J.~B.~Hartle, T.~Hertog, O.~Janssen and Y.~Vreys,
``Damped perturbations in the no-boundary state,''
Phys. Rev. Lett. \textbf{121} (2018) no.8, 081302
doi:10.1103/PhysRevLett.121.081302
[arXiv:1804.01102 [gr-qc]].

\bibitem{Matsui:2024bfn}
H.~Matsui,
``No smooth spacetime: Exploring primordial perturbations in Lorentzian quantum cosmology,''
Phys. Rev. D \textbf{110} (2024) no.2, 023503
doi:10.1103/PhysRevD.110.023503
[arXiv:2404.18609 [gr-qc]].

\bibitem{Mondal:2025dsc}
V.~Mondal and S.~Mukohyama,
``Anisotropic quantum universe in Ho{\v{r}}ava-Lifshitz gravity,''
Phys. Rev. D \textbf{112} (2025) no.8, 083542
doi:10.1103/6wp2-b17k
[arXiv:2505.16266 [gr-qc]].

\bibitem{Lehners:2023yrj}
J.~L.~Lehners,
``Review of the no-boundary wave function,''
Phys. Rept. \textbf{1022} (2023), 1-82
doi:10.1016/j.physrep.2023.06.002
[arXiv:2303.08802 [hep-th]].

\bibitem{Abbott:1980hw}
L.~F.~Abbott,
``The Background Field Method Beyond One Loop,''
Nucl. Phys. B \textbf{185} (1981), 189-203
doi:10.1016/0550-3213(81)90371-0

\bibitem{Yamada:2025rld}
M.~Yamada,
``New regularization scheme for the wave function of the Universe in the Lorentzian path integral,''
Phys. Rev. D \textbf{113} (2026) no.6, 063547
doi:10.1103/z59k-v3l8
[arXiv:2511.09621 [gr-qc]].

\bibitem{Khandekar:1986ib}
D.~C.~Khandekar and S.~V.~Lawande,
``Feynman Path Integrals: Some Exact Results and Applications,''
Phys. Rept. \textbf{137} (1986), 115-229
doi:10.1016/0370-1573(86)90029-3

\bibitem{Sabir:1991PrJ}
M.~Sabir and S.~Rajagopalan,
``Path integral analysis of harmonic oscillators with time-dependent mass,''
Pramana \textbf{37} (1991) no.3, 253-260
doi:10.1007/BF02847479

\bibitem{McKane:1995vp}
A.~J.~McKane and M.~B.~Tarlie,
``Regularization of functional determinants using boundary perturbations,''
J. Phys. A \textbf{28} (1995), 6931-6942
doi:10.1088/0305-4470/28/23/032
[arXiv:cond-mat/9509126 [cond-mat]].

\bibitem{Kleinert:1998rz}
H.~Kleinert and A.~Chervyakov,
``Simple explicit formulas for Gaussian path integrals with time dependent frequencies,''
Phys. Lett. A \textbf{245} (1998), 345-357
doi:10.1016/S0375-9601(98)00380-6
[arXiv:quant-ph/9803016 [quant-ph]].

\bibitem{Kleinert:1999Cher}
H.~Kleinert and A.~Chervyakov,
``Functional determinants from Wronski Green functions,''
Journal of Mathematical Physics \textbf{40} (1999) no.11, 6044–6051
doi:10.1063/1.533069
[arXiv:physics/9712048]

\bibitem{Matsui:2025guo}
H.~Matsui,
``Functional determinants for constrained path integrals in minisuperspace Jackiw-Teitelboim gravity,''
Phys. Rev. D \textbf{113} (2026) no.6, 6
doi:10.1103/jspm-lswp
[arXiv:2512.21549 [gr-qc]].

\bibitem{DiTucci:2019dji}
A.~Di Tucci and J.~L.~Lehners,
``No-Boundary Proposal as a Path Integral with Robin Boundary Conditions,''
Phys. Rev. Lett. \textbf{122} (2019) no.20, 201302
doi:10.1103/PhysRevLett.122.201302
[arXiv:1903.06757 [hep-th]].

\bibitem{Polchinski:1988ua}
J.~Polchinski,
``The phase of the sum over spheres,''
Phys. Lett. B \textbf{219} (1989), 251-257
doi:10.1016/0370-2693(89)90387-0

\bibitem{Mazur:1989ch}
P.~O.~Mazur and E.~Mottola,
``ABSENCE OF PHASE IN THE SUM OVER SPHERES,''
LA-UR-89-2118.

\bibitem{Baumgart:2019clc}
M.~Baumgart and R.~Sundrum,
``De Sitter Diagrammar and the Resummation of Time,''
JHEP \textbf{07} (2020), 119
doi:10.1007/JHEP07(2020)119
[arXiv:1912.09502 [hep-th]].

\bibitem{Honda:2023unh}
M.~Honda, R.~Jinno, L.~Pinol and K.~Tokeshi,
``Borel resummation of secular divergences in stochastic inflation,''
JHEP \textbf{08} (2023), 060
doi:10.1007/JHEP08(2023)060
[arXiv:2304.02592 [hep-th]].

\bibitem{Cespedes:2023aal}
S.~C{\'e}spedes, A.~C.~Davis and D.~G.~Wang,
``On the IR divergences in de Sitter space: loops, resummation and the semi-classical wavefunction,''
JHEP \textbf{04} (2024), 004
doi:10.1007/JHEP04(2024)004
[arXiv:2311.17990 [hep-th]].

\bibitem{Huenupi:2024ztu}
J.~Huenupi, E.~Hughes, G.~A.~Palma and S.~Sypsas,
``Note on loop resummation in de Sitter spacetime with the wave function of the universe approach,''
Phys. Rev. D \textbf{112} (2025) no.4, 043543
doi:10.1103/nxjc-8qw2
[arXiv:2412.01891 [hep-th]].

\bibitem{Kamenshchik:2024ybm}
A.~Kamenshchik and P.~Petriakova,
``IR finite correlation functions in de Sitter space, a smooth massless limit, and an autonomous equation,''
JHEP \textbf{04} (2025), 127
[erratum: JHEP \textbf{06} (2026), 201]
doi:10.1007/JHEP04(2025)127
[arXiv:2410.16226 [hep-th]].

\bibitem{Huenupi:2024ksc}
J.~Huenupi, E.~Hughes, G.~A.~Palma and S.~Sypsas,
``Regularizing infrared divergences in de Sitter spacetime: Loops, dimensional regularization, and cutoffs,''
Phys. Rev. D \textbf{110} (2024) no.12, 123536
doi:10.1103/PhysRevD.110.123536
[arXiv:2406.07610 [hep-th]].

\bibitem{Xue:2011hm}
W.~Xue, K.~Dasgupta and R.~Brandenberger,
``Cosmological UV/IR Divergences and de-Sitter Spacetime,''
Phys. Rev. D \textbf{83} (2011), 083520
doi:10.1103/PhysRevD.83.083520
[arXiv:1103.0285 [hep-th]].

\bibitem{Narain:2018rif}
G.~Narain and N.~Kajuri,
``Non-local scalar field on deSitter and its infrared behaviour,''
Phys. Lett. B \textbf{791} (2019), 143-148
doi:10.1016/j.physletb.2019.02.030
[arXiv:1812.00947 [hep-th]].

\bibitem{Kamenshchik:2025ses}
A.~Kamenshchik and P.~Petriakova,
``From the Fokker-Planck equation to perturbative QFT{\textquoteright}s results in de Sitter space,''
JHEP \textbf{08} (2025), 063
doi:10.1007/JHEP08(2025)063
[arXiv:2504.20646 [hep-th]].

\bibitem{Abdalla:2026mxn}
A.~I.~Abdalla, S.~Antonini, R.~Bousso, L.~V.~Iliesiu, A.~Levine and A.~Shahbazi-Moghaddam,
``Consistent Evaluation of the No-Boundary Proposal,''
[arXiv:2602.02682 [hep-th]].

\bibitem{Lehners:2026tej}
J.~L.~Lehners,
``Connected Geometries in Gravitational Path Integrals, and Normalized No-Boundary Probabilities,''
[arXiv:2610.08596 [hep-th]].

\bibitem{Lehners:2021mah}
J.~L.~Lehners,
``Allowable complex metrics in minisuperspace quantum cosmology,''
Phys. Rev. D \textbf{105} (2022) no.2, 026022
doi:10.1103/PhysRevD.105.026022
[arXiv:2111.07816 [hep-th]].

\bibitem{Jonas:2022uqb}
C.~Jonas, J.~L.~Lehners and J.~Quintin,
``Uses of complex metrics in cosmology,''
JHEP \textbf{08} (2022), 284
doi:10.1007/JHEP08(2022)284
[arXiv:2205.15332 [hep-th]].

\bibitem{michel1901maniere}
Michel Petrovitch, M. Sur une manière d'étendre le théorème de la moyenne aux équations différentielles du premier ordre. 
{\em Mathematische Annalen}. 
\textbf{54} pp. 417-436 (1901)

\bibitem{Hertog:2023vot}
T.~Hertog, O.~Janssen and J.~Karlsson,
``Kontsevich-Segal Criterion in the No-Boundary State Constrains Inflation,''
Phys. Rev. Lett. \textbf{131} (2023) no.19, 191501
doi:10.1103/PhysRevLett.131.191501
[arXiv:2305.15440 [hep-th]].

\bibitem{Lehners:2023pcn}
J.~L.~Lehners and J.~Quintin,
``A small Universe,''
Phys. Lett. B \textbf{850} (2024), 138488
doi:10.1016/j.physletb.2024.138488
[arXiv:2309.03272 [hep-th]].

\bibitem{BenettiGenolini:2026raa}
P.~Benetti Genolini, O.~Janssen and S.~Murthy,
``Allowable complex metrics and the gravitational index of AdS$_5$ black holes,''
J. Phys. A \textbf{59} (2026) no.32, 325401
doi:10.1088/1751-8121/ae94c1
[arXiv:2601.23197 [hep-th]].

\bibitem{Krishna:2026rma}
V.~Krishna and F.~Larsen,
``Allowable complex Black Holes in the Euclidean Gravitational Path Integral,''
JHEP \textbf{06} (2026), 026
doi:10.1007/JHEP06(2026)026
[arXiv:2602.05979 [hep-th]].
\end{thebibliography}
\end{document}